\documentclass[fleqn,usenatbib]{mnras}

\usepackage{newtxtext,newtxmath}

\newcommand{\ba}[0]{\textsc{BeAtlas}~} 

\usepackage[utf8]{inputenc}

\usepackage[T1]{fontenc}

\DeclareRobustCommand{\VAN}[3]{#2}
\let\VANthebibliography\thebibliography
\def\thebibliography{\DeclareRobustCommand{\VAN}[3]{##3}\VANthebibliography}

\usepackage{graphicx}	
\usepackage{amsmath}	

\usepackage{newtxmath}

\title[Recent Outbursts and Properties of HD\,6226]{Outbursts and stellar properties of the classical Be star HD\,6226}

\author[N. D. Richardson et al.]{Noel D. Richardson,$^{1}$\thanks{E-mail: noel.richardson@erau.edu}
Olivier Thizy,$^{2}$
Jon E. Bjorkman,$^{3}$ 
Alex Carciofi,$^{4}$
\newauthor Amanda C. Rubio,$^{4}$
Joshua D. Thomas,$^{5}$
Karen S. Bjorkman,$^{3}$ 
Jonathan Labadie-Bartz,$^{4}$
\newauthor Matheus Genaro,$^{4}$
John P. Wisniewski,$^{6}$ 
Luqian Wang,$^{7}$
Douglas R. Gies,$^{7}$
\newauthor S. Drew Chojnowski,$^{8}$
Andrea Daly,$^{1}$
Thompson Edwards,$^{5}$
Carlie Fowler,$^{5}$
\newauthor Allison D. Gullingsrud,$^{3}$
Nolan Habel,$^{3}$
David J. James,$^{9}$
Emily Kehoe,$^{5}$
\newauthor Heidi Kuchta,$^{3}$
Alexis Lane,$^{1}$
Anatoly Miroshnichenko$^{10,11,12}$
Ashish Mishra,$^{3}$
\newauthor Herbert Pablo,$^{13}$
Maurice Peploski,$^{5}$
Joshua Pepper,$^{14}$
Joseph E. Rodriguez,$^{15}$
\newauthor Robert J. Siverd,$^{16}$
Keivan G. Stassun,$^{17}$
Daniel J. Stevens,$^{18,19,20}$
Jesica L. Trucks,$^{21}$ 
James Windsor,$^{22}$
\newauthor Mackenna Wood,$^{5}$
\'{E}tienne Bertrand,$^{2}$
Jean-Jacques Broussat,$^{2}$
Erik Bryssinck,$^{2}$
\newauthor Christian Buil,$^{2}$
St\'{e}phane Charbonnel,$^{2}$
Arnold de Bruin,$^{2}$
Joe Daglen,$^{2}$
Valerie Desnoux,$^{2}$
\newauthor James Dull,$^{23}$
Olivier Garde,$^{2}$
Keith Graham,$^{2}$
Kevin Gurney,$^{2}$
Alun Halsey,$^{2}$
\newauthor Patrik Fosanelli,$^{2}$
Joan Guarro Fl\'{o},$^{2}$
Franck Houpert,$^{2}$
Foster James,$^{2}$
Christian Kreider,$^{2}$
\newauthor Robin Leadbeater,$^{2}$
Tim Lester,$^{2}$
Dong Li,$^{2}$ 
Alain Maetz,$^{2}$
Albert Stiewing,$^{2}$
\newauthor Peter Somogyi,$^{2}$
Jean-No\"{e}l Terry,$^{2}$
St\'{e}phane Ubaud,$^{2}$
and Ulrich Waldschlaeger$^{2}$\\
$^{1}$Department of Physics and Astronomy, Embry-Riddle Aeronautical University, 3700 Willow Creek Road, Prescott, AZ 86301, USA\\
$^{2}$ARAS\\
$^{3}$Ritter Observatory, Department of Physics and Astronomy, The University of Toledo, Toledo, OH 43606-3390, USA\\
$^{4}$Instituto de Astronomia, Geof\'isica e Ci\^encias Atmosf\'ericas, Universidade de S\~ao Paulo, Rua do Mat\~ao 1226\\
$^{5}$Department of Physics, Clarkson University, 8 Clarkson Ave, Potsdam, New York 13699, USA\\
$^{6}$Homer L. Dodge Department of Physics and Astronomy, University of Oklahoma, 440 W. Brooks St., Norman, OK 73019, USA\\
$^{7}$Center for High Angular Resolution Astronomy and Department of Physics and Astronomy, Georgia State University, P.O. Box 5060, Atlanta, GA 30302-5060, USA\\
$^{8}$Apache Point Observatory and New Mexico State University, PO Box 59, Sunspot, NM 88349-0059, USA\\
$^{9}$ASTRAVEO, LLC, PO Box 1668, Gloucester, MA 01931, USA \\
$^{10}$Department of Physics and Astronomy, University of North Carolina at Greensboro, P.O. Box 26170, Greensboro, NC 27402-6170, USA\\
$^{11}$ Main Astronomical Observatory of the Russian Academy of Sciences, Saint-Petersburg, Pulkovskoye Shosse, 65, 196140, Russia\\
$^{12}$ Fesenkov Astrophysical Institute, Observatory, 23, Almaty, 050020, Kazakhstan\\
$^{13}$ American Association of Variable Star Observers, 49 Bay State Road, Cambridge, MA 02138, USA\\
$^{14}$ Department of Physics, Lehigh University, 16 Memorial Drive East, Bethlehem, PA 18015, USA \\
$^{15}$ Center for Astrophysics \textbar \ Harvard \& Smithsonian, 60 Garden St, Cambridge, MA 02138, USA \\
$^{16}$ Gemini Observatory/NSF’s NOIRLab, 670 N. A’ohoku Place, Hilo, HI, 96720, USA \\
$^{17}$ Department of Physics and Astronomy, Vanderbilt University, Nashville, TN 37235, USA \\
$^{18}$ Department of Astronomy \& Astrophysics, The Pennsylvania State University, 525 Davey Lab, University Park, PA 16802, USA \\
$^{19}$ Eberly Research Fellow  \\
$^{20}$ Center for Exoplanets and Habitable Worlds, The Pennsylvania State University, 525 Davey Lab, University Park, PA 16802, USA\\
$^{21}$Abrams Planetarium, Michigan State University, East Lansing, MI, USA\\
$^{22}$Department of Astronomy and Planetary Science, Northern Arizona University, Flagstaff, AZ, 86011, USA \\
$^{23}$The College of Idaho, Caldwell, ID 83605, USA \\
}

\date{Accepted XXX. Received YYY; in original form ZZZ}

\pubyear{2021}

\begin{document}
\label{firstpage}
\pagerange{\pageref{firstpage}--\pageref{lastpage}}
\maketitle
\clearpage

\begin{abstract}

The bright and understudied classical Be star HD\,6226 has exhibited multiple outbursts in the last several years during which the star grew a viscous decretion disk. We analyze 659 optical spectra of the system collected from 2017-2020, along with a UV spectrum from the {\it Hubble Space Telescope} and high cadence photometry from both {\it TESS} and the KELT survey. We find that the star has a spectral type of B2.5IIIe, with a rotation rate of $74$\% of critical. The star is nearly pole-on with an inclination of $13\fdg4$. We confirm the spectroscopic pulsational properties previously reported, and report on three photometric oscillations from KELT photometry. The outbursting behavior is studied with equivalent width measurements of H$\alpha$ and H$\beta$, and the variations in both of these can be quantitatively explained with two frequencies through a Fourier analysis. One of the frequencies for the emission outbursts is equal to the difference between two photometric oscillations, linking these pulsation modes to the mass ejection mechanism for some outbursts. During the {\it TESS} observation time period of 2019 October 7 to 2019 November 2, the star was building a disk. With a large dataset of H$\alpha$ and H$\beta$ spectroscopy, we are able to determine the timescales of dissipation in both of these lines, similar to past work on Be stars that has been done with optical photometry. HD\,6226 is an ideal target with which to study the Be disk-evolution given its apparent periodic nature, allowing for targeted observations with other facilities in the future.

\end{abstract}




\section{Introduction}

\label{sec:intro}

Classical Be stars are non-supergiant B stars that are rotating at {nearly} critical rates \citep{2003A&A...411..229R}. The rapid rotation, perhaps coupled with other triggers such as pulsations or binarity, causes the formation of equatorial decretion disks. These disks are seen in hydrogen emission lines (the `e' in Be) and sometimes other spectral lines, free-free excess emission in the infrared and radio, and through polarization of the continuum light \citep[see review in ][]{2013A&ARv..21...69R}. Long-baseline interferometry has verified the disk hypothesis and shown that the disk material rotates in a Keplerian manner \citep[e.g., ][]{2012ApJ...744...19K}. The Be phenomenon tends to peak near the early-type B stars \citep{2005ApJS..161..118M} although this may be in part due to the observational biases where the observables of later type Be stars are harder to observe, \citep[e.g.,][]{2018A&A...609A.108S, 2017MNRAS.464.3071V}. 

Amongst the Be stars, the under-studied HD\,6226 (V442 And, BD+46 245, SAO\,36891) is the target of this study. It was noticed by \citet{1998A&A...330..222B} that the photometry of the star showed semi-regular outbursts, and HD\,6226 was one of the first Be stars predicted to be a Be star from its photometric behavior. They suspected that the star was a previously unrecognized Be star. The outbursts differed from one cycle to another, but a time-series of the times of maximum light hinted at quasi-regular outbursts for the system. In the year 2000, H$\alpha$ emission was detected that verified the Be nature of the system, indicating an outburst period of $\sim600$ d \citep{Bozic}. During most of this photometric cycle, the star was observed in a low-state with little variability followed by outbursts of variable duration and amplitude. 

\citet{Bozic} also analyzed a collection of 35 spectra of HD\,6226 taken over a span of $\sim6$ years to further constrain the nature of the variability. The emission profile for H$\alpha$ was typically double-peaked during the outbursts, although the strength varied greatly. The outbursts can last for weeks or months, but each outburst has a different behavior compared to previous emission episodes. During most of the photometric cycle, the star is in a low-state with little variability followed by outbursts of variable duration and amplitude. However, \citet{2019RNAAS...3..151G} showed a blue spectrum obtained shortly after the end of the data analyzed by \citet{Bozic}, taken during a time period where there should be no emission with their derived 600-d period, and this spectrum shows clear H$\beta$ emission from a disk. Similarly, \citet{2005Ap&SS.296..179S} found H$\alpha$ emission at a time when it was predicted to be absent by \citet{Bozic}.

In addition to these disk variations, the star's photospheric absorption profiles for He I and metal lines shifted in radial velocity with a strict period of 2.615 d \citep{Bozic}. The period was phased accurately over the course of $>6$ years, and recent spectra (Section 4) verify the periodicity in the position of the line center. This period is similar to that of the pulsations observed in the slowly pulsating B stars \citep[see, e.g., ][]{2010aste.book.....A}. \citet{Bozic} considered this to be a rotational property of the star, despite its similarity to pulsational behavior of other Be stars.

It has been shown that the variability of HD\,6226 bears similarities to the outbursting Be star $\omega$\,CMa (28\,CMa) {and $\mu$ Centaurii. $\mu$ Cen was the first Be star where pulsations were a reasonable explanation for the disk outbursts} \citep{1998A&A...333..125R}. In the cases of $\omega$ CMa \citep{2018MNRAS.479.2214G} {and $\lambda$ Eri} \citep{1998A&A...330..631M}, the outbursts have been able to be predicted through photometry, and the stars show line profile variations similar to those observed in the spectrum of HD\,6226 \citep{Bozic}.
One specific outburst of $\omega$\,CMa was monitored with photometry during the dissipation of the disk. Models of the dissipation using HDUST and the viscous decretion disk model \citep[][]{1991MNRAS.250..432L,2011IAUS..272..325C} provided the first ever measurement of the viscosity parameter in the disk, $\alpha=1$, and a disk mass injection rate of $\dot{M}=(3.5\pm1.3) \times 10^{-8}\, M_\odot \,{\rm yr}^{-1}$ \citep{2012ApJ...744L..15C}. The mass injection rate is also an order of magnitude above the mass loss rate of the wind. More work was done on $\omega$ CMa by \citet{2018MNRAS.479.2214G}, who analyzed the photometric behavior over a period of nearly 40 years. They found that each outburst lasts for many years and that the outbursts do not seem to be periodic. They also found that the viscosity parameter, $\alpha$ varies from 0.1 to 1.0. 

In 2017 August, amateur spectroscopists noted new H$\alpha$ emission in spectra of HD\,6226. This prompted us to begin an intensive observing campaign that incorporates a large collection of optical spectra, {\it TESS} photometry, and the first high-quality ultraviolet spectrum of the star. These observations are detailed in Section 2. 
In Section 3, we discuss the fundamental parameters of the star from the optical and UV spectroscopy, along with a model fit to a disk-free optical spectrum and the ultraviolet spectral energy distribution. Section 4 details the pulsational mode seen in the optical spectroscopy as well as photometric oscillations from ground-based photometry. Section 5 details the star's outbursting properties, while Section 6 presents our analysis of the {\it TESS} data that were serendipitously taken while the star was building a disk. Section 7 discusses our results and we conclude this study in Section 8.

\section{Observations}


\subsection{Optical Spectroscopy}
We collected many optical spectra that are archived in the  BeSS archive\footnote{http://basebe.obspm.fr} \citep{2011AJ....142..149N}. These spectra had moderate to high spectral resolving power ($R \sim 7,000 - 25,000$), and spanned the time between 2017 January and 2020 March. While some studies \citep[e.g., ][]{2010ApJ...709.1306W} have found offsets between differing resolutions of spectroscopy for measurements such as the equivalent width of lines, we find that these data provide measurements similar to each other {if the resolving power was greater than $\sim5000$}.  Thus, we rejected lower resolution spectra from the BeSS archive for this reason. All spectroscopic data were reduced using standard techniques, with corrections for bias, dark, and flats being taken into account. We note that more than half of the spectra we obtained from the BeSS archive were taken with the eshel spectrograph\footnote{https://www.shelyak.com/description-eshel/?lang=en}, which is an echelle spectrograph manufactured by the Shelyak company. We utilize these spectra for studying the pulsational properties of the star as they have uniform resolution and wavelength coverage for different observers and locations. 

In addition to the BeSS archive, we obtained some data from facilities within our collaboration, including the 1-m telescope at Ritter Observatory (Toledo, Ohio, USA), the 0.8-m telescope at the Three College Observatory (North Carolina, USA), the 0.3-m telescope Reynolds Observatory (Potsdam, New York, USA), and the 3.5-m telescope at Apache Point Observatory (Sunspot, New Mexico, USA). All these data were reduced with standard techniques employing use of biases, darks (when needed), flats, and comparison lamp spectra. All spectroscopic data utilized in this project are summarized in Table \ref{specobs-table} that lists the observer/observatory with the telescope aperture, the number of spectra and the resolving power of these data, as well as the HJDs of the time-series.

All spectroscopic data were normalized to the local continuum consistently around features of interest, and H$\alpha$ spectra were corrected for telluric contamination by means of division by a telluric template from the atlas given by \citet{2000vnia.book.....H}. We analyzed 659 H$\alpha$ spectra, 363 H$\beta$ spectra, 372 spectra covering the He I 4471 and Mg II 4481 lines, 374 spectra covering the He I 4713 line, 362 spectra covering the He I 4921 line, and 369 spectra covering the He I 5876 line.

\begin{table*}
\begin{minipage}{160mm}
\centering
\caption{List of contributed spectra, in order of number of spectra. All spectra cover H$\alpha$, with the datasets highlighted in bold including lines such as He I $\lambda$ 5876, H$\beta$, and Mg II $\lambda$ 4481 that were used in further analysis. 
\label{specobs-table}}
\begin{tabular}{l c c c c c c}
\hline \hline
Observer	&	 $N_{\rm spectra}$ 	&	HJD$_{\rm first}$	&	 HJD$_{\rm end}$	&	Spectrograph 	&	Aperture    & $R$	\\
 or Observatory       	&	 	                &	$-2450000$      	&	$-2450000$	        &                   &   (m)         &   	\\ \hline
{\bf Thizy}	&	245	&	7987.5452	&	8794.3721	&	eshel	&	0.28	&	11,000	\\
{\bf Guarro Flo}	&	76	&	7993.4328	&	8880.3072	&	eshel	&	0.40	&	9,000	\\
{\bf Charbonnel}	&	57	&	8315.5591	&	8721.6023	&	eshel	&	0.50	&	11,000	\\
Li	&	54	&	8000.157	&	8771.038	&	LHIRES	&	0.28	&	18,000	\\
Stiewing	&	49	&	8770.7	&	8906.6573	&	LHIRES	&	0.28	&	18,000	\\
Houpert	&	38	&	8057.4159	&	8847.3196	&	LHIRES	&	0.28	&	15,000	\\
{\bf Garde}	&	37	&	7802.2705	&	8865.2392	&	eshel	&	0.40	&	11,000	\\
Daglen	&	20	&	7777.6348	&	8855.6228	&	LHIRES	&	0.35	&	17,000	\\
Bryssinck	&	20	&	8881.3059	&	8881.3059	&	LHIRES	&	0.28	&	9,000	\\
Ritter Obs.	&	16	&	8176.5764	&	8208.5546	&	Low-Dispersion Spectrograph	&	1.00	&	4,300	\\
Kreider	&	15	&	8199.3425	&	8887.3368	&	LHIRES	&	0.43	&	11,000	\\
 Fosanelli 	&	9	&	7811.3601	&	8351.4384	&	LHIRES	&	0.30	&	12,000	\\
Desnoux	&	9	&	7956.5319	&	8441.2923	&	LHIRES	&	0.23	&	14,000	\\
Clarkson	&	9	&	7998.7549	&	8176.5078	&	LHIRES	&	0.30	&		\\
Bertrand	&	7	&	8051.48	&	8402.4273	&	LHIRES	&	0.20	&	15,000	\\
Foster	&	7	&	8486.7017	&	8860.6762	&	LHIRES	&	0.43	&	13,000	\\
Leadbeater	&	7	&	8769.3878	&	8816.2459	&	LHIRES	&	0.28	&	5,400	\\
de Bruin	&	6	&	8017.4124	&	8884.2778	&	L200 	&	0.28	&	6,200	\\
Graham	&	5	&	8104.5126	&	8775.6183	&	LHIRES	&	0.30	&	13,000	\\
{\bf Buil}	&	4	&	7988.4413	&	8074.4039	&	eshel	&	0.25	&	11,000	\\
Buil	&	4	&	8011.5289	&	8767.4588	&	VHIRES	&	0.25	&	39,000	\\
Terry	&	4	&	8018.4112	&	8352.4009	&	LHIRES	&	0.28	&	13,000	\\
Maetz	&	4	&	8718.4373	&	8871.3256	&	LHIRES	&	0.21	&	12,000	\\
Halsey	&	4	&	8784.5018	&	8868.3617	&	LHIRES	&	0.23	&	12,000	\\
{\bf Three College Obs.}	&	3	&	8134.5832	&	8193.5283	&	eshel	&	0.81	&	11,000	\\
{ de Bruin}	&	3	&	8818.4372	&	8865.307	&	L200	&	0.28	&	18,000	\\
Lester	&	1	&	\multicolumn{2}{c}{7770.5334}	&	LHIRES	&	0.31	&	8,700	\\
Apache Point Obs.	&	1	&	\multicolumn{2}{c}{8175.5618}	&	ARCES	&	3.50	&	38,000	\\
Ubaud	&	1	&	\multicolumn{2}{c}{8358.6368}	&	LHIRES	&	0.28	&	19,000	\\
Broussat	&	1	&	\multicolumn{2}{c}{8373.4894}	&	LHIRES	&	0.23	&	13,000	\\
Gurney	&	1	&	\multicolumn{2}{c}{8441.4463}	&	LHIRES	&	0.25	&	13,000	\\
Somogyi	&	1	&	\multicolumn{2}{c}{8818.3911}	&	LHIRES	&	0.30	&	18,000	\\
Dumont 	&	1	&	\multicolumn{2}{c}{8848.3183}	&	LHIRES	&	0.50	&	24,000	\\
Dull	&	1	&	\multicolumn{2}{c}{8880.3072}	&	LHIRES	&	0.35	&	16,000	\\

\hline \hline
\end{tabular}
\end{minipage}
\end{table*}



\subsection{Ultraviolet Spectroscopy with HST}

We obtained an ultraviolet spectrum of HD\,6226 with the {\it Hubble Space Telescope} and the Space Telescope Imaging Spectrograph (STIS) on 2018 March 5 over the course of two consecutive orbits. We used the E230M and E140M gratings to cover the range of 1150--3000\AA\ with a spectral resolving power of at least 40,000 \citep{1998ApJ...492L..83K}. The 0.2$\times$0.05ND slit was used to obtain the spectra as the object is too bright to observe without a neutral density filter. The first orbit consisted of two E230M observations, with exposures of 572 s with the central wavelength of 2707\AA, and an exposure of 862 s centered at 1978\AA. The second orbit used the E140M mode, centered at 1425\AA, and we exposed for 2919 s. Overall, the data are of high quality, with a typical signal-to-noise ratio of 15--20 per resolution element for most of the spectrum. The FUV channel had a higher signal-to-noise ratio than that of the NUV channels, allowing for stellar classification using the methods of \citet{1991ApJ...369..515R, 1993NASRP1312.....R}.

\subsection{KELT and KWS photometry}

The Kilodegree Extremely Little Telescope (KELT) is a
photometric survey using two small-aperture (42 mm) widefield ($26^\circ \times 26^\circ$) telescopes, with a northern location at Winer
Observatory in Arizona in the United States, and a southern location at the South African Astronomical Observatory
near Sutherland, South Africa. The KELT survey covers over
60\% of the sky and is designed to detect transiting exoplanets
around stars in the magnitude range 8 < V < 10, but obtains photometry for stars between 7 < V < 13. \citet{2017AJ....153..252L} included HD\,6226 (V442 And) in their survey of light curves of Be stars. We utilized these data, as well as some newer observations. These data are reduced with the same techniques as \citet{2017AJ....153..252L}, using the techniques described in detail by \citet{2007PASP..119..923P, 2012PASP..124..230P}. The KELT pipeline produces  both raw and detrended versions of the light curves. The detrending process is built into the KELT pipeline, and uses the Trend Filtering Algorithm (TFA) \citep{2005MNRAS.356..557K} as implemented in the VARTOOLS package described by \citet{2012ascl.soft08016H}. We used these data in our pulsational analysis after removing outliers and subtracting long-term trends on scales of more than 6-days. 

We also use some lower-precision photometry taken with the Kamogata/Kiso/Kyoto Wide-field Survey (KWS)\footnote{http://kws.cetus-net.org/~maehara/VSdata.py}. These data were taken in $V$-band, and have a typical error listed of 0.008 mag, but subsequent exposures show differences of $\sim 0.02$ mag. This discrepency in the errors and subsequent exposure measurements led us to only use these data for qualitative purposes in Section 7.

\subsection{{\it TESS} photometry}

The {\it Transiting Exoplanet Survey Satellite} \citep[{\it TESS}, ][]{2016SPIE.9904E..2BR} observed HD\,6226 (TIC 196501216) for 25.02 days with a two-minute cadence. The resulting light curve contains 18,012 points taken during the satellite's Sector 17 campaign (2019 October 7--2019 November 2). We utilized the light curve reduced and available from MAST for the pulsational analysis, which included data taken with a 2-minute cadence. We also use the available full-frame images, which contains long-term trends for the analysis of the disk and represent a 30-minute cadence for the time-series.

\citet{2016SPIE.9913E..3EJ} describe the {\it TESS} science processing operations centre (SPOC) pipeline that produces the 2-min light curves. We downloaded these 2-min data and used ‘PDCSAP’ flux, which generally produces smooth light curves without any large drifts in flux from systematic effects on the detector. We performed no additional processing on these light curves. We also note that these data were taken during a time when the star had a changing disk, and the long-term trends that are likely associated with the outbursts \citep[e.g.,][]{Bozic} are therefore removed from this light curve, but variability on the time scales of a few days remains intact.

\section{Fundamental Parameters of HD 6226}

\begin{figure*}
\begin{center} 
\includegraphics[angle=90, width=18cm]{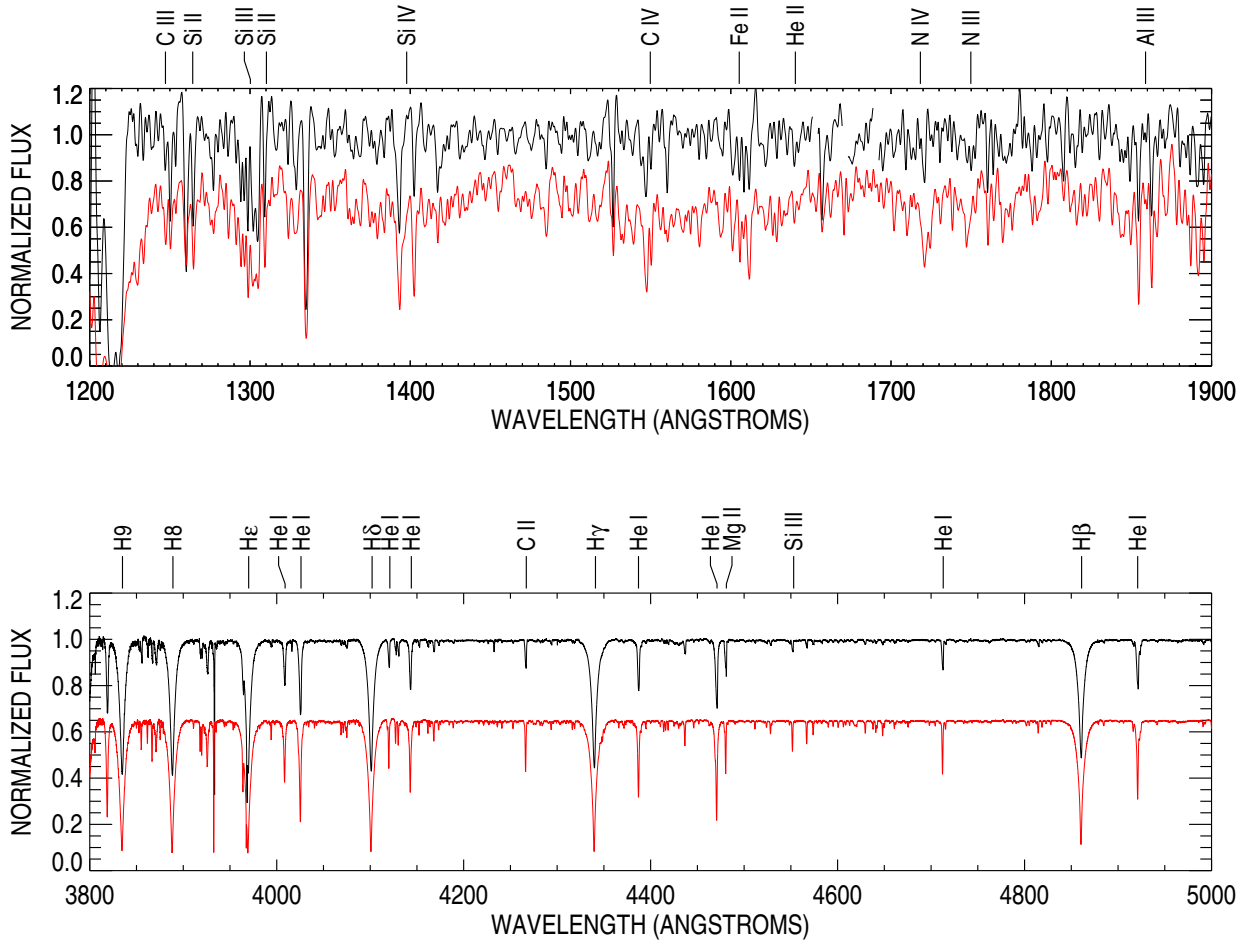}
\end{center} 
\caption{A comparison of the spectrum of HD\,6226 (black spectrum, top) and the spectral standard star $\pi^2 Cyg$ (red spectrum, bottom). The UV spectrum of HD\,6226 is from our {\it HST} program, while the spectrum of $\pi^2$ Cyg was obtained from the {\it IUE} archive. The {\it HST} data were smoothed to the resolution of the {\it IUE} data. The optical spectra were taken with the ESPaDOnS spectrograph on the CFHT, and from the archive. The optical spectrum showed no signature of H$\alpha$ or any other Balmer emission. Observations have been smoothed for clarity in the figure. We shifted the spectrum of $\pi^2$ Cyg to match the radial velocity of HD\,6226.} 
\label{classification} 
\end{figure*} 

\subsection{Spectral classification in the ultraviolet and optical}

One of our goals for the ultraviolet observations with {\it HST} was to ensure that the stellar parameters were correctly measured, including the spectral type. This is especially important for Be stars, where the disk can contaminate the flux and change line ratios, or introduce emission in some lines, altering the measured spectral type. We anticipate that even though Balmer emission was present at the time of our {\it HST} observations, the disk has minimal effect at these shorter wavelengths.

Ultraviolet spectroscopic classification for B stars is described in detail by \citet{1991ApJ...369..515R, 1993NASRP1312.....R}. Ultraviolet classification is done for B stars in the range of 1200--1900\AA. The main lines used in UV classification, include 
He II $\lambda$ 1640, 
C III $\lambda$ 1247,
C IV $\lambda\lambda$ 1548,1551, 
N III $\lambda\lambda$ 1748,1751, 
N IV $\lambda$ 1718, 
Al III $\lambda\lambda$ 1854,1863, 
Si II $\lambda \lambda$ 1264,1310, 
Si III $\lambda$ 1300, 
Si IV $\lambda\lambda$ 1393,1402, 
and Fe II $\lambda\lambda$ 1600--1610.
All these lines are included in the {\it HST} spectrum, with sufficient signal-to-noise to permit easy detection of lines if they are present. 

This portion of the ultraviolet spectrum between 1200--1900\AA, shown in Figure \ref{classification}, has some important features. He II $\lambda$ 1640 is very weak, with 
N III $\lambda$ 1748--1751 and N IV $\lambda$ 1718 absent. Si II $\lambda 1264$ is stronger than C III $\lambda 1247$, with Si II $\lambda$ 1310 weaker than Si III $\lambda$ 1300. These features are consistent with a B2.5 spectral type. The presence of C IV $\lambda\lambda 1548, 1551$ in the spectrum indicates a luminosity class III. We also show in Fig.~\ref{classification} a comparison of the {\it HST} ultraviolet spectrum with that of the B2.5III standard star $\pi^2$ Cyg from the work of \citet{1968ApJS...17..371L} and \citet{1993NASRP1312.....R}. The spectrum of HD\,6226 appears to be very similar to that of the standard star with the major classification lines matching quite well.

\begin{figure*}
 \centering
 \includegraphics[width=\linewidth]{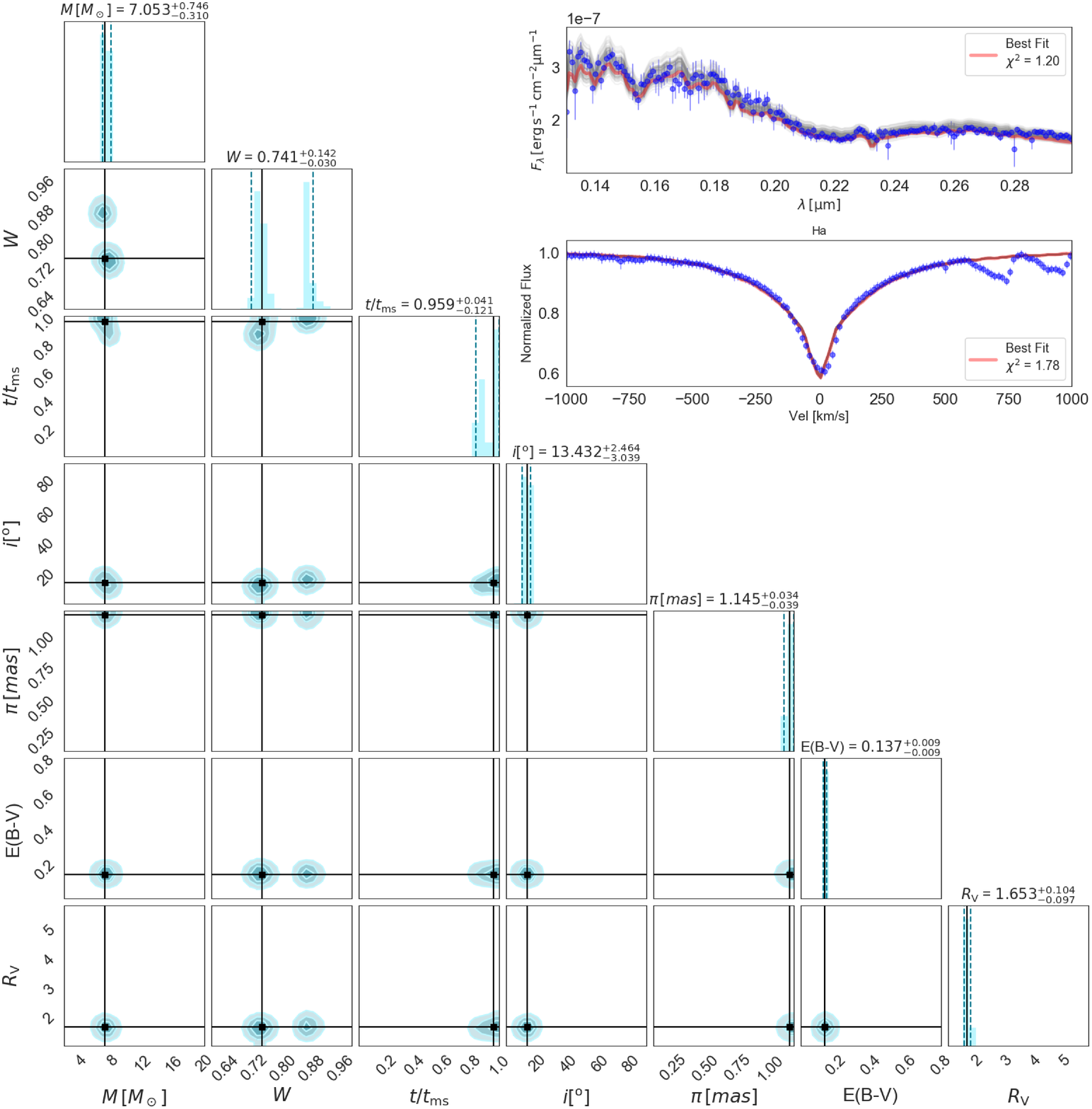}
  \caption{Corner plot for a MCMC simulation of HD\,6226. The main diagonal shows the probability density function for each fitted parameter (see text for details). The vertical dashed lines mark the limits of the HDR, while the solid black line is the mean value in the region, and the most likely value for each parameter. The off-diagonal plots indicate the two-on-two correlations of the probabilities.
  The inset shows the UV SED (upper part) and H$\alpha$ spectrum (bottom part). The data are shown as the blue points with error bars. The red line represents the model obtained from the most likely values for each of the parameters, and the grey lines in the SED panel are the models proposed in the last position of each of the walkers in the \ba simulation. Note that the model is only used to calculate the hydrogen lines, so the C II doublet is not modeled in the shown profile.} 
  \label{fig:hd6226_corner}
\end{figure*}

The MiMeS survey \citep{2016MNRAS.456....2W} observed a large number of O and B stars searching for magnetic fields through optical spectropolarimetry with the ESPaDOnS spectrograph on the CFHT\footnote{The ESPaDOnS data were downloaded from the Polarbase archive \citep{1997MNRAS.291..658D,2014PASP..126..469P}, http://polarbase.irap.omp.eu/}. HD\,6226 was included in the sample, and the spectrum from 2011 June 17 shows no obvious signs of Balmer emission (see Fig.~\ref{classification}). We assume that the star was disk-free at the time of the ESPaDOnS observation, although we have no photometric record around that time period. The Stokes $I$ spectrum from this observation was very high quality, with a signal to noise ratio of several hundred in the continuum. The ESPaDOnS archive also contained a spectrum of the standard star $\pi^2$ Cyg. The spectra of the two stars are indistinguishable in the Balmer lines. The He I and metal lines matched in ratio, but the lines of $\pi^2$ Cyg are narrower and deeper than the lines of HD\,6226. This is consistent with the previous observations of \citet{Bozic} showing that the projected rotational velocity $v \sin i$ of HD\,6226 is 70 km s$^{-1}$, while the results of \citet{2002ApJ...573..359A} show that $\pi^2$ Cyg has a $v \sin i$ of 35 km s$^{-1}$. While optical classification of Be stars can be difficult given possible disk contamination, it seems that the quiescent optical spectrum of HD\,6226 yields a classification that is consistent with that from the ultraviolet spectroscopy.

\subsection{The Ultraviolet SED and underlying parameters}

{The ultraviolet portion of the spectral energy distribution is not affected by the free-bound and free-free emission that causes the infrared excess observed in Be stars. Thus, the UV SED, as measured with our {\it HST} observations, and the optical line profiles in the absence of emission, can be used to determine the stellar parameters. We used the \ba Markov Chain Monte Carlo (MCMC) method, described in Mota et al.~(in prep) and \citet{mota2019}.  \ba consists of a grid of thousands of radiative transfer models for Be stars, calculated with the HDUST code \citep{2006ApJ...639.1081C, 2008ApJ...684.1374C}, that covers the range of stellar parameters relevant to the Be phenomenon (i.e., spectral subclasses O9 to A0). To explore the grid and sample the parameter space, the python implementation of the affine-invariant MCMC ensemble sampler of \citet{goodman2010}, \textsc{emcee}, is used. Using the Monte Carlo approach has the distinct advantage of providing us not just a best fit model, but rather a probability density function for each of the parameters, and the correlations between them, which makes a more statistically robust inference for the derived parameters.

The \ba simulation takes as input the grid of models, the observational data for the star and whatever prior information is available in the literature. As input data, we use the ultraviolet {\it HST} observation described in Section 2.2, and the archival ESPaDOnS H$\alpha$ line profile, which was made when the star had no signs of a disk in the optical Balmer profiles. As prior information, we use the Gaia DR2 parallax \citep{2018A&A...616A...1G}  of $\pi = 0.639 \pm 0.077$ mas. The stellar parameters covered in \ba are the mass, $M_\textrm{star}$, the inclination angle of the star's spin axis with respect to the line of sight, $i$, the rotation rate, $W = {v_{\rm rot} \over v_{\rm orb}}$, as defined in \citet{2013A&ARv..21...69R}, and the fractional age of the star in units of main sequence lifetime ($t/t_\textrm{MS}$). In addition to these, fitted parameters are the parallax, $\pi$, and two parameters that define the interstellar reddening, namely $E(B-V)$ and the total-to-selective extinction ratio, $R_V$. We note that the $E(B-V)$ and $R_V$ parameters could be slightly contaminated by circumstellar disk that was present at the time of the ultraviolet observations, based on our H$\alpha$ and H$\beta$ analysis (Section 5). The code also derives values for the equatorial and polar radii as well as the gravity darkening exponent $\beta$ \citep[originally defined by ][]{1924MNRAS..84..665V}. Other parameters needed for the \ba HDUST runs (e.g., $R_\textrm{star}$) are calculated using the Geneva grid of models for fast spinning stars \citep{georgy2013}.

The simulation was run with 300 walkers and 5000 steps, and had an acceptance fraction of 0.33, which is within the range defined by \citet{2013PASP..125..306F} for examining the parameter space that will describe the stellar parameters. The walkers represent the number of separate models exploring parameter space during the Monte Carlo run, whereas the steps are the number of iterations that each walker takes during the process. 
The results are shown in Fig.~\ref{fig:hd6226_corner} and the outcome is summarized in Table~\ref{tab:hd6226}. We define the uncertainty range in the determination of each parameter as the range of the highest density (or probability) region (HDR) of each probability density function (PDF) that contains 68$\%$ of the density. The most likely (or ``best fit'') value for each parameter is represented by the mean of the HDR.

Both the ultraviolet SED and the H$\alpha$ line profile are fit well by the models. The $E(B-V)$ parameter is quite well constrained by the $2200$\AA\ band associated with interstellar matter \citep{1989ApJ...345..245C}. The high uncertainties in $W$ are explained by the double-peaked shape of the PDF, which indicates that while the highest probability lies around the values of 0.74, but a much larger value for $W$ cannot be excluded. The star is seen at an almost pole-on orientation ($i= 13.4^{+2.5}_{-3.0}\circ$). The estimated value for $v \sin i =  57^{+42}_{-18}$ km s$^{-1}$ is in agreement with the value of 70 km s$^{-1}$ of \citet{Bozic}. 

The inferred parallax is about 50\% larger (about 4-$\sigma$), than the Gaia DR2 value. To investigate this issue we performed two other \ba simulations, one without the parallax prior, another two forcing the parallax to lie within $1 \sigma$ of the Gaia DR2 or EDR3 values.
In the first of these simulations, our results are very similar to those of our derived model using the {\it Gaia} parallax, with a slightly different mass (7.3$M_\odot$), and a similar goodness of fit. In the second model, the derived mass is then much larger (10.3 $M_\odot$), but the quality of the fit is much poorer, as the reduced $\chi^2$ is increased by roughly a factor of 3 for the UV SED, and about a factor of 1.5 for the H$\alpha$ line. Finally, in a third model, we also use the newer {\it Gaia} EDR3 parallax as a prior. This model suffered from unrealistic values of $W$ and $v\sin i$. Since we see that the value for $v\sin i$ must be higher than that of the spectral standard $\pi^2$ Cyg (Section 3.1), the derived value of 26 km s$^{-1}$ from the third model led us to adopt the model that has a free parallax and has a distance consistent with the {\it Gaia} DR2 measurement. The inconsistencies from these models seem to lie with the {\it Gaia} parallax, which may be explained with an unresolved binary companion in the system. If the orbital inclination of a binary companion is similar to the rotational inclination, $\sim 13^\circ$, then we suspect that any orbital-related photometric or spectroscopic variations would be too small to detect. 

We note that \citet{Bozic} did find a $\sim 25$ d period that could be associated with such a binary, but their data were too scarce to make a strong statement about the cause of that periodicity. Many Be stars are thought to have a stripped star companion, such as the case of $\phi$ Per \citep{1998ApJ...493..440G, 2015A&A...577A..51M} and several other Be stars \citep[see, e.g.,][and references therein]{2018ApJ...853..156W}. If the putative binary companion is a stripped star, then its luminosity would be much smaller than that of the Be star, and it would not influence our analysis of the UV spectrum. However, if the companion were a main-sequence star of comparable mass to HD\,6226, it could help reconcile the UV spectrum with the {\it Gaia} distance. The presence of a companion star may cause additional uncertainty in the derived parameters such the main-sequence lifetime. 

The proper motion of HD\,6226, combined with the high radial velocity of the star makes the star a runaway star according to \citet{2018MNRAS.477.5261B}.
It is situated well out of the plane of the Galaxy,
and the mean radial velocity indicates that it is
on its way back to the plane. We used the {\tt GalPy} package \citep{2015ApJS..216...29B} and the {\it Gaia} proper motion to 
to estimate the time of flight since ejection from
the disk.  Using a mean radial velocity of $-$55 km s$^{-1}$,
the result is a time of flight of 31 Myrs
which compares to the evolutionary age of 51 Myrs for a 7$M_\odot$ model from \citet{2013A&A...558A.103G}. It may
be a star that was spun up by binary mass transfer
and then ejected when the companion experienced
a supernova explosion.  Many OB runaways are indeed
rapid rotators that have experienced spin up.
In that case, HD\,6226 may have been rejuvenated by
mass transfer, and the current age estimate
corresponds to the time since mass transfer was
completed.

\begin{table} 
\centering 
\begin{tabular}{lll} 
\hline 
Parameter  & Value & Type \\ 
\hline 
$M\,[M_\odot]$& $7.05^{+0.75}_{-0.31}$ & Free \\ 
$W$& $0.74^{+0.14}_{-0.03}$ & Free \\ 
$t/t_\mathrm{ms}$& $0.96^{+0.04}_{-0.12}$ & Free \\ 
$i\,[\mathrm{^o}]$& $13.4^{+2.5}_{-3.0}$ & Free \\ 
$\pi\,[mas]$& $1.14^{+0.03}_{-0.04}$ & Free \\ 
E(B-V)& $0.14^{+0.01}_{-0.01}$ & Free \\ 
$R_\mathrm{V}$& $1.65^{+0.10}_{-0.10}$ & Free \\ 
$R_{\rm eq}/R_{\rm p}$ & $1.27^{+0.12}_{-0.02}$ & Derived  \\ 
$R_{\rm eq}\,[R_\odot]$ & $7.4^{+1.3}_{-1.1}$ & Derived  \\ 
$\log(L)\,[L_\odot]$ & $3.58^{+0.19}_{-0.17}$ & Derived  \\ 
$\log(g)\,[cgs]$    & $3.55^{+0.18}_{-0.17}$ & Derived  \\ 
$\beta$ & $0.18^{+0.00}_{-0.02}$ & Derived \\ 
$v \sin i\,\rm[km/s]$ & $57^{+42}_{-18}$ & Derived  \\ 
$T_{\rm eff}\,[K]$ & $17000^{+3500}_{-2600}$ & Derived  \\ 
\hline 
\end{tabular} 
\caption{Fitted and derived parameters for HD\,6226 using the \ba MCMC sampler.} \label{tab:hd6226}
\end{table} 
}

\section{Pulsational Properties}

\citet{Bozic} found that HD\,6226 had a spectroscopic variability signature with a period of 2.61507$\pm$0.00013 days, which they interpreted as the rotational period. They found some evidence of line profile variations with a period of $\sim 24-29$d in the He I $\lambda$6678 transition, but they did not have sufficient spectral cadence to document adequately such long timescale variability. 

\begin{figure} 
\begin{center} 
\includegraphics[angle=90, scale=0.35]{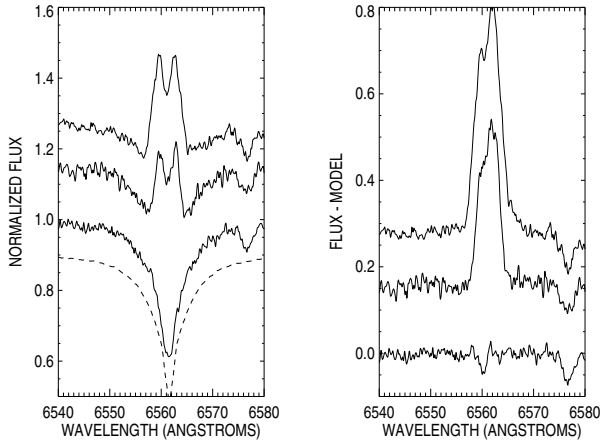}
\end{center} 
\caption{H$\alpha$ line profiles from the first well-observed outburst in our spectroscopic dataset, showing three representative profiles. The weak emission profile (pure absorption) on the bottom was taken on HJD 2458077.2, the moderate emission was taken during the decay phase on HJD 2458018.5, and the strong profile (top) was taken on HJD 2457999.5. The spectra are offset for clarity, and we show the model calculated in Section 3.2 as a dashed line. We subtracted the model from these profiles to obtain the profiles shown in the right panel and found the disk profiles to be single-peaked. Note that the subtracted model is only for the hydrogen line, leaving the weak C II $\lambda \lambda$ 6578,6583 absorption lines in the subtracted region. } 
\label{Ha-profiles} 
\end{figure}

\begin{figure*} 
\begin{center} 
\includegraphics[angle=0, width=8cm]{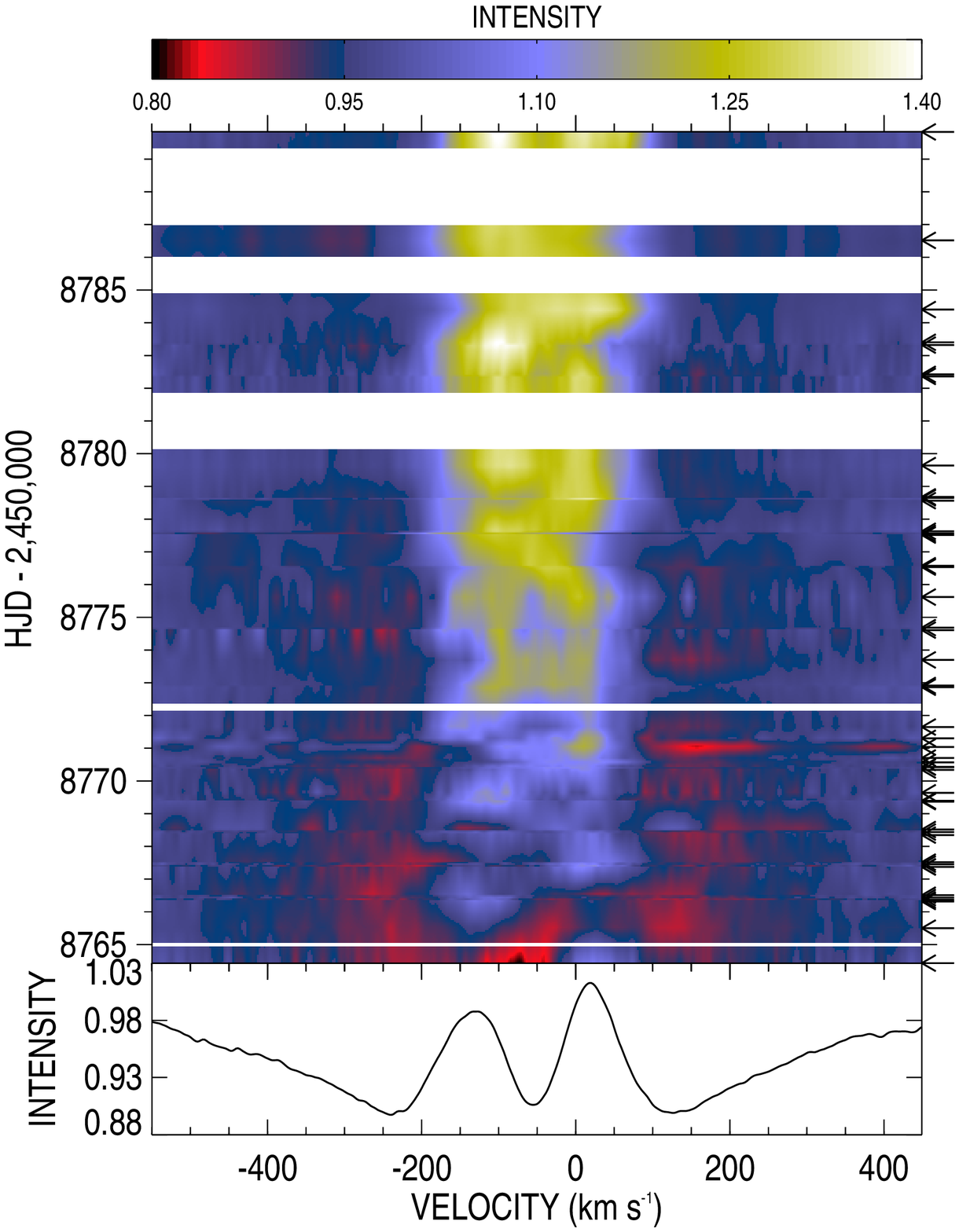}
\includegraphics[angle=0, width=8cm]{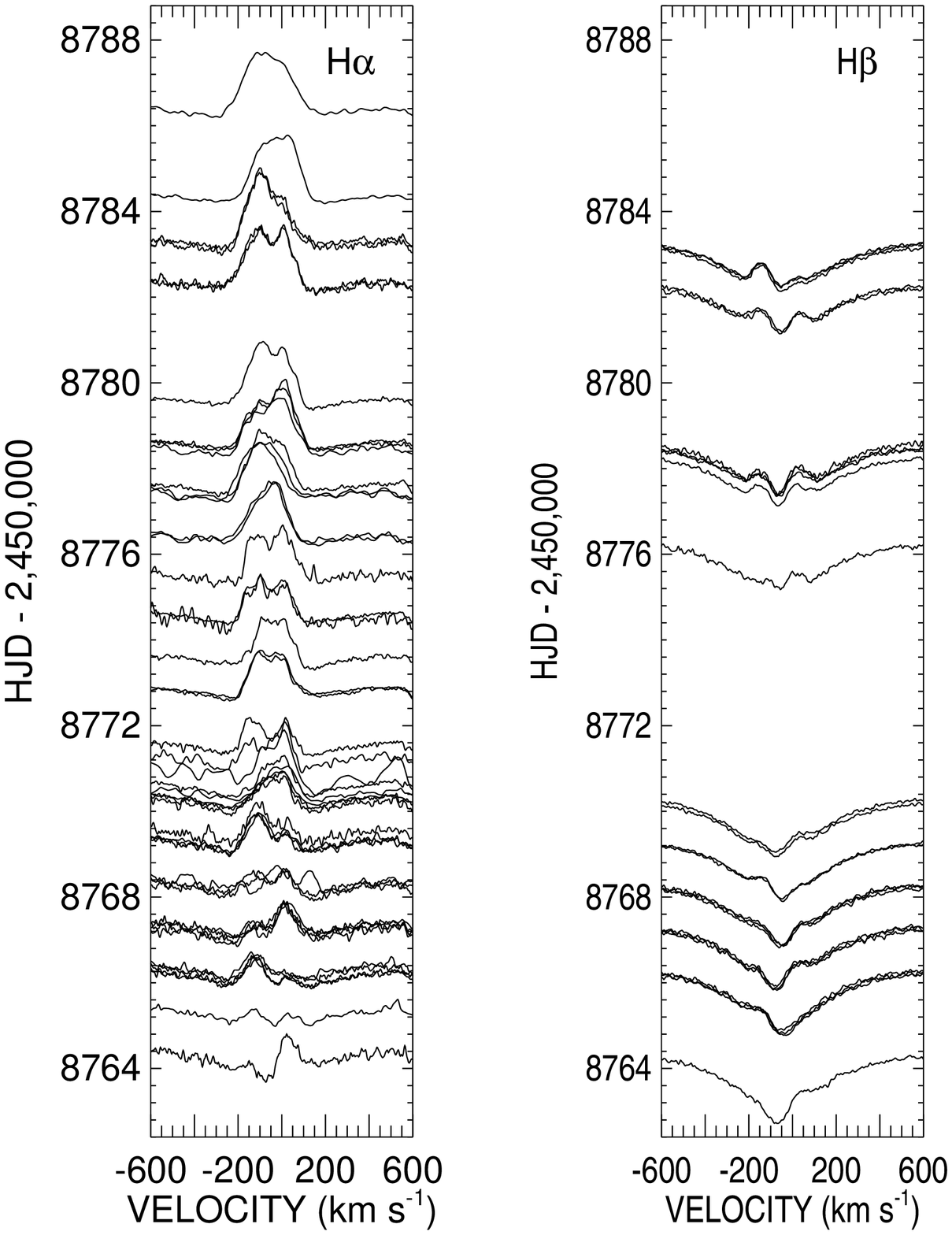}
\end{center} 
\caption{A dynamical representation of the H$\alpha$ spectra from the time of the {\it TESS} photometry is shown on the left, with an average profile shown in the bottom panel. On the right, we show the line profiles of H$\alpha$ (left panel) and H$\beta$ (right panel). The time of the observations is noted on the y-axis. We note that some evidence of the pulsational signature (2.615 d) is seen in the dynamical spectrum. } 
\label{tessprofiles} 
\end{figure*}

We confirmed this spectroscopic signature of pulsations with our time-series spectroscopy, concentrating on data taken with eshel spectrographs for consistency between the different observational setups. We considered the lines of He I $\lambda$ 4471, 4713, 4921, and 5876, along with Mg II $\lambda$ 4481. We chose these lines as they had reasonable signal-to-noise ratios (S/N $\sim$ 100 per resolution element), and were sufficiently far from the edge of the orders. By limiting our analysis to the eshel data, we had more than 350 observations per line. There is evidence of similar variations in the hydrogen lines (see, e.g., Figs.~\ref{Ha-profiles} and \ref{tessprofiles} for a dynamical representation of the data during the {\it TESS} observation time frame), but the frequent contamination by the disk emission prevented us from analyzing the exact behavior. {The behavior is similar to that of the pulsating Be stars discussed by \citet{2003A&A...411..229R}, and we interpret these variations in that context here.}

We performed a cross-correlation of our spectra against a template line profile from the TLUSTY BSTAR2006 grid \citep{TLUSTY} using the parameters of $T_{\rm eff} = 17,000$ K, $\log g = 3.0$, and $v \sin i = 70$ km s$^{-1}$, similar to that of \citet{Bozic} and our modeling results. This is also consistent with fundamental parameters of other normal B giants, such as the B3III star HD\,209008 that was found to have $T_{\rm eff} = 15,800$ K by \citet{2014A&A...566A...7N}. We used this rather than our model from Section 3, as the model in Section 3 only calculates the hydrogen line profiles. The resulting velocities are included in the Appendix. Many of these lines were used by \citet{2003A&A...411..229R}, allowing us to qualitatively compare our results to that of other Be stars. 

\begin{figure} 
\begin{center} 
\includegraphics[angle=90, width=8cm]{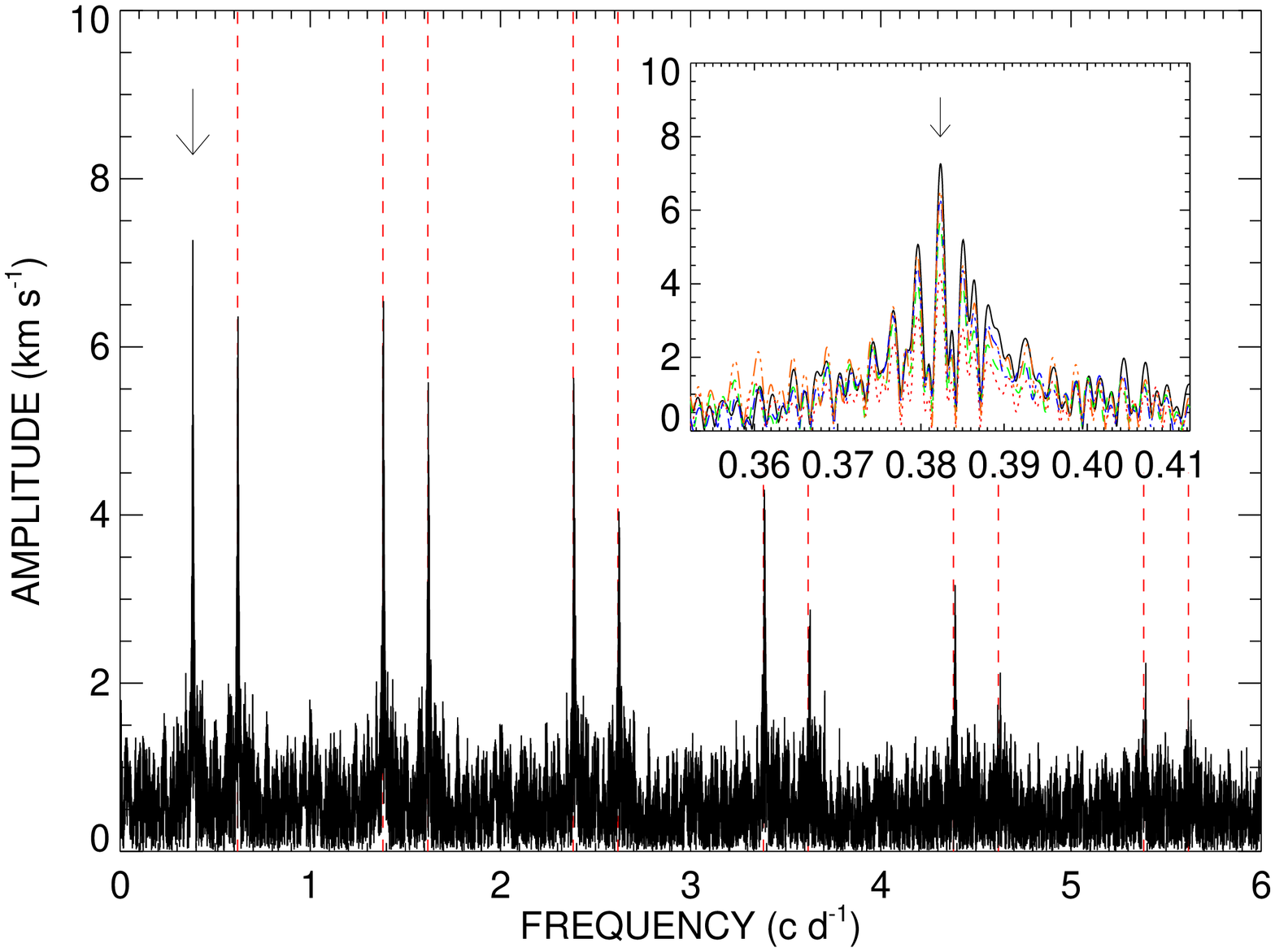}
\includegraphics[angle=90, width=8cm]{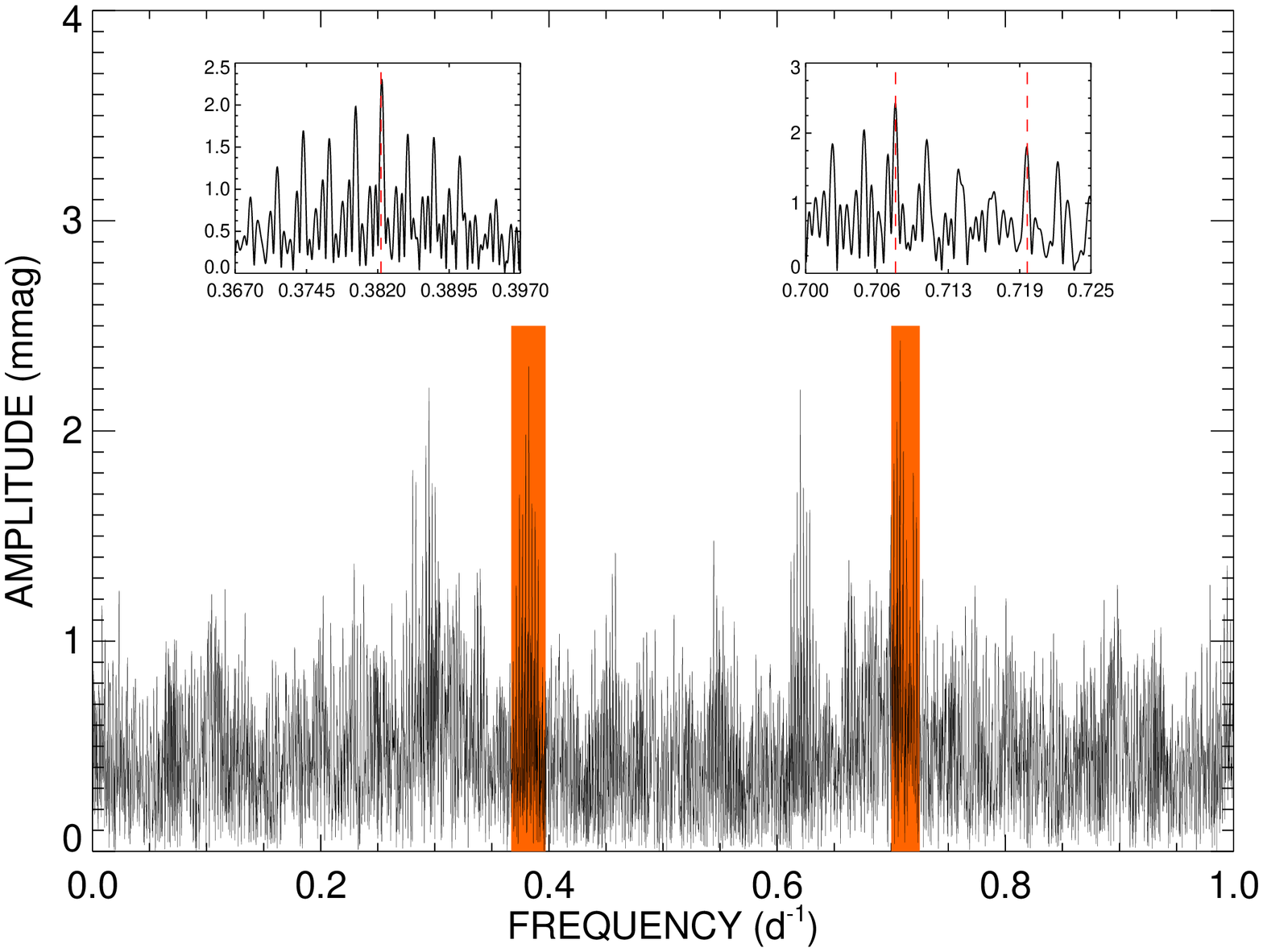}
\end{center} 
\caption{Fourier Amplitude spectra of radial velocity measurements of He I $\lambda 4471,4713,4921,5876$ (red, blue, green, and orange, respectively) and Mg II $\lambda 4481$ (black) are shown in the top panel. The dominant frequency is indicated with a large arrow, while diurnal aliases are indicated with vertical red, dashed lines. The bottom panel shows the Fourier amplitude spectrum of the KELT photometry. The orange highlighted regions represent the regions that are zoomed in for the two inset panels, where we show the derived frequencies as vertical red dashed lines. All other peaks are aliases of the derived frequencies similar to the derived frequencies from the spectroscopic measurements. The neighboring peaks for both time-series are spaced close to the expectation for 365 d, originating from the Earth-based data sets.} 
\label{fourier-puls} 
\end{figure}

We calculated Fourier amplitude spectra (Fig. \ref{fourier-puls}) for the radial velocity of each spectral line with the {\tt Period04} software \citep{2005CoAst.146...53L}. In Fig. \ref{fourier-puls}, we show the Fourier amplitude spectrum of the Mg II radial velocities. All lines show similar behavior in the resulting Fourier analysis, and the inset plot shows the resulting peak frequency, which is common for all lines, of 0.382332$\pm$0.000026 d$^{-1}$, corresponding to a period of 2.61553$\pm$0.00018 d. We note that the inset panel shows the Fourier spectrum for all line transitions, and these were combined in the derived frequency. This value is very similar to the 2.61507$\pm$0.00013 d period derived by \citet{Bozic}, and the differences could be that our time-series has a shorter baseline and lower spectral resolution than that of \citet{Bozic}, but also has a larger number of spectra. Our maximum radial velocity occurs at the epoch HJD 2,458,041.67168$\pm$0.00025, which is within errors of the maximum radial velocity epoch measured by \citet{Bozic}.

\begin{figure*} 
\begin{center} 
\includegraphics[angle=0, width=5.5cm]{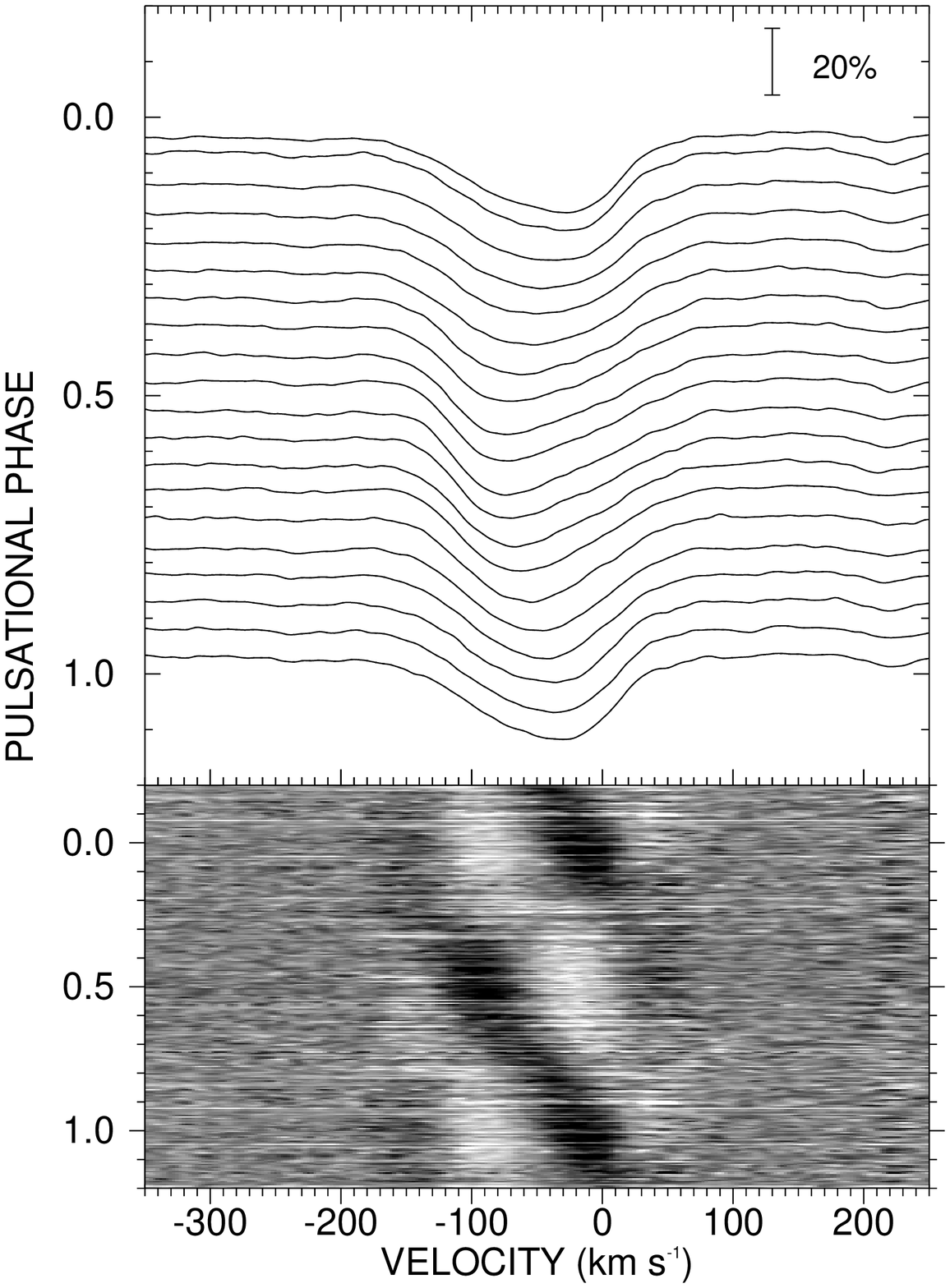}
\includegraphics[angle=0, width=5.5cm]{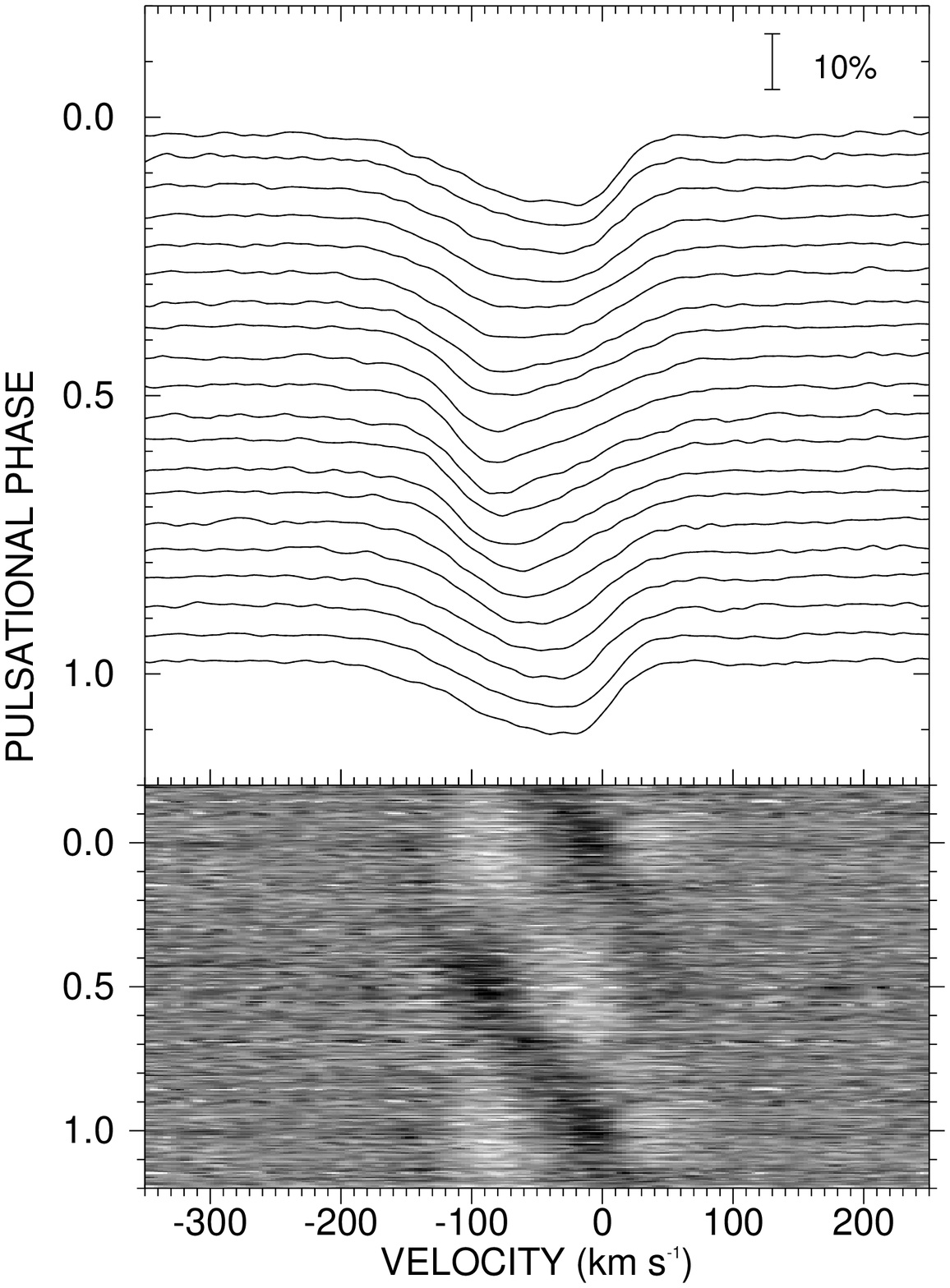}
\includegraphics[angle=0, width=5.5cm]{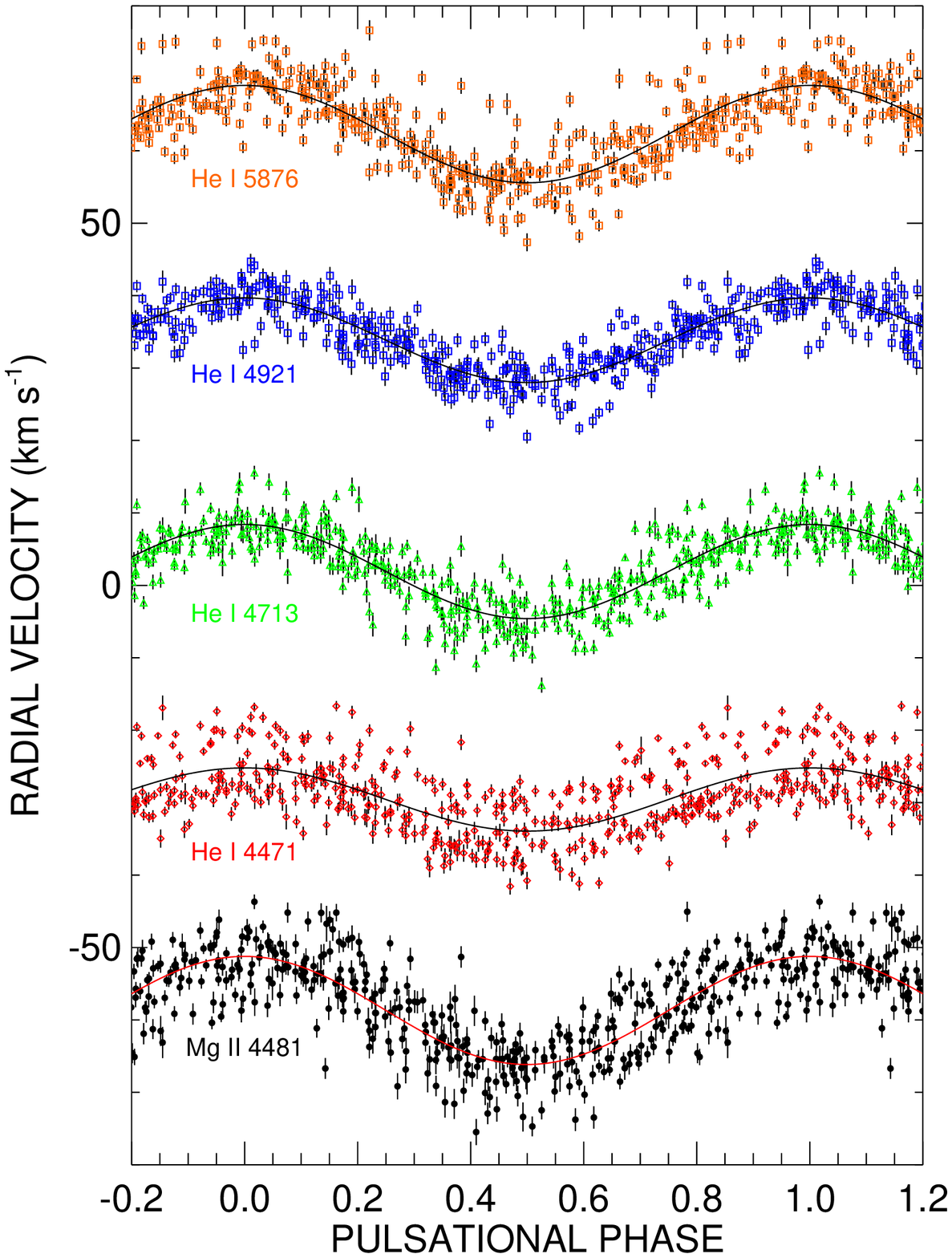}
\end{center} 
\caption{Dynamical representation of HD\,6226 He I $\lambda 5876$ (left) and Mg II $\lambda 4481$ (center) spectra during our campaign phased on the 2.6 d pulsational period. The grey scales show the differences from an average profile and the phasing is referenced in Table \ref{puls-table}. The black/white contrast indicates a $\pm$4\% deviation in each plot. The depth of the individual line profiles is shown in continuum units in each panel. Each shown line profile represents an average of all observations in a 0.05 phase interval. Representations of the He I $\lambda \lambda 4471, 4713, 4921$ profiles are shown in the Appendix. On the right panel, we show the radial velocity of the absorption lines measured (Mg II (black), He I 4471 (red), He I 4713 (green), He I 4921 (blue), and He I 5876 (orange)). Each point was measured via cross correlation with a TLUSTY stellar atmosphere model with similar stellar parameters. 
Measurements are offset for clarity and color coded. All measurements are phased to the same reference time and period.} 
\label{dynam-puls} 
\end{figure*}

Our Fourier transform spectra shown in Fig.~\ref{fourier-puls} all showed several aliases of the dominant frequency due to  diurnal sampling. These included frequencies of the form $n\pm f$, where $n$ is an integer between 1 and 6. These aliases are highlighted with vertical red dashed lines. Once we fit the dominant period and subtracted the trends from the data, these aliases did not show up in Fourier amplitude spectra of the residuals. The variance of the residuals between the sinusoidal fit and the data are comparable to typical errors in the data. 
The amplitude of the spectral variations varied depending on the line measured. The fit parameters are given in Table \ref{puls-table}, with the fits shown in Fig.~\ref{dynam-puls}. 

\begin{table*}
\begin{minipage}{170mm}
\centering
\caption{Derived pulsational frequencies from spectroscopic and photometric measurements\label{puls-table}}
\begin{tabular}{l c c c c c c}
\hline \hline
Line		&	Frequency 		&	Period 	&	Amplitude		& $V_0$ &	r.m.s.	 & $N_{\rm measurements}$  \\ 
	 		&	(c d$^{-1}$) 	& 	(d)		&	(km s$^{-1}$)	&	(km s$^{-1}$)	& (km s$^{-1}$)	&		 	\\ \hline
He I 4471 &	0.382313(36) &	2.61566	&	4.36(33)	&	$-59.6$ &	4.00	& 372 \\
He I 4713 &	0.382325(20) &	2.61557	&	6.49(22)	& 	$-58.1$	&	3.07	& 374 \\
He I 4921 &	0.382310(18) &	2.61568	&	5.84(21)	&	$-56.2$	&	2.52	& 362  \\
He I 5876 &	0.382342(21) &	2.61546	&	6.72(25)	&	$-57.7$	&	3.50	& 369 \\
Mg II 4481 &	0.382368(22) &	2.61528	&	7.46(32)	&	$-58.7$	&	3.99	& 372 \\

\hline
Combined	 	& 0.382332(26)	& 	2.61553(18) &	\ldots	&	\ldots	&	\ldots	& \ldots  \\
\citet{Bozic} 	& 0.38240(02)	&	2.61507(13)	& 	\ldots	&	\ldots	&	\ldots	& 35 \\
\hline \hline
Photometric Band    &	Frequency 		&	Period 	&	Amplitude		&   &   &   \\
	 		&	(c d$^{-1}$) 	& 	(d)		&	(mmag)	&	&   &  	 	\\ \hline
\hline
KELT ($\sim R$)     &   0.38243(21)     & 2.6149(47)      &     2.35  &   &   & \\
KELT ($\sim R$)     &   0.70787(21)     & 1.41269(42)      &     2.61  &   &   & \\
KELT ($\sim R$)     &   0.71940(26)     & 1.39013(50)      &     2.16  &   &   & \\

\hline

\end{tabular}
\end{minipage}
\end{table*}

We subtracted average line profiles of each transition from all observations and show the resulting residual variations in Fig.~\ref{dynam-puls}, which shows the variability of He I 5876 and Mg II 4481. The observations show a bright or dark part of the profile moving from blue to red across the profile over the period. The {\it TESS} photometry, which will be discussed in Section 6, also show some evidence of the pulsational period, although at lower amplitude given the redder passband of the {\it TESS} instrument than the KELT data. The results of the other lines were similar and are shown in the online appendix.

Overall, we suspect that this periodic line profile variation corresponds to a low-order non-radial mode. We re-examined the KELT photometry of HD\,6226 \citep[Fig.~\ref{fourier-puls}; ][]{2017AJ....153..252L}, and saw that there is a weak associated peak in the Fourier spectrum with an amplitude of $2.3$ millimag. With such a weak associated light variation, we suspect that it should be at least an $l=2$ mode based on comparisons to the pulsations discussed by \citet{2003A&A...411..229R}. Qualitatively, our greyscales in Fig.~\ref{dynam-puls} look similar to $\omega$\,CMa that were observed by \citet{2003A&A...411..229R}. However, the case of HD\,6226 provides a much longer period than in any of the Be stars studied by \citet{2003A&A...411..229R}. Given the long period, we suspect that the star is a slowly pulsating B star (SPB), and is not a pulsating $\beta$\,Cephei star.

The KELT data were detrended to remove variability on time-scales longer than 6 days so as to remove the outbursting trends from the data as well as any systematic effects associated with yearly or lunar cycles. We found that the frequency detected in the spectroscopic absorption lines was also seen in the photometry, with an amplitude of 2.35 millimag.
The KELT data also show evidence of a pair of frequencies at 0.70786 d$^{-1}$ and 0.71936 d$^{-1}$. These correspond to periods of 1.41 and 1.39 days, with amplitudes of 2.6 mmag and 2.2 mmag. We show the Fourier transform from the KELT data in the second panel of Fig.~\ref{fourier-puls}. The 1.4 d timescale was also seen in the {\it TESS} data, but the period varied over the time of the {\it TESS} data, which we discuss in terms of the outbursting disk in Section 6.

There are three detected pulsational modes between the analysis of the spectroscopy and the KELT photometry. These modes are consistent with those of slowly pulsating B stars. Given the model presented for the Be star based on the data available in Fig.~\ref{fig:hd6226_corner} and Table \ref{tab:hd6226}, we see that the mass is 7.1 $M_\odot$, which is near the high end for what we expect for an SPB star \citep[see Chapter 2 of ][]{2010aste.book.....A}. Indeed, the star would seem to lie closer to the expected instability strip for the $\beta$ Cephei stars. {However, \citet{2007MNRAS.375L..21M} find that the instability strip for SPB stars extends itself to higher luminosities and has a large overlap with the instability strip for the $\beta$ Cephei stars.}

We conclude that HD\,6226 is a slowly pulsating B star with narrow lines and P=2.6 d. We see a higher amplitude radial velocity variation in the Mg II line than in the He I lines, and \citet{Bozic} found that the metal lines had higher amplitude variations. 
{We suspect that the larger amplitude of the metallic radial velocity shifts are due to the intrinsic atomic broadening in each case. Metal transitions have narrower features that create more contrasting features in the Doppler-shift distribution of the profiles. Indeed the bumps look sharper in the Mg II case in Figure \ref{dynam-puls}. This results in a larger amplitude when measured using cross-correlation methods.}
The star also exhibits two photometric periods of $P$=1.39 and 1.41 d providing the constructive modes and episodic ejections, which will be discussed by explaining the H$\alpha$ and H$\beta$ outburst behavior in the following section. 

\section{Spectroscopic Measurements of the Outbursting Disk Activity}

Be star outbursts are typically considered to be discrete events with time-scales of days to decades, representing a time when material from the star is transferred to an equatorial decretion disk, governed by gravity and viscosity. 
During the outburst events, the emission and polarization can change as the inner-most portions of the disk increase in density \citep{2003A&A...411..229R}. The appearance of these outbursts depend on many system properties, including the inclination of the disk, as an edge-on disk will absorb light and cause the star to appear dimmer, whereas a nearly pole-on star will appear brighter \citep{2013ApJ...765...41S}. In the case of HD\,6226, \citet{2017AJ....153..252L} examined the star with KELT photometry, observing photometric outbursts and classifying the star as having intermediate periodicity with periodic variability with time-scales greater than 2 days. They also found outburst variability with semi-regular outbursts, and reported a period of 60.3 d.

\begin{figure} 
\begin{center} 
\includegraphics[angle=0, width=8cm, clip=false]{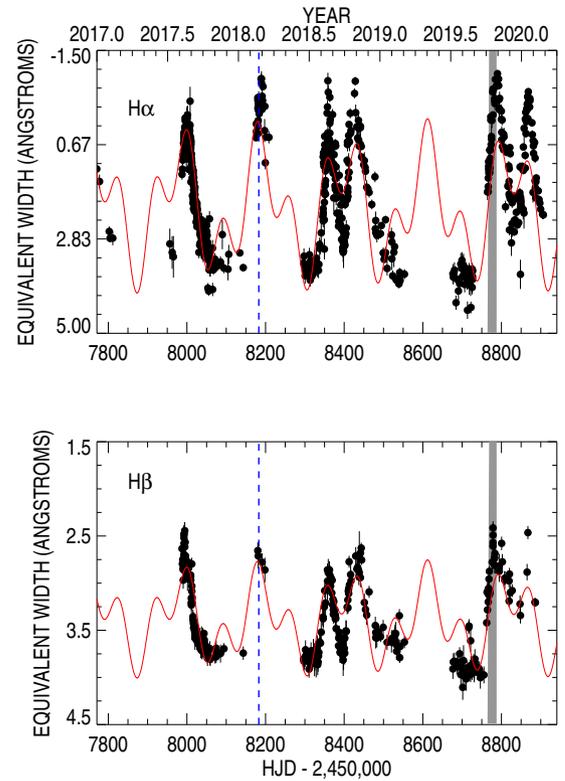}
\end{center} 
\caption{H$\alpha$ (top) and H$\beta$ (bottom) equivalent widths. We show the time of the {\it HST} observation with a vertical blue dashed line. The timing of the {\it TESS} observations is shown as a grey rectangular region. We overplotted our two-frequency fit to the data as a red solid line. The dashed blue line represents the epoch of the {\it HST} observation. } 
\label{ew} 
\end{figure} 

The star went through dramatic episodes of disk building and loss through the time period of 2017--2020 (see, e.g., Fig. \ref{ew}). With our extensive spectroscopic dataset, we can study these disk episodes through measurements of sensitive lines, in particular H$\alpha$ and H$\beta$. 

Our study of HD\,6226 truly began in late August 2017 when one of us (Thizy) observed the star was growing a disk shortly after it had been observed in a disk-less state. We began our analysis of the stellar disk with measurements of the equivalent width of the H$\alpha$ and H$\beta$ lines. In Fig.\ \ref{Ha-profiles}, we show three of the H$\alpha$ profiles from the time surrounding the first outburst from our campaign. We subtracted the model described in Section 3.2 from the profiles and found that the disk does not show double peaks. This further supports the results of Section 3.2, in that we observe the disk close to a pole-on orientation, and we then see only a small separation of the peaks observed in the raw spectra in Fig.~\ref{Ha-profiles}. 

\begin{figure} 
\begin{center} 
\includegraphics[angle=0, width=8cm]{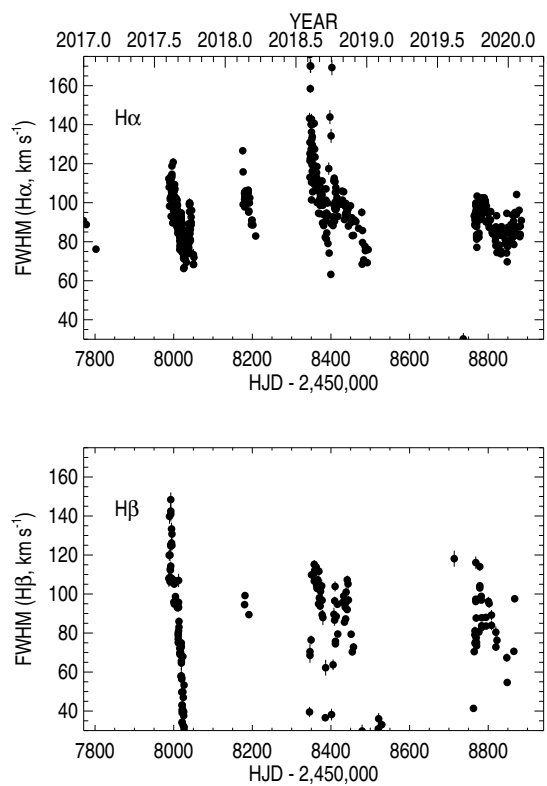}
\end{center} 
\caption{The FWHM of the H$\alpha$ (top) and H$\beta$ (bottom) emission features, calculated from the residual of the of the observed minus photospheric model profiles (\S 3.2). } 
\label{fwhm} 
\end{figure}

We measured the equivalent width of the H$\alpha$ and H$\beta$ profiles, as shown in Fig.\ref{ew}. We tabulate these values in the Appendix online. The H$\alpha$ equivalent width shows evidence of six outbursts during the time period we examined, with some additional emission present at the time preceding our campaign. Overall, these variations are seen to be similar, but the amplitude of the H$\beta$ variability is smaller than the H$\alpha$ variations, as expected. The H$\alpha$ variations are better sampled, as more H$\alpha$ spectra exist in BeSS, but are often seen to have larger errors, which can be explained as the H$\beta$ spectra are all from eshel spectrographs, and are therefore more consistent between different observers. Errors were estimated based on the signal-to-noise ratio for the observation as well as the spectral resolution. This was done based on the derivation by \citet{2006AN....327..862V} with the formula 
$$\sigma = \frac{{n \Delta \lambda - W_\lambda}}{SNR}. $$
Here, $\sigma$ represents the error, $n$ is the number of data points across the region of integration, $\Delta \lambda$ is the pixel separation, $W_\lambda$ is the measured equivalent width, and the $SNR$ is the signal-to-noise per resolution element.
 
We analyzed the data with the same kinds of Fourier techniques used in our pulsational analysis and found evidence of two frequencies in the data. We note that Fourier analysis is not a perfect way to go about understanding the outbursts as they are inherently non-sinusoidal. For example, we know that the light curves of Be stars are often modeled to have exponential decay \citep[e.g., ][]{2018MNRAS.479.2214G}. However, the Fourier analysis shows that the timing of the maxima and minima in the equivalent width variability can be reproduced with two periods. The details of the two found periods and the fit are given in Table \ref{halpha-freq-table}. We note that the variable in Table \ref{halpha-freq-table} of $W_{\lambda0}$ refers to the point at which the equivalent widths oscillate around, which is an ``average" equivalent width, so that the variability shown in Fig. \ref{ew} can be reproduced by the equation 
$$W_\lambda = W_{\lambda0} + \sum_{i=0}^{n} A_i \sin(2\pi (f_i t + \phi_i)).$$ In this formulation, $A_i$ is the amplitude of the particular sine wave, $f_i$ is the frequency, $t$ is the time, and $\phi_i$ is a phase introduced to shift the maxima to the correct time points. We note that there was no indication of the 2.615 d pulsational period (Section 4) in the analysis of the equivalent widths.

\begin{table*}
\begin{minipage}{170mm}
\caption{Derived Frequencies for the Equivalent Widths \label{halpha-freq-table}}
\begin{tabular}{l c c c c c}
\hline \hline
Line		&	Frequency 		&	Period 	    &	Amplitude	& $W_{\lambda0}$ &	phase (relative to)   \\ 
	 		&	(c d$^{-1}$) 	& 	(d)		    &	(\AA)	    &	(\AA) & (HJD - 2,450000)	\\ \hline
H$\alpha$   &	0.004742(19)    &	211.9(8)	&	1.157(41)	&   2.073(31)   &   0.9180(72)     \\
H$\alpha$   &   0.011499(29)    &	87.0(6)	    &	0.845(40)   &   \ldots   &      0.7138(83) \\
H$\beta$    &   0.004742(67)    &	211.9(8)	&	0.342(37)   &   3.479(35)   &   0.918(17)    \\
H$\beta$    &   0.011499(80)    &	87.0(6)	    &	0.284(37)   &   \ldots   &      0.714(20) \\
\hline
\end{tabular}
\end{minipage}
\end{table*}

Be star disk lines are often analyzed with the ratio of the violet-to-red emission peak heights ($V/R$). With HD\,6226, this is a difficult measurement to consider as the disk peak is quite narrow compared to most Be stars, likely due to a pole-on orientation. The raw profiles show a double-peaked emission, but they appear as single-peaked when the model spectrum is subtracted (see Fig.~\ref{Ha-profiles}). In our online supplement, we show individual line profiles and dynamical representations of the data in Appendix A, along with a 3-D printable version of the first outburst in H$\alpha$ in Appendix B.

One advantage of the analysis of $V/R$ measurements is that one can determine the radial velocity difference between the two peaks. In the case of $\mu$ Cen \citep{1998A&A...333..125R}, the onset of a disk outburst shows a larger separation of the radial velocity of $V$ and $R$ peaks due to the disk material being located closest to the star at that time. As time continues during the outburst, the material moves further out, causing the velocities to lower and the peaks to move closer together.

Since we are unable to measure the $V$ and $R$ peaks directly, we decided to fit the residual profile resulting from subtracting the model spectrum from the individual observations. Our fit was comprised of a simple Gaussian, yielding three parameters: the central peak velocity, the peak height, and the width, which are tabulated in Appendix C. The full width at half maximum is a good proxy for the difference in the radial velocity difference between the $V$ and $R$ peaks as it measures the difference of the velocity between the two sides of the emission that is roughly Gaussian shaped. In order to make these measurements, we first culled our data to observations when the equivalent width was less than 2.75\AA\ for H$\alpha$ or less than 3.6\AA\ for the H$\beta$ profile so that we only measure spectra with a disk profile that is easily seen with the typical S/N of the data. The FWHM measurements are shown in Fig.~\ref{fwhm}, and we see a general decline in each of the outbursts. For the H$\beta$ observations, the data are more uniform, and we see the width becomes very narrow near the end of the outbursts. The peak height measurements show almost identical curves to the equivalent width curves, so are not shown in our plots. We estimated the errors of these measurements as the error from the fit of the Gaussian, which is likely an underestimate of the actual error.

\section{TESS Photometry}

\begin{figure*} 
\begin{center} 
\includegraphics[angle=0, width=16cm]{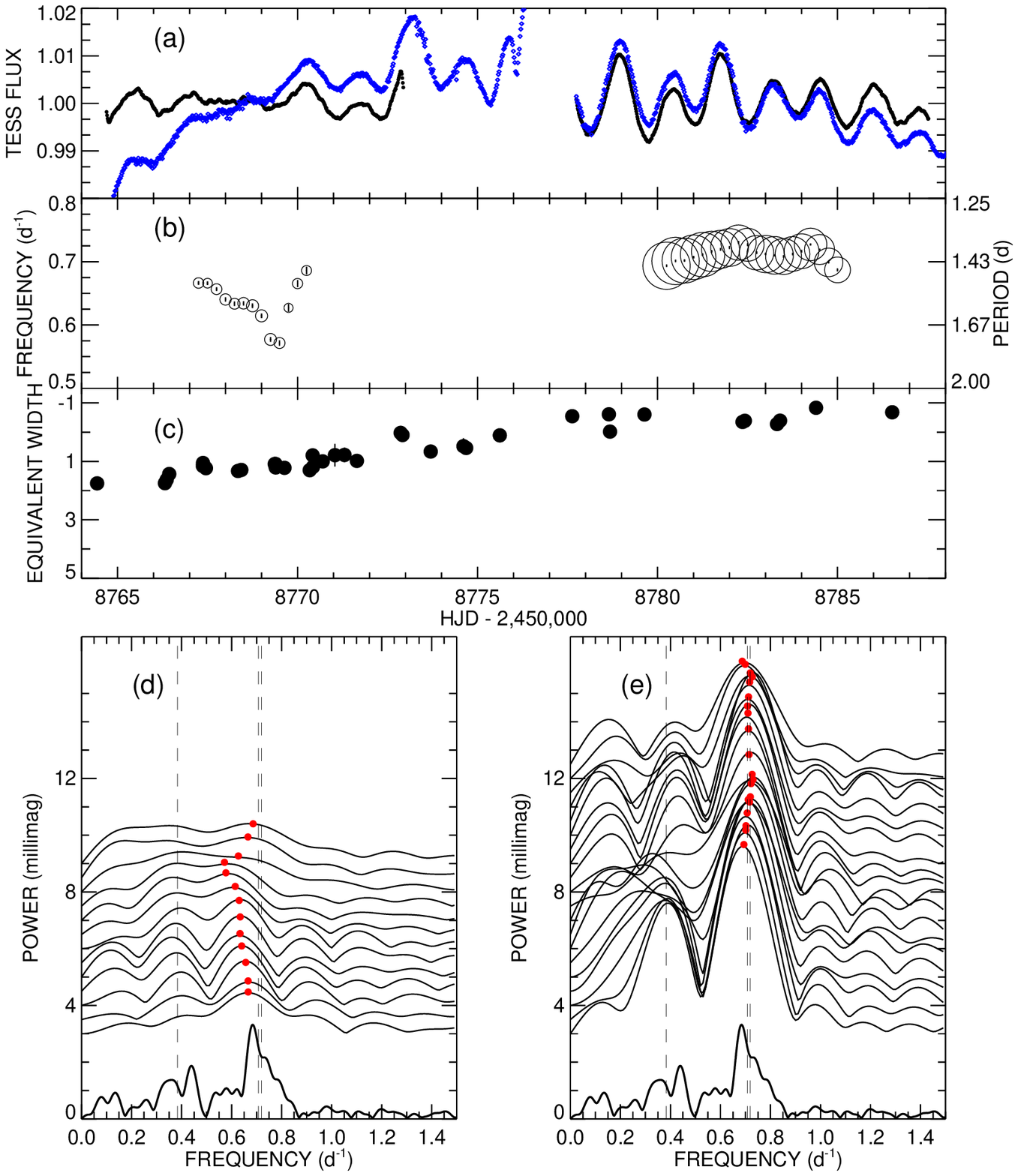}
\end{center} 
\caption{Panel (a): The {\it TESS} light curve is plotted (black: 2 minute cadence data; blue: full-frame image light curve).
Panel (b): The measured frequencies for five-day increments of the time-string is shown. The errors on the measured frequencies are shown, and the circles representing the measurement are proportional to the measured amplitude of the measured frequencies. 
Panel (c): The measured H$\alpha$ equivalent width during the time of the {\it TESS} observations, which is seen to be increasing during the first half of this tiem period. 
Panels (d) and (e) show the Fourier spectra of the time-series. The bottom-most spectrum in bold is of the full time-series, and each spectrum above is representative of a five-day increment in the {\it TESS} data. Panel (d) shows the data from the first half of the {\it TESS} observations, while panel (e) shows the second half. In both panels (d) and (e), the red points indicate the peak frequency and amplitude of the individual Fourier spectra. The vertical dashed lines indicate the frequency of the pulsational signals discussed in Section 4.} 
\label{tess} 
\end{figure*} 

The {\it TESS} observations were obtained as the disk was beginning to grow according to the H$\alpha$ and H$\beta$ activity, as shown in the grey-highlighted portion of Fig.~\ref{ew}. This is supported by the full-frame image light curve that shows an increase in flux through the beginning part of the {\it TESS} observations. The flux then decreases even while the H$\alpha$ equivalent width continues to grow due to the different formation radii for the broad-band optical flux and hydrogen emission. 
The H$\alpha$ and H$\beta$ line profiles increase in strength throughout the time period of the TESS observations, and the line profile peak oscillates from being blueshifted to redshifted and back during this time-scale (Fig.~\ref{tessprofiles}).

We show the {\it TESS} light curve in the top panel of Fig.~\ref{tess}. The light curve shows variations with a time scale of $\sim 1.4$ d. The extracted 30-minute cadence light curve shows the effect of the disk growth during this time. This looks to be similar to the time scale of the shifting of the profile morphology in Fig.~\ref{tessprofiles}. We performed a Fourier analysis of the {\it TESS} photometry. In the bottom panels of Fig.~\ref{tess}, we show the combined Fourier transform of the entire 2-minute cadence {\it TESS} light curve. We divided the light curve into two halves based on the gap in coverage. The Fourier spectrum of the entire data set shows the strongest peak to be a broad feature near 0.7 d$^{-1}$, which is confirmed visually in the second half of the light curve. We also indicate the pulsational frequency seen spectroscopically as well as the two additional frequencies derived from the KELT data in the panels for the Fourier spectra. The KELT data were too sparse in this time period for inter-comparison.

In order to further examine the variability in the observed light curve, we took 5-day segments of the 2-minute cadence light curve and then performed a Fourier analysis on each segment. The resulting strongest peak is indicated in each of the Fourier spectra shown in Fig.~\ref{tess} with a red point. The peak frequencies are plotted in the top-middle panel, where the symbol size is proportional to the amplitude of the signal. The H$\alpha$ equivalent width measurements from the same time period are also plotted for convenience, showing an increase in emission during this time. 

The beginning of the {\it TESS} observations is dominated by a lower-amplitude signal that becomes smaller with time. As the disk reaches a maximum near the time of the {\it TESS} data gap (so that the satellite can send data back to Earth), the signal became strong and then stayed roughly constant in frequency and amplitude for the remainder of the {\it TESS} observations. {This is consistent with a constructive interference of the two KELT periods being the driving mechanism for the disk growth for the HD\,6226 system.}

\section{Discussion}


The KELT study of the photometric behavior of 610 Be stars found 21\% of the Be stars show semi-regular outbursts \citep{2017AJ....153..252L, 2018AJ....155...53L}. However, these are not necessarily periodic. Amongst the semi-regular outbursting Be stars, $\lambda$ Eri has been shown to have periodic photometric outbursts with a period of either 469 d or 939 d \citep{1998A&A...330..631M}, which are harmonics of each other. If we consider the longer period harmonics, $\lambda$ Eri shows signs of multiple outbursts that have different strengths each cycle. \citet{1998A&A...330..631M} also showed that HR\,2142 had a periodic timescale of 344 d in its photometry.

The periodic nature of the outbursts can provide the context in which we can predict upcoming outbursts of disk activity in order to better observe and then model the formation scenarios of Be star disks. Our data presented here on HD\,6226 have shown the strong activity cycles for pulsations and outbursting activities in this fairly bright Be star, similar to those seen for $\lambda$ Eri and HR\,2142. 
We anticipate that much of the predictable nature of the equivalent widths (see Fig.~\ref{ew}) can only be used to predict the time periods when outbursts should occur. We suspect that the recurrence timescales derived may be a sort of beating mode of multiple stellar pulsations, where the `difference/combination frequencies' exist when there is some non-linear coupling coinciding with a maximum time of these pulsations ejecting material into the disk.

Given the two frequencies needed to reproduce the timing of the H$\alpha$ and H$\beta$ variations in Fig. \ref{ew} and Table \ref{halpha-freq-table}, we can speculate that the disk is grown through coherent addition of several various stellar phenomena. We see a slow pulsational period of 2.6 d (Section 4, Table \ref{fourier-puls}) that is reminiscent of the slowly pulsating B stars and is not atypical amongst the classical Be stars. 
SPB stars have longer pulsational periods than the $\beta$ Cephei stars and likely are the result of $g-$modes rather than $p-$modes \citep[see, e.g., ][]{2007CoAst.151...48M}. Our results on the spectroscopic time series are indeed consistent with the mode calculations of \citet{2007CoAst.151...48M} for this star being a SPB star with an $l$=1, 2, or 3 mode. {While our spectroscopic data do not directly show any additional modes, it has been well established that these stars often show multiple pulsational modes \citep[see recent analysis of ][]{2020A&A...633A.122F}, which could include the two photometric frequencies we discussed in Section 4.}

It is reasonable to assume that the pulsational modes could be coupled to cause a disk outburst on these time scales, and in fact our analysis points to that being one of the mechanisms behind the disk outburst episodes. The secondary period used in the fit shown in Fig. \ref{ew} has a frequency of 0.0115 cycles day$^{-1}$, which corresponds to the dominant difference frequency of the two photometric oscillations we see in the KELT data at 0.7194 and 0.7079 cycles day$^{-1}$. Given the results of the fundamental parameters, we anticipate the that these frequencies are similar to those seen in the {\it SMEI} and {\it BRITE} photometry of 28 Cyg \citep{2018A&A...610A..70B}. Pulsations have long been thought to be a driving force behind the mass ejections in Be stars, but in the case of HD\,6226, we see strong evidence that the coupling of two modes can build a disk from the analysis of the KELT photometry and BeSS spectroscopy. 
These two frequencies have amplitudes of 2.2 and 2.6 millimags. Given the difference in luminosity classes for these objects, a sub-giant for 28 Cyg and a giant for HD\,6226, these modes may be $g-$modes for the star, although no spectroscopic signature is seen at these frequencies with our data for mode identification. 

The $\sim$0.7 d$^{-1}$ frequency observed in HD\,6226 was the strongest signal present in the short 25-day observation from {\it TESS}, which is similar to the two pulsational frequencies found with KELT at 0.70787 and 0.71940 d$^{-1}$. The {\it TESS} observations were fortuitously taken as the disk was building (Fig. \ref{tessprofiles}) and near in time to when the beating of the frequencies were near the maximum of constructive interference (see Fig. \ref{tess}). We note that the frequency measured from {\it TESS} data is likely at a phase where the combination of the two pulsational modes measured from the KELT data are adding together in a manner that the power and derived frequency would appear to change over this time period. 

We can begin to understand more about the decay of the disk from the data in hand. In recent years, the viscous decretion disk (VDD) model has been successfully used to describe a large suite of multi-wavelength observational results of Be stars \citep[e.g., ][]{2012ApJ...744L..15C, 2015A&A...584A..85K}. The results of this model have been used to calculate the changes in photometry \citep{2012ApJ...756..156H} and polarization of the disk during an outburst \citep{2014ApJ...785...12H, 2011ApJ...728L..40D}. 

\citet{2016ASPC..506..157R} show that the photometry of the dissipation is generally given in the form of 
$$m(t) = m_0 - (\Delta_{\rm mag}) e^{{-{\xi_{\rm band}{({t-t_0})} \over {\tau} }}}. $$
{ In this derivation, $m(t)$ is the time-dependent magnitude, $m_0$ is the quiescent magnitude, $\Delta_{\rm mag}$ is the amplitude of the decay, $\xi_{\rm band} \over {\tau}$ dictates the timescale of the decay of the light curve, and $t_0$ is a zero-point in time. From \citet{2012ApJ...756..156H}, we note} that the parameter $\tau$ is inversely proportional to the viscosity parameter $\alpha$ of the disk. The models are built from the results of the {\tt HDUST} models for the VDD that have been well described by \citet{2006ApJ...639.1081C, 2008ApJ...684.1374C}.

\begin{figure} 
\begin{center} 
\includegraphics[angle=0, width=8cm]{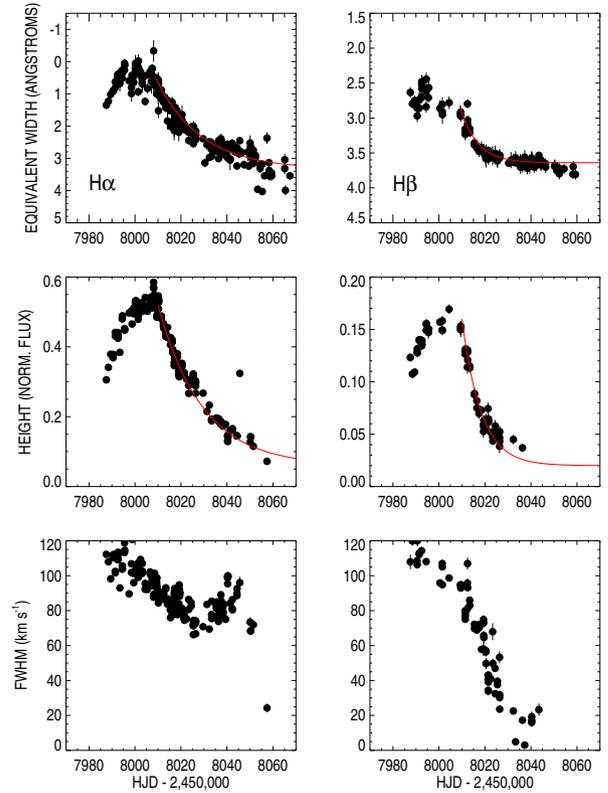}
\end{center} 
\caption{H$\alpha$ (left) and H$\beta$ (right) equivalent widths, peak height and the width of the emission feature during the best-observed outburst. We show similar fits for four more outbursts in Appendix A. } 
\label{outburst1} 
\end{figure}

Our results for the H$\alpha$ and H$\beta$ line strength, shown in Fig.~\ref{ew} and in detail for the first outburst in Fig.~\ref{outburst1}, are similar to the photometric decay of Be stars such as $\omega$ CMa \citep{2018MNRAS.479.2214G} and therefore offer an exceptional dataset with which to fit the exponential decay of the disk dissipation. Of the six outbursts we observed, the first one, peaking near HJD 2,458,000), was the best observed so we concentrate our efforts on the discussion of this particular outburst. There is some evidence that the disk did not fully dissipate at the end of this episode, as we observe larger equivalent widths in mid-2019, but these changes are very small and likely will not influence our results.

In Figure \ref{outburst1}, we show the measurements of equivalent width, the Gaussian peak height, and the Gaussian full-width-at-half-maximum (FWHM) for both H$\alpha$ and H$\beta$ during the first, and best-observed, outburst. These measurements were described in Section 5 and are tabulated online in the Appendix. The equivalent width variability can be understood as a proxy for the light emitted from the disk in these particular lines. The equivalent width is similar to the product of the height and FWHM, modified by a constant.

\begin{table*}
\begin{minipage}{170mm}
\caption{Estimated Exponential Decay for the Spectroscopic Measurements \label{exponential}}
\begin{tabular}{l c c c c c c c c}
\hline
\multicolumn{9}{c}{Outburst 1} \\
\hline
Line	&	$W_{\lambda,0}$ &	$W_\lambda ({\rm peak})$    &	$\Delta t$	&	$t_0$		       &	$H_{0}$	    &	$H({\rm peak})$	&	$\Delta t$	&	$t_0$           \\
        &   (\AA)           &   (\AA)                       & (d)           &   (HJD-2,450,000)    &   (norm.)      &   (norm.)         &   (d)         &   (HJD-2,450,000)\\ \hline
H$\alpha$	&	3.34	    &	0.6	                        &	19.5	    &	8007.9              &	0.04	    &	0.52	        &	19.3	    &	8010.0	\\
H$\beta$	&	3.64    	&	2.84	                    &	7.4	        &	8010.0	            &	0.02	    &	0.16	        &	8.3	        &	8010.0.  \\ \hline
\multicolumn{9}{c}{Outburst 3}	        \\
\hline
H$\alpha$	&   3.34        &   0.64                        &   23.1        &   8366.3              &    0.06       &    0.52           &    23.1       &  8366.3     \\
H$\beta$	&   3.70        &   3.00                        &   13.2        &   8366.3              &    0.02     &         0.14        &     11.2         &   8368.3	\\
\hline
\multicolumn{9}{c}{Outburst 4}	        \\
\hline
H$\alpha$	&   3.34        &   -0.50                       &   37.5        &   8427.5              &    0.06       &    0.56           &    31.9       &  8443.7     \\
H$\beta$	&   3.70        &   2.80                        &   23.4        &   8440.6              &    0.02     &    0.18       &     23.4         &   8440.6	\\
\hline
\multicolumn{9}{c}{Outburst 5}	        \\
\hline
H$\alpha$	&   3.34        &   -0.50                       &   30.5        &   8792.8              &    0.20       &    0.69          &    28.1       &  8795.2     \\
H$\beta$	&   \multicolumn{4}{c}{Unable to fit}                                                   &    0.08     &    0.21       &    10.5         &   8798.4		\\
\hline
\multicolumn{9}{c}{Outburst 6}	        \\
\hline
H$\alpha$	&   3.34        &   -0.27                       &   19.5        &   8875.4              &     \multicolumn{4}{c}{Unable to fit}      \\
H$\beta$	&   \multicolumn{8}{c}{Unable to fit}                                                    	\\
\hline

\end{tabular}
\end{minipage}
\end{table*}

We did a basic fit of the exponential decay with an equation of form
$$W_\lambda(t) = W_{\lambda,0} + (W_\lambda ({\rm peak}) - W_{\lambda,0}) e^{-({t - t_{\rm peak}}) \over {\Delta t}} .$$
This decay curve can be used for both the equivalent width as well as the peak height from the Gaussian fit of the residuals of the data and the TLUSTY model. This is estimated with an equivalent width or height, $W_{\lambda,0}$, the peak and quiescent emission, $W_\lambda ({\rm peak, 0})$, a time in which we estimate the curve has fully decayed, $\Delta t$, and the time of the peak emission prior to the decay, $t_{\rm peak}$. We show estimates of the fits for the equivalent widths and peak heights of both H$\alpha$ and H$\beta$ in Fig.~\ref{outburst1}. The parameters of these fits are shown in Table \ref{exponential}, and we expect the errors of these parameters to be on the order of 5\%. We also performed similar fits to at least some of the data from the other four outbursts with reasonable observational coverage of the decay from the campaign, including the coefficients in Table \ref{exponential} and including the similar figures in Appendix A.

We see from the data shown in Fig.~\ref{outburst1} and the estimates of the exponential decay in Table \ref{exponential} that the H$\beta$ line decays much quicker than the H$\alpha$ emission. 
The optical depth at line center is much smaller for H$\beta$ than for H$\alpha$. If the disk is conceptually seen as a pseudo-photosphere of given radius \citep{2015MNRAS.454.2107V}, then the size of the H$\alpha$ pseudo-photosphere would be larger than the H$\beta$ one.
Therefore, we see evidence that the disk clears from the inner to outer regions, as has been seen for other Be stars \citep[e.g., ][]{2010ApJ...709.1306W}. 

We see from these simple fits to the data that the time scale for $\Delta t$ to vary from $\sim19$ to $\sim 37$ days for H$\alpha$, and from $\sim 7$ to $\sim 23$ days for H$\beta$. The timescales of the disk show that the H$\alpha$ emission remains $\sim 1.4 - 2.7$ times longer than the H$\beta$ line, which depends on the outburst. With a variety of timescales needed for fitting the decay of the disk in these two lines, we hypothesize that the viscosity parameter, $\alpha$, could also be variable if the mass injection for the disk stops after the peak. This has been seen in the similar Be star $\omega$ CMa, which shows variability in the disk's viscosity coefficient from 0.1-1.0 according to the modeling of the optical light curve by \citet{2018MNRAS.479.2214G}. 
Alternatively, the disk injection could be lower after the peak, in which case the viscosity parameter $\alpha$ could be constant, with a variable disk injection rate. 

Unfortunately, the KELT data did not cover this or other outbursts in our campaign with enough cadence to fit an exponential decay to the emergent flux in the KELT bandpass. We were also unable to have reasonable cadence from surveys such as ASAS-SN. A future campaign will need to have high cadence observations of photometry and spectroscopy during an outburst to fully characterize the viscosity parameters of the disk. However, we did find some lower-precision photometry taken with the Kamogata/Kiso/Kyoto Wide-field Survey (KWS)\footnote{http://kws.cetus-net.org/~maehara/VSdata.py}.

\begin{figure} 
\begin{center} 
\includegraphics[angle=90, width=8cm]{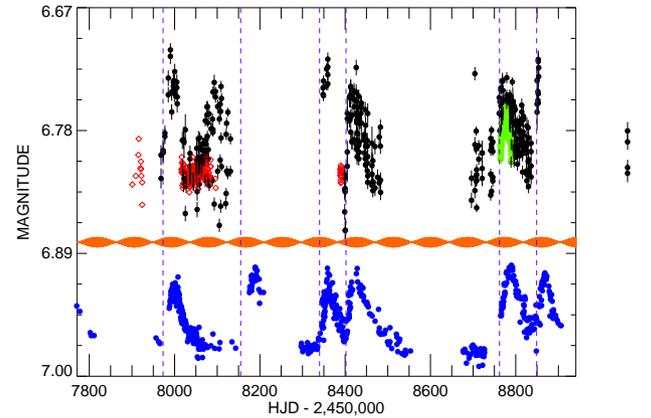}
\end{center} 
\caption{The KWS photometry is plotted in black, with the KELT data over plotted in red, and the 30-minute light curve from {\it TESS} shown in green. We show a scaled H$\alpha$ equivalent width curve in blue at the bottom. Our derived two-frequency fit for the KELT pulsations are shown in the orange curve, showing the clear beating pattern. We estimated the start times of the H$\alpha$ outbursts, and have marked those with vertical, purple dashed lines. } 
\label{beat} 
\end{figure} 

In Fig.~\ref{beat}, we plot the KWS photometry, along with our limited KELT photometry, and the H$\alpha$ equivalent width variability seen in Fig.~\ref{ew}. In addition to the photometry and equivalent widths, we have overplotted the summation of the two pulsation modes found with KELT with frequencies near 0.7 d$^{-1}$. We see these two pulsation modes providing a beating pattern in the light curve, with the beat period that is comparable to the second frequency used in our Fourier analysis of the H$\alpha$ equivalent widths. We then estimated the timing of the outbursts in the H$\alpha$ equivalent widths, i.e., the time that the equivalent width begins to grow, and compared these times to the beating pattern. Most of the H$\alpha$ outbursts are accompanied by photometric outbursts (although we are unable to fit the KWS data with an exponential decay like we did with H$\alpha$ and H$\beta$ (see Table \ref{exponential}). This plot shows that as the two modes begin beating constructively, the star sometimes begins to build its disk. It appears that sometimes the constructive interference of pulsational modes can build the Be star disk in the HD\,6226 system. 

We also see a change in the FWHM of the disk emission during the outbursts, as shown in Figs.~\ref{fwhm} and \ref{outburst1}. One can envision the disk as optically thick out to some radius, the pseudo-photosphere as described above, where the disk profile would be double-peaked at a more moderate inclination. In the case of HD\,6226, our results are consistent with the disk clearing from the inside out during the outbursts. Near the end of these outbursts, some of the material may be falling back toward the star, causing the FWHM to grow at the end of these outbursts. 

When the star begins an ejection of material, this feeds the disk and the material spreads outward in the disk. When the injection mechanism is turned off, likely when pulsational modes no longer add constructively for matter ejection near the stellar equator, the inner part of the disk begins to drain of material more quickly than the outer part. This causes the H$\beta$ line to decay faster than H$\alpha$, and then the FWHM of the emission will also become narrower. That is seen in detail in Fig.~\ref{outburst1} and with the exponential decay fits in Table \ref{exponential}.

\section{Conclusions}

Our study of HD\,6226 has found many interesting properties of the Be star, which we summarize here:
\begin{itemize}
    \item The star shows a consistent spectral type with both optical spectra taken during the quiescent phase as well as with an ultraviolet spectrum taken with {\it HST} near the peak of an outburst. The spectral type is B2.5IIIe, based on the similarity of the B2.5III standard star $\pi^2$ Cygni. A TLUSTY model with $T_{\rm eff} = 17,000$ K, $\log g =3.0$, and $v \sin i = 70$ km s$^{-1}$ matches our observations well. The system is seen near pole-on based on the inclination derived from our model, and is similar to these parameters from the TLUSTY model.
    
    \item HD\,6226 is seen to have one pulsation period of 2.61553 d that is seen in spectroscopy. The variations are similar to those of a low-order mode, perhaps an $l = 2$ mode. There is some marginal power near this frequency in the Fourier analysis of the {\it TESS} data, but was very weak.
    
    \item The star shows a periodic behavior in the H$\alpha$ and H$\beta$ emission. Of the two periods (212 and 87 d) found for the H$\alpha$ behavior, we find that the 87 d period is a difference frequency of two pulsational frequencies observed in the KELT data. The predicted curve, shown in Fig.~\ref{ew}, predicts the times of maximum and minimum equivalent width. We encourage observers to take frequent spectra of HD\,6226 in the month or more prior to predicted times of outburst so that the build-up of disk material can be well observed.  Thus far, no outburst has been seen in detail from the beginning of the material build-up. 
    
    \item When the star begins an outburst, we see a wide emission profile superimposed on the stellar photospheric line. The peak was fit with a Gaussian profile, which we see having a wide full-width-at-half-maximum and then becoming more narrow with time. This can be interpreted as material moving from the interior to the outer parts of the disk. The disk then dissipates leaving the profile too weak to detect the signature of material falling back onto the star. This general picture agrees with the hydrodynamic simulations of \citet{2012ApJ...756..156H}.


    \item The H$\alpha$ and H$\beta$ variability during the outbursts is similar to the photometric behavior that has been modeled for other Be stars during their outbursts \citep[e.g.,][]{2017MNRAS.464.3071V}. The decay curves can be used for future modeling with the viscous decretion disk codes used for Be star disks \citep[e.g., ][]{2012ApJ...756..156H, 2018MNRAS.479.2214G}. No general modeling for the viscous disk parameters has been done for the entire population of Be stars, so the exact parameters of the disk decay are unconstrained. The time scale for each of the outbursts is different, indicating that the viscosity parameter is variable with each outburst or that the mass injection rate is variable. 
    
    \item Two of the modes seen in the KELT photometry have a beating frequency that is equal to one of the two frequencies found in the Fourier analysis of the H$\alpha$ and H$\beta$ equivalent widths. The beating properties indicate that the constructive interference of the pulsational modes could be the driving mechanism behind the outbursts for this Be star.

\end{itemize}

The continued study of the star HD\,6226 will be necessary for a better understanding of how disks are built and dissipate for Be stars. The most needed observations in future are a confirmation of periodic behavior of the outbursts, and potentially a study of the polarization of the disk during an outburst. It's unclear if we should expect a signature of polarization because of the near pole-on orientation of the star. However, if this is caught very early in an outburst, a strong polarization signal could be seen as the material is ejected from a particular spot on the stellar surface.  The fact that the outbursts appear to be periodic allows us to consider mounting observational campaigns to catch the outburst with multiple techniques such as photometry, spectroscopy, polarimetry, or even long baseline interferometry. We had also considered modeling the phase variations of the line profiles with current asteroseismology tools such as FAMIAS \citep{famias}. However, these tools assume a non-rotating star, and Be stars are known to be rotating at nearly critical speeds. Furthermore, our data are of lower S/N and spectral resolution than needed for the code, given the typical data quality found in BeSS. The complete analysis of this pulsational mode will require a dedicated campaign in the future. 

The unique time frame and data set collected during the time that {\it TESS} observed the star represents one of the best data sets collected showing the growth of a disk of a Be star. The complementary spectroscopy and {\it TESS} photometry will undoubtedly prove essential to future modeling efforts of Be stars and their disks. Other Be stars have been observed by both {\it TESS} and {\it BRITE-Constellation}, but very few with such extensive spectroscopic coverage. We hope this study will help to excite amateur spectroscopists to provide similar data sets for both HD\,6226 and other Be stars in the future.  

\section*{Data Availability}

The data for this paper are largely available online, with links given for various surveys used throughout the text. {\it HST} and {\it TESS} data are available freely from MAST. The optical spectroscopy are largely available through BeSS, and most of the KELT data were from \citet{2017AJ....153..252L}. Any author requesting any spectra that are not in BeSS or MAST is encouraged to contact the authors for these data. 

\section*{Acknowledgements}
Some spectra were obtained with the Ritter 1-m telescope. Ritter Observatory recently celebrated its fiftieth anniversary of first light during the course of this project, and we dedicate this paper to the hard work and determination of the retired observatory directors Adolf Witt, Bernie Bopp, and Nancy Morrison. Observations at Ritter Observatory would not have been possible without the support and hard work of the observatory technician, Michael Brown. Further, many of the Ritter observations were supported by an enthusiastic team of undergraduate students who volunteered their time, whom we thank for their dedication. We also thank Peter
Wysocki for their help with the {\tt GalPy} calculations in Section 3. 

This work has made use of data from the European Space Agency (ESA) mission
{\it Gaia} (\url{https://www.cosmos.esa.int/gaia}), processed by the {\it Gaia}
Data Processing and Analysis Consortium (DPAC,
\url{https://www.cosmos.esa.int/web/gaia/dpac/consortium}). Funding for the DPAC
has been provided by national institutions, in particular the institutions
participating in the {\it Gaia} Multilateral Agreement.
This work has also made use of the BeSS database, operated at LESIA, Observatoire de Meudon, France:  http://basebe.obspm.fr. The professional astronomers thank this international group of enthusiastic amateur spectroscopists for their time and personal investments in this project, as their curiosity was contagious amongst our team. We also thank Justin Barnes for assistance in creating the 3D STL file presented in Appendix B and to Kristy Richardson for assistance in photographing the 3D-printed model.  

Support for {\it HST} Program number 15432 was provided by NASA through a grant from the Space Telescope Science Institute, which is operated by the Association of Universities for Research in Astronomy, Incorporated, under NASA contract NAS 5--26555. This paper includes data collected by the TESS mission through the TESS-GI program G022064. Funding for the TESS mission is provided by the NASA Explorer Program.

NDR acknowledges previous postdoctoral support by the University of Toledo and by the Helen Luedtke Brooks Endowed Professorship. 
JEB and KSB acknowledge support of the NSF through grant AST-1412135. Participation in this project by A. Daly and A. Lane was supported by Embry-Riddle Aeronautical University's Undergraduate Research Institute. DRG is supported through NSF grant AST-190826.
ACC acknowledges support from CNPq (grant 311446/2019-1) and FAPESP (grant 2018/04055-8). 
ACR acknowledges the support of FAPESP grants 2017/08001-7 and 2018/13285-7. MG acknowledges the support of FAPESP grant 2018/05326-5.
This work made use of the computing facilities of the Laboratory of Astroinformatics (IAG/USP, NAT/Unicsul), whose purchase was made possible by the Brazilian agency FAPESP (grant 2009/54006-4) and the \mbox{INCT-A}. 
D.J.S. is supported as an Eberly Research Fellow by the Eberly College of Science at the Pennsylvania State University. The Center for Exoplanets and Habitable Worlds is supported by the Pennsylvania State University, the Eberly College of Science, and the Pennsylvania Space Grant Consortium. 




\bibliographystyle{mnras}
\bibliography{bibtex} 





\appendix

\section{Additional Figures}
\begin{figure*} 
\begin{center} 
\includegraphics[angle=0, width=8cm]{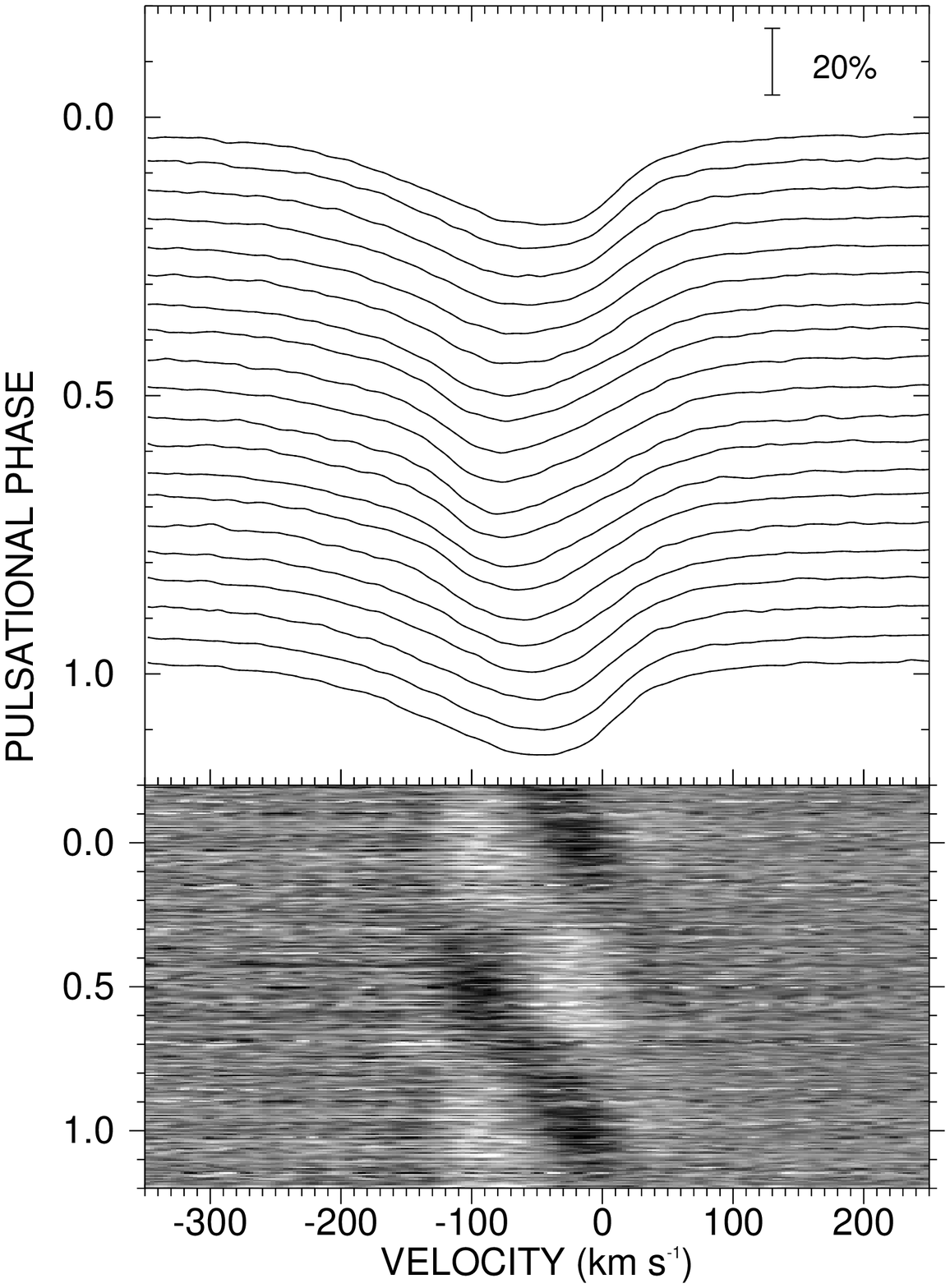}
\includegraphics[angle=0, width=8cm]{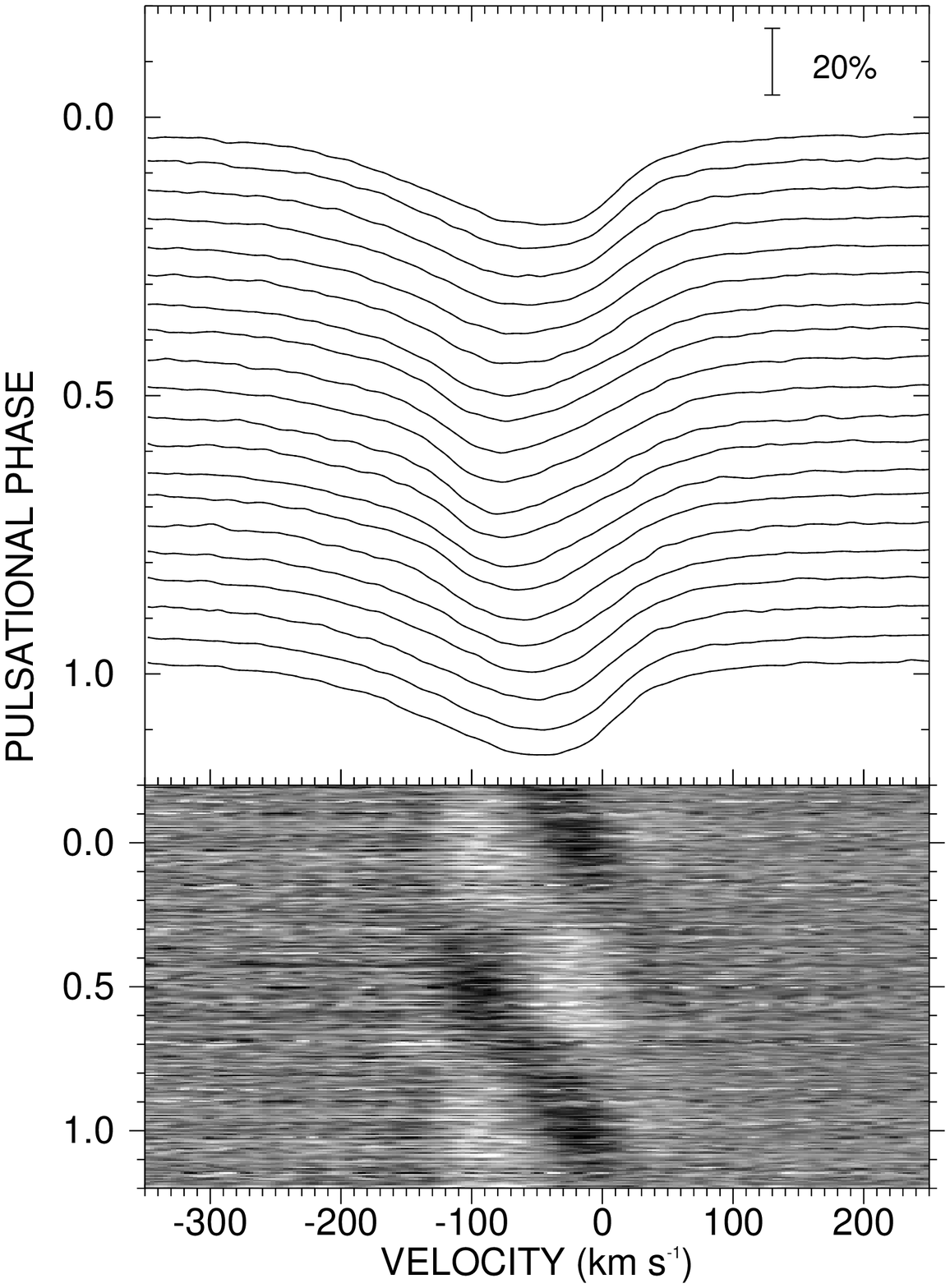}
\includegraphics[angle=0, width=8cm]{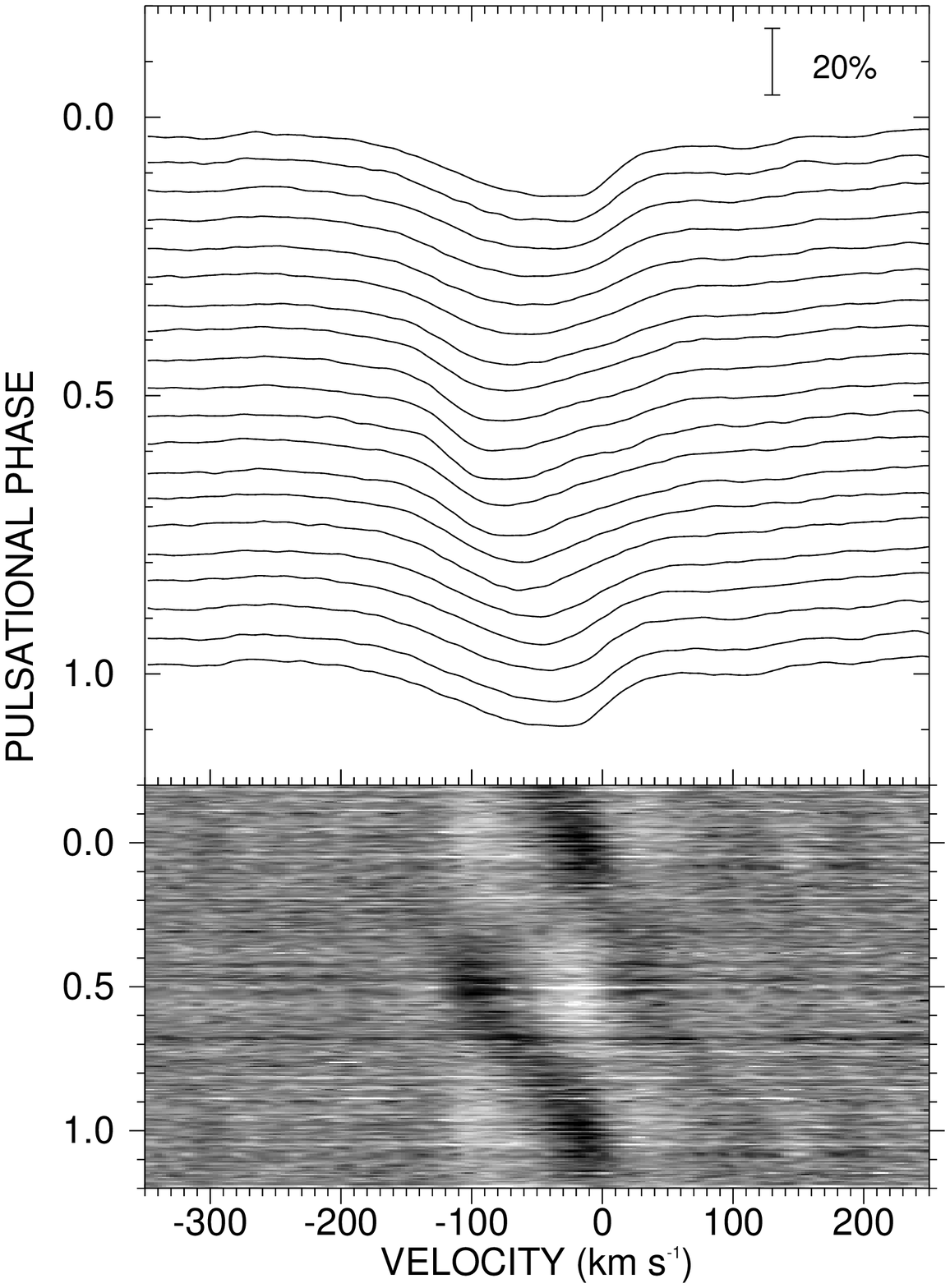}

\end{center} 
\caption{Dynamical representation of HD\,6226 He I $\lambda \lambda 4471$ (top left), 4713 (top right), 4921 (bottom) profiles spectra during our campaign for the 2.6 d pulsational period. The grey scales show the differences from an average profile and the phasing is the same as in Fig.~\ref{dynam-puls}. The black/white contrast indicates a $\pm$4\% deviation in each plot. The depth of the individual line profiles is shown in continuum units in each panel. Each shown line profile represents an average of all observations in a 0.05 phase interval. Representations of the He I $\lambda 5876$ and Mg II 4481 profiles are shown in Fig. \ref{dynam-puls}.}
\label{dynam-puls-appendix} 
\end{figure*} 
\clearpage

\begin{figure*} 
\begin{center} 
\includegraphics[angle=0, width=5cm]{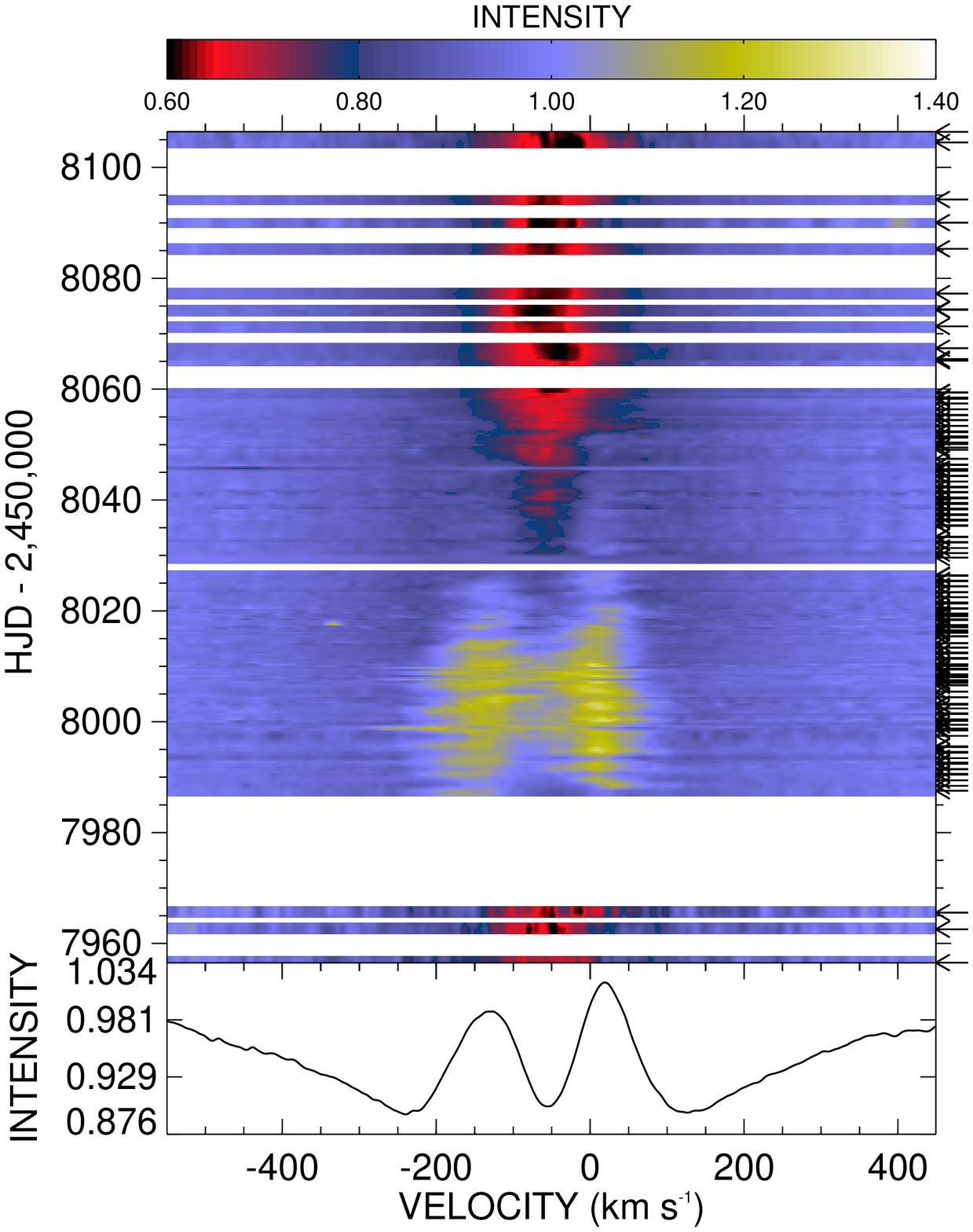}
\includegraphics[angle=0, width=5cm]{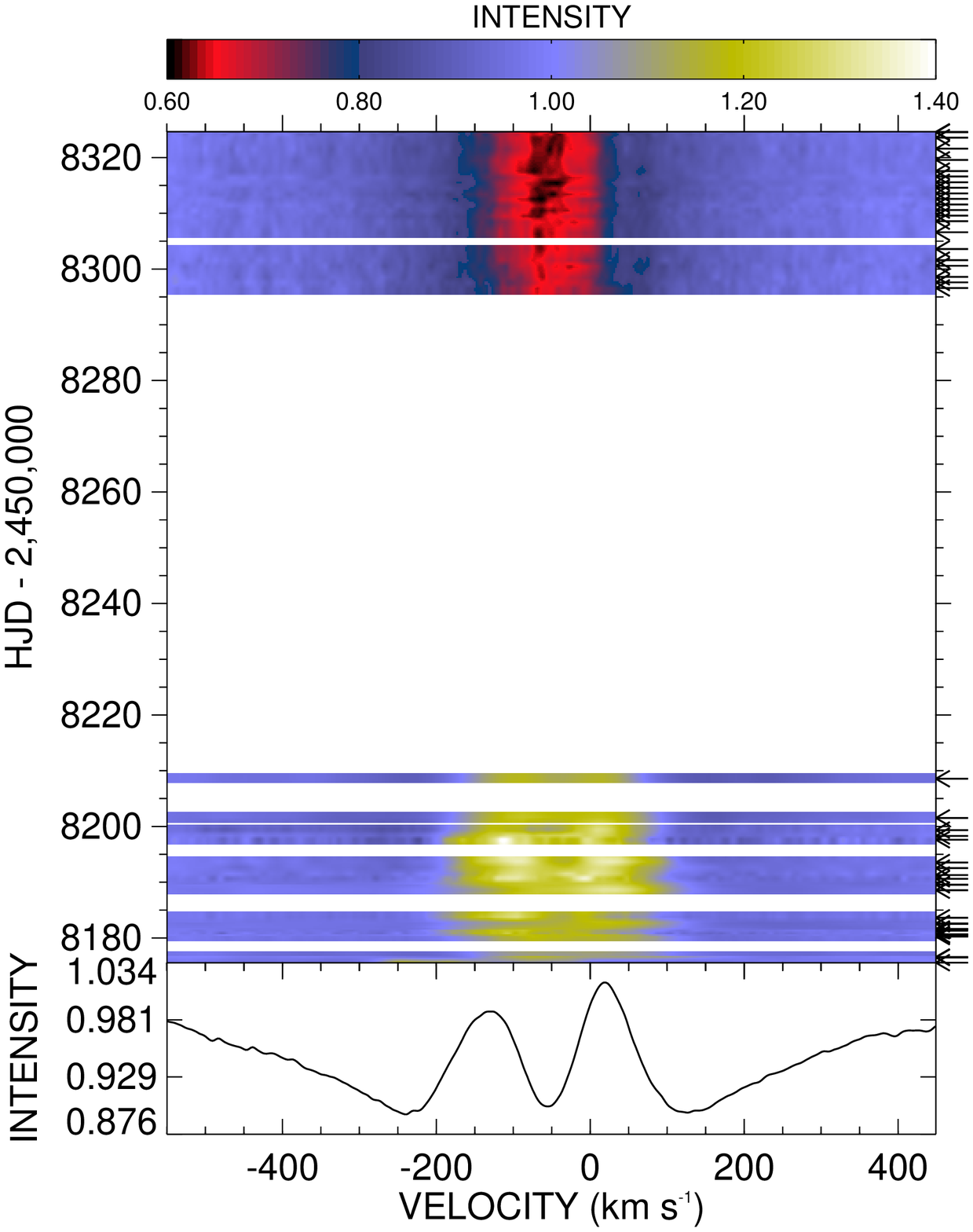}
\includegraphics[angle=0, width=5cm]{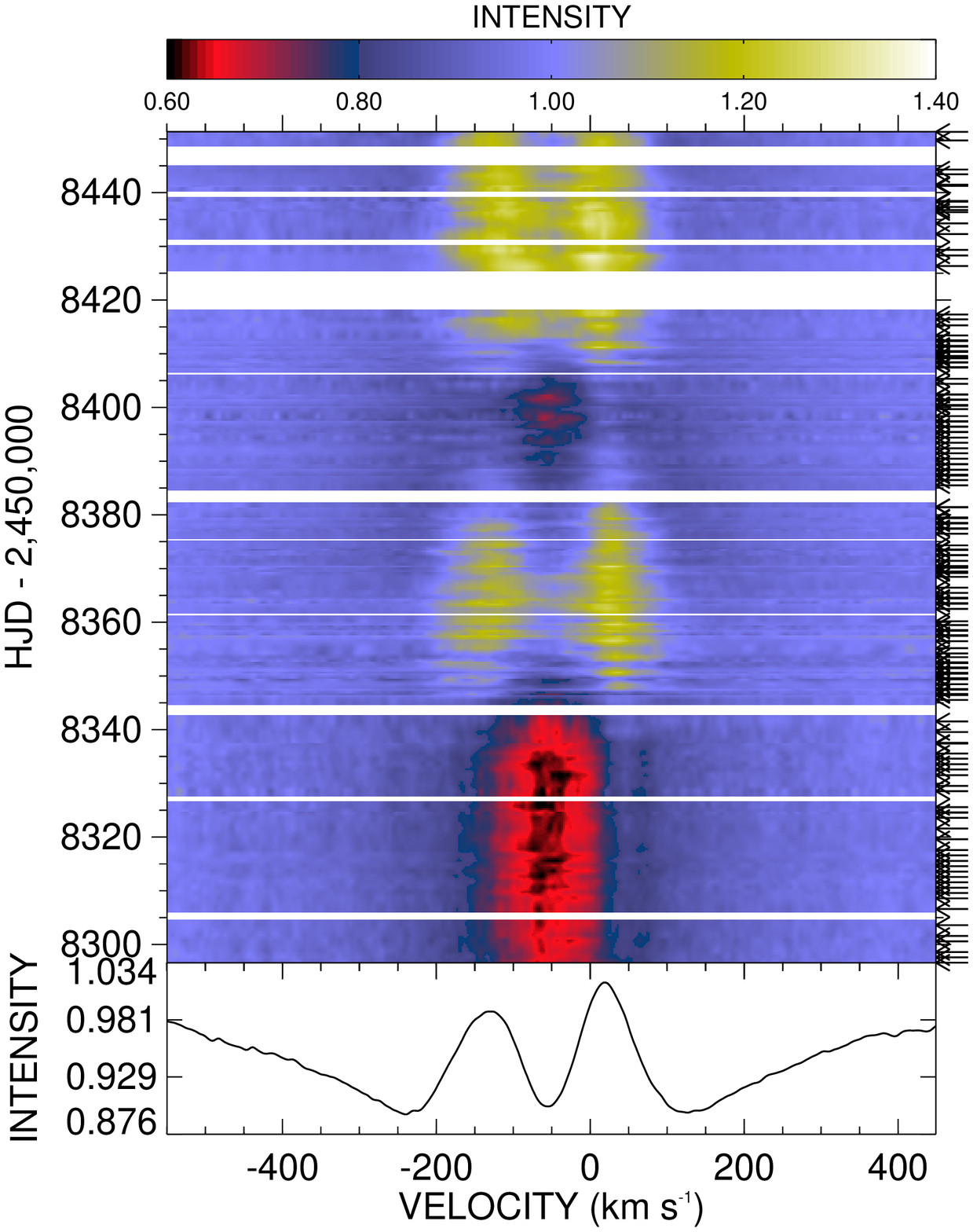}
\includegraphics[angle=0, width=5cm]{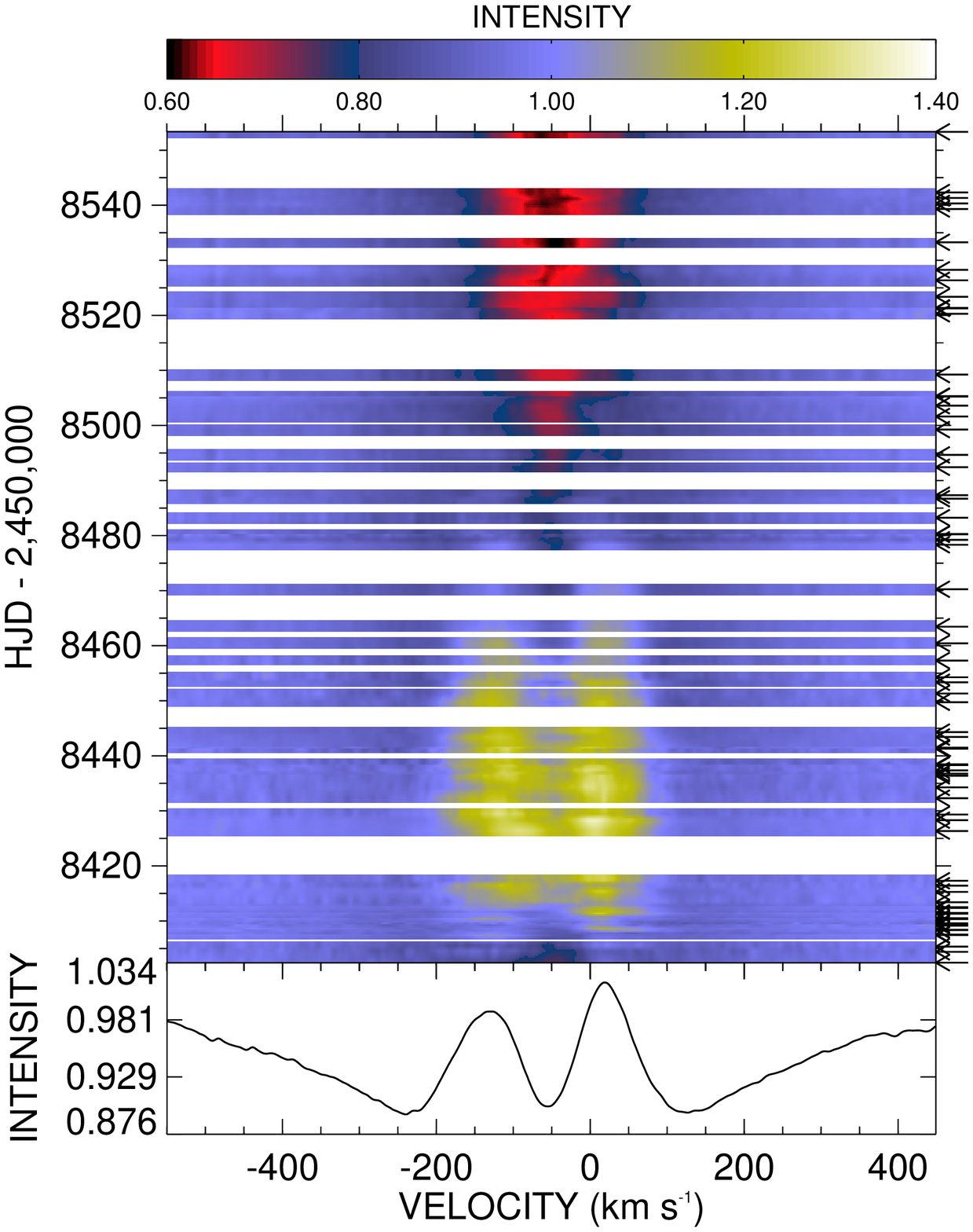}
\includegraphics[angle=0, width=5cm]{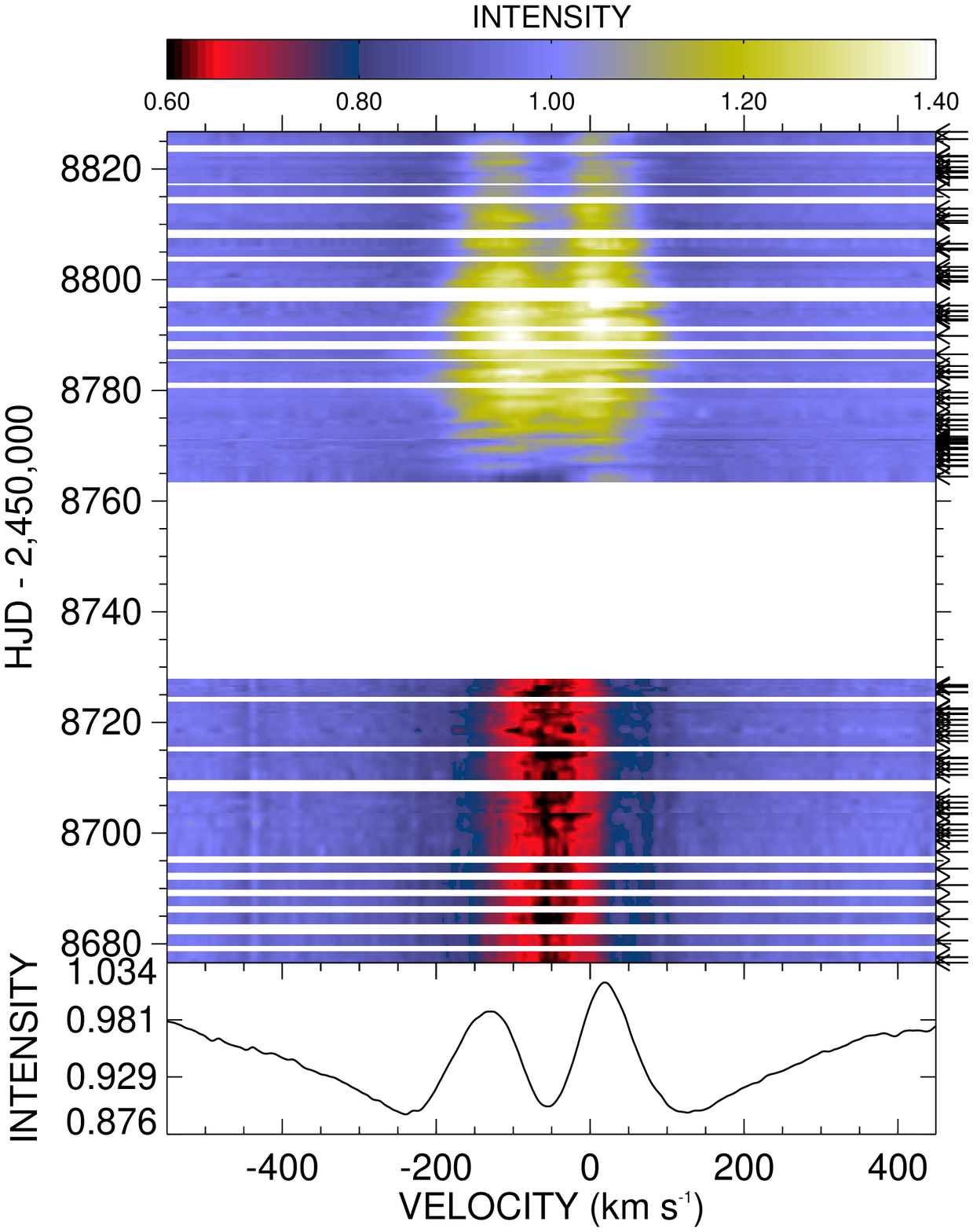}
\includegraphics[angle=0, width=5cm]{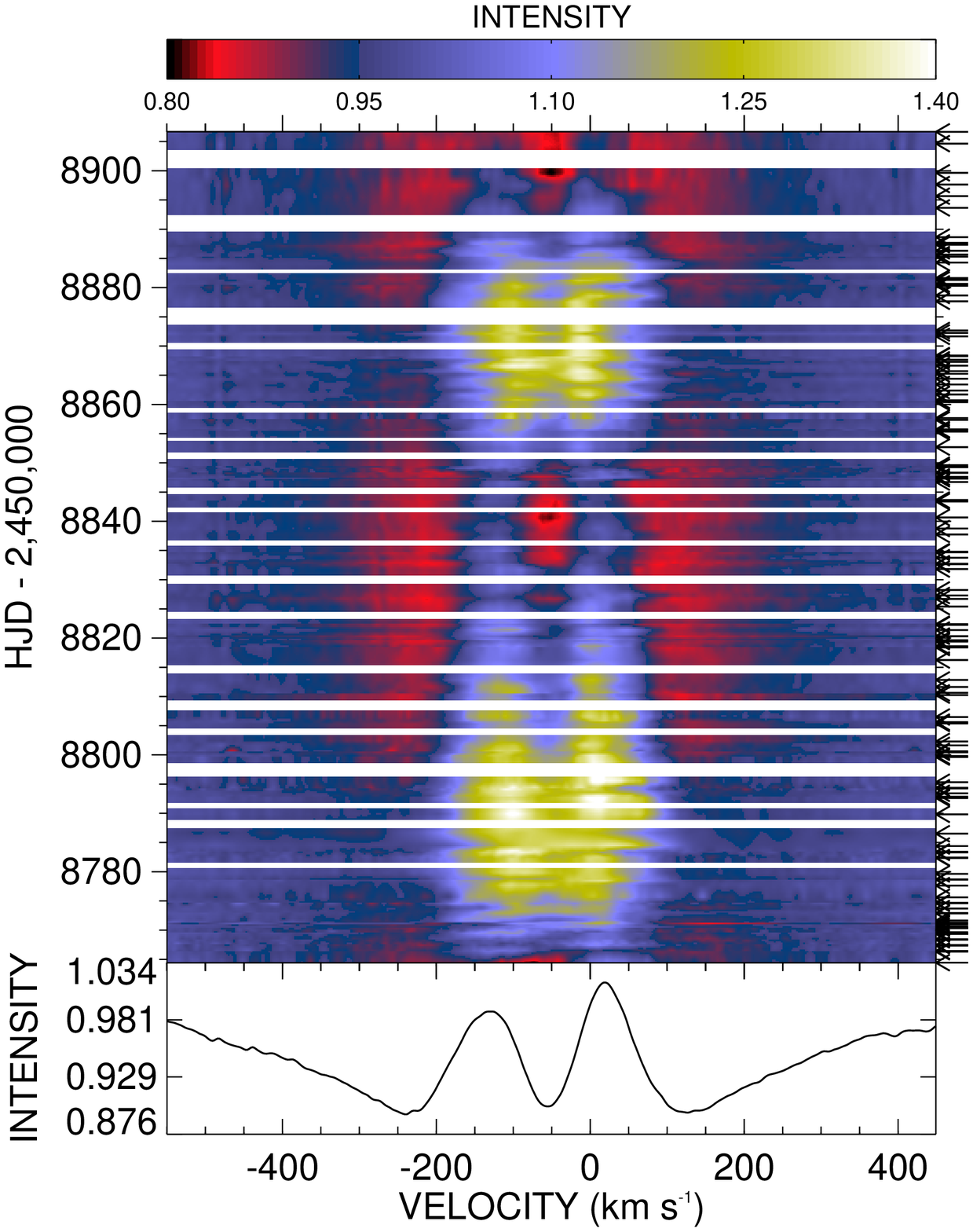}
\end{center} 
\caption{Dynamical representation of HD\,6226 H$\alpha$ profiles spectra during our campaign. The average profile for the campaign is shown in the bottom panel of each plot and the vertical scale of each plot is made to be $\sim150$ d for each panel, allowing for some overlap between some of the dynamical spectra. Gaps are made to represent gaps in data $\gtrsim 2$ d.}
\label{dynam-halpha} 
\end{figure*} 
\clearpage

\begin{figure*} 
\begin{center} 
\includegraphics[angle=0, width=18cm]{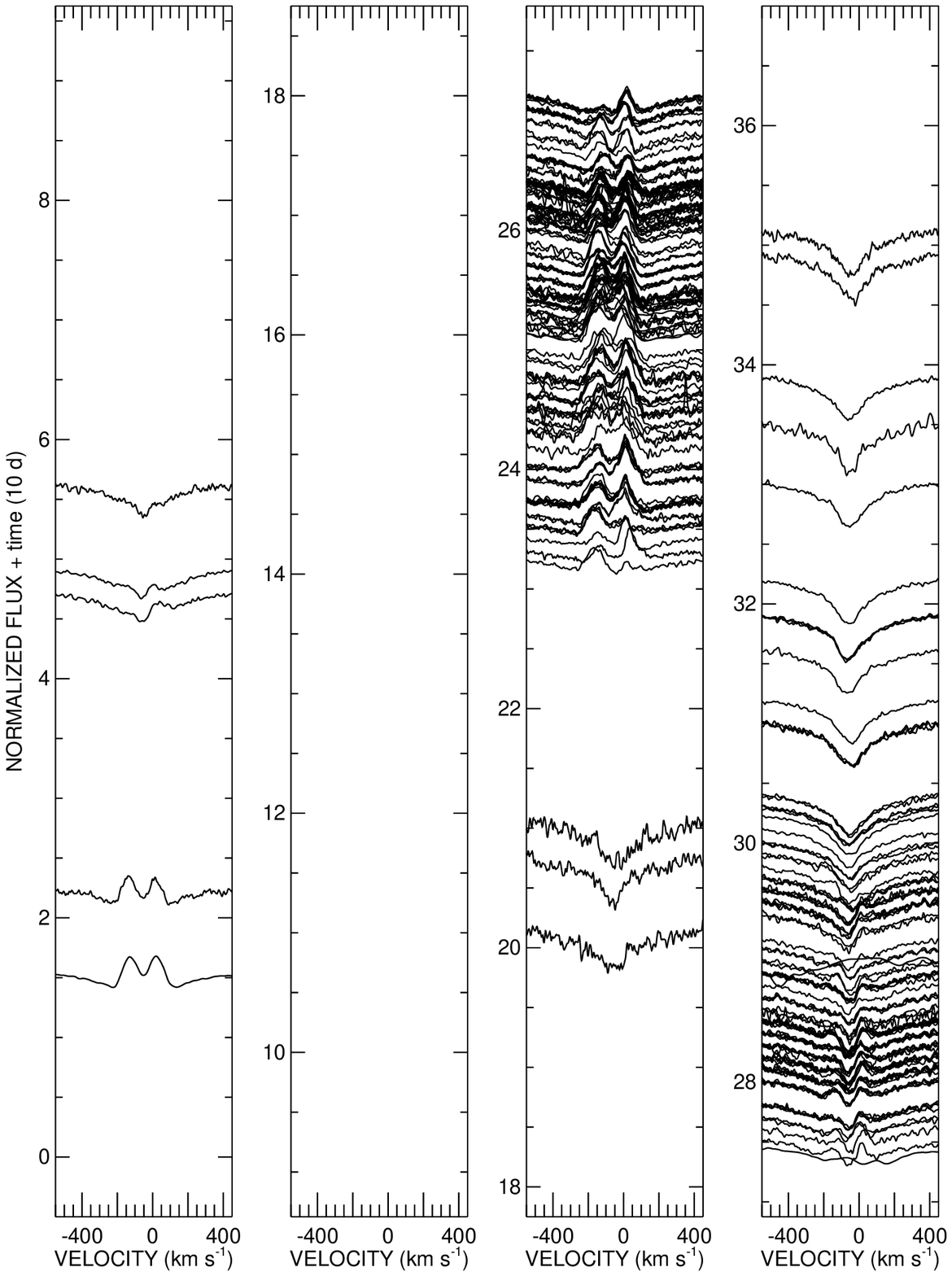}
\end{center} 
\caption{Line profiles of HD\,6226 H$\alpha$ profiles spectra during 2017. The offset for the spectra is representative of 10 d per each vertical value of 1. The time point of 0 represents 2017 January 1.}
\label{halpha-2017} 
\end{figure*} 
\clearpage

\begin{figure*} 
\begin{center} 
\includegraphics[angle=0, width=18cm]{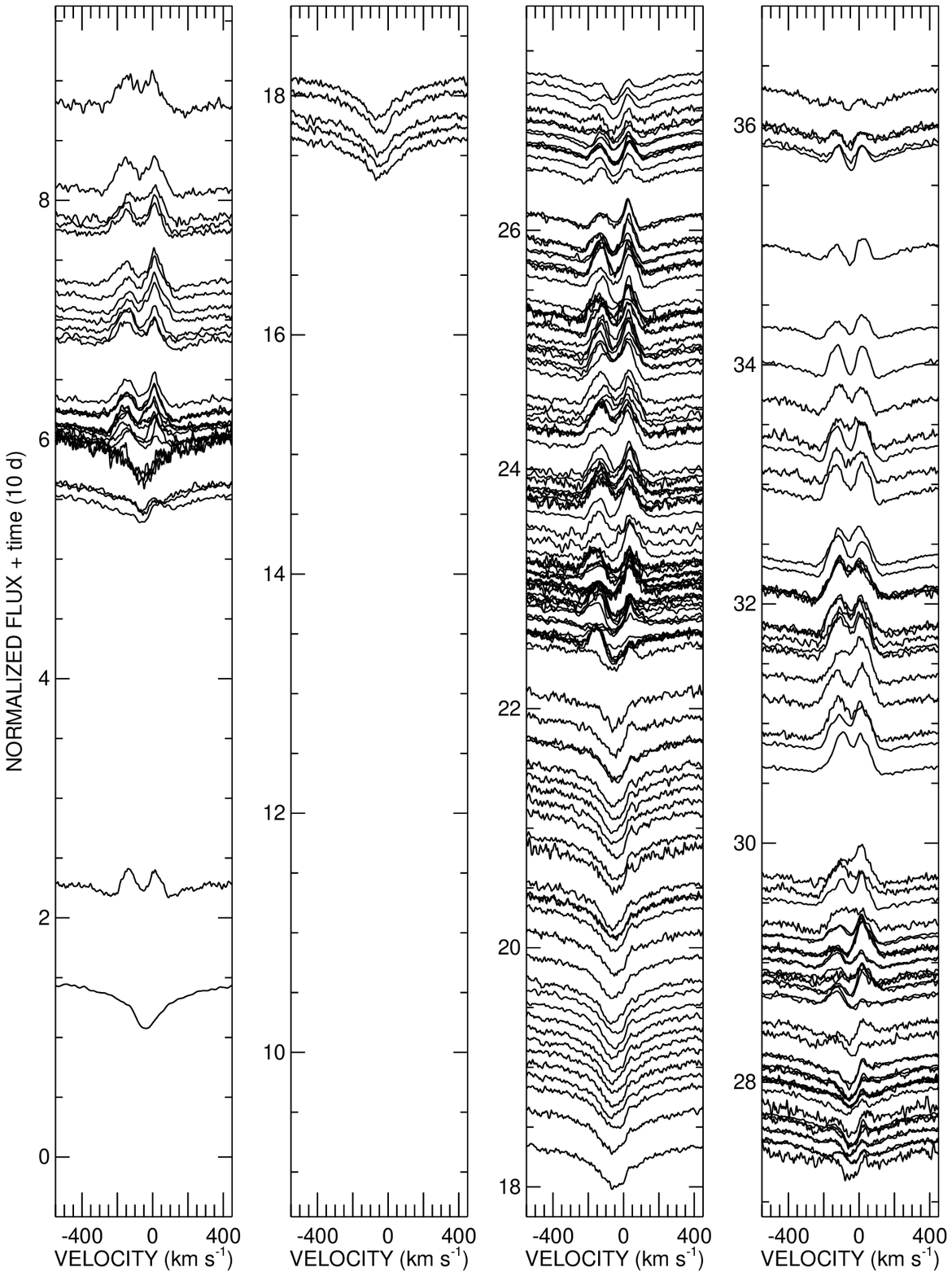}
\end{center} 
\caption{Line profiles of HD\,6226 H$\alpha$ profiles spectra during 2018. The offset for the spectra is representative of 10 d per each vertical value of 1. The time point of 0 represents 2018 January 1.}
\label{halpha-2018} 
\end{figure*} 
\clearpage

\begin{figure*} 
\begin{center} 
\includegraphics[angle=0, width=18cm]{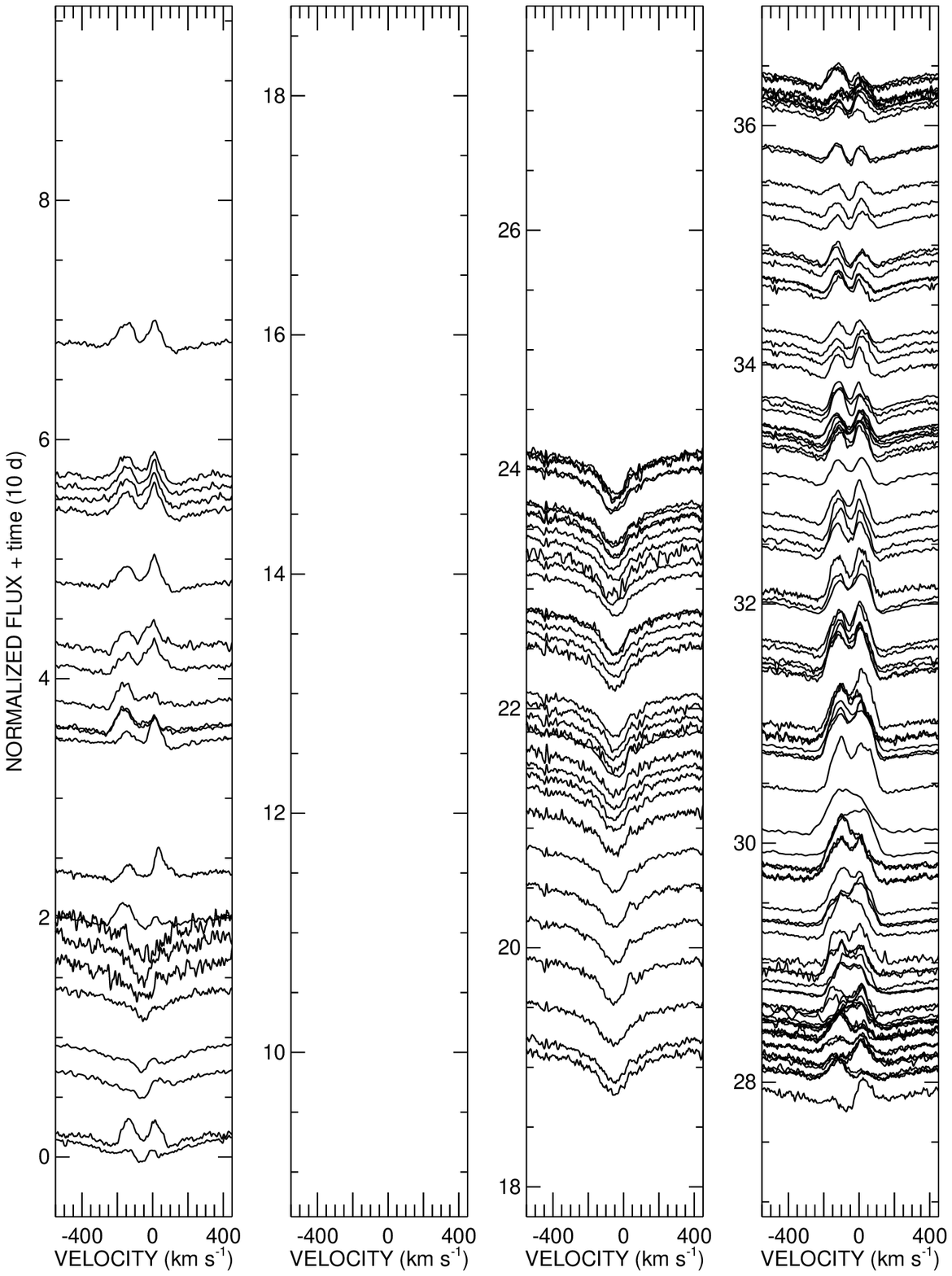}
\end{center} 
\caption{Line profiles of HD\,6226 H$\alpha$ profiles spectra during 2019. The offset for the spectra is representative of 10 d per each vertical value of 1. The time point of 0 represents 2019 January 1.}
\label{halpha-2019} 
\end{figure*} 
\clearpage

\begin{figure*} 
\begin{center} 
\includegraphics[angle=0, width=18cm]{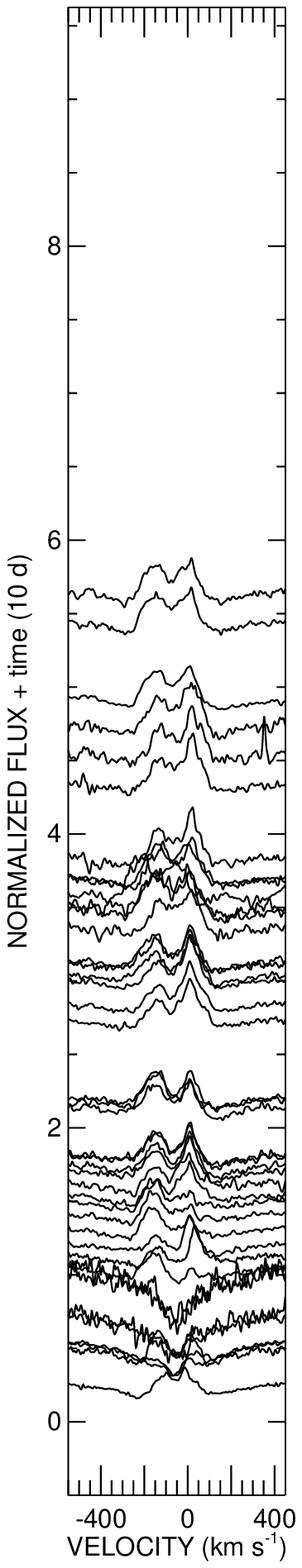}
\end{center} 
\caption{Line profiles of HD\,6226 H$\alpha$ profiles spectra during 2020 that were used in this analysis. The offset for the spectra is representative of 10 d per each vertical value of 1. The time point of 0 represents 2020 January 1.}
\label{halpha-2020} 
\end{figure*} 
\clearpage

\begin{figure*} 
\begin{center} 
\includegraphics[angle=0, width=5cm]{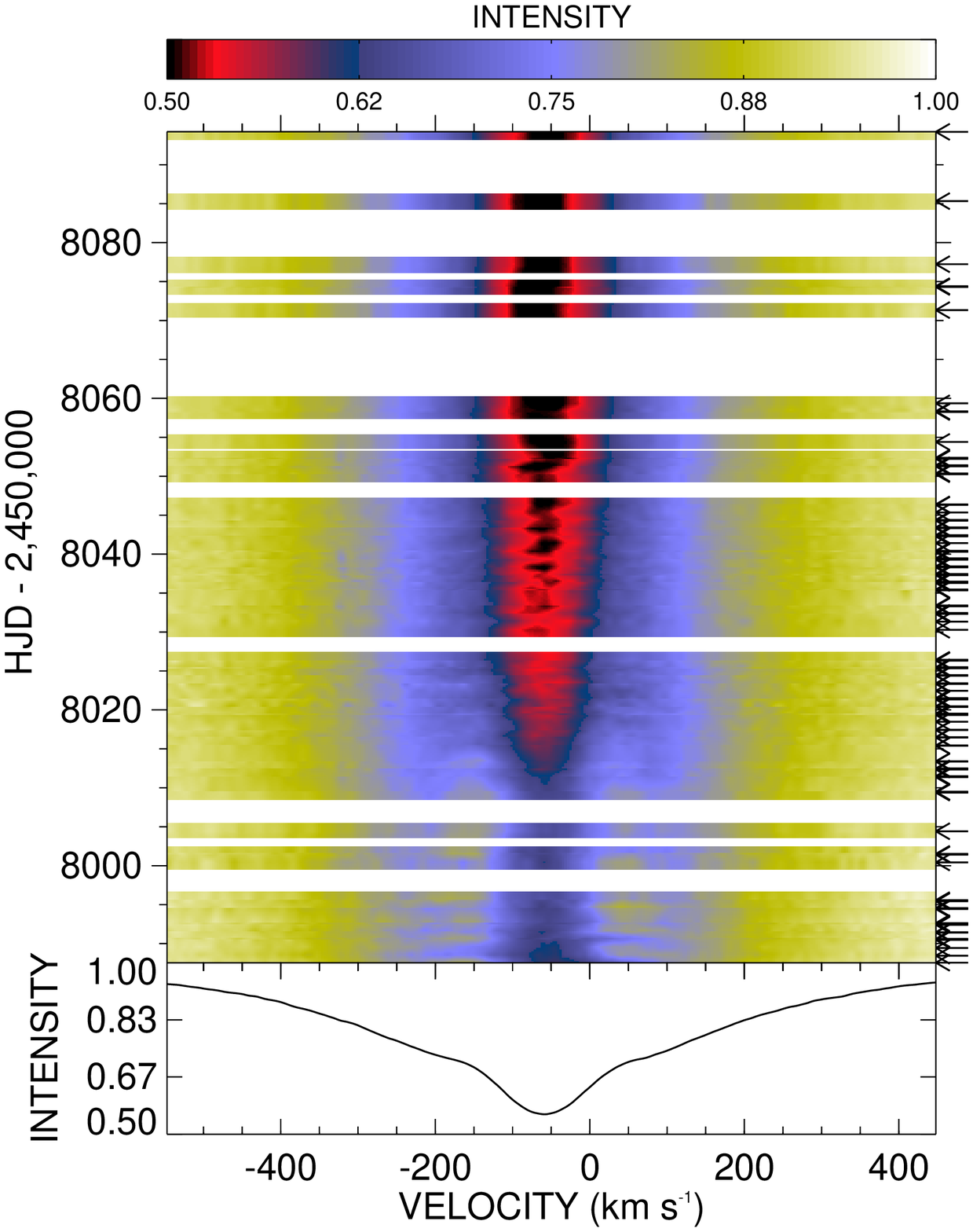}
\includegraphics[angle=0, width=5cm]{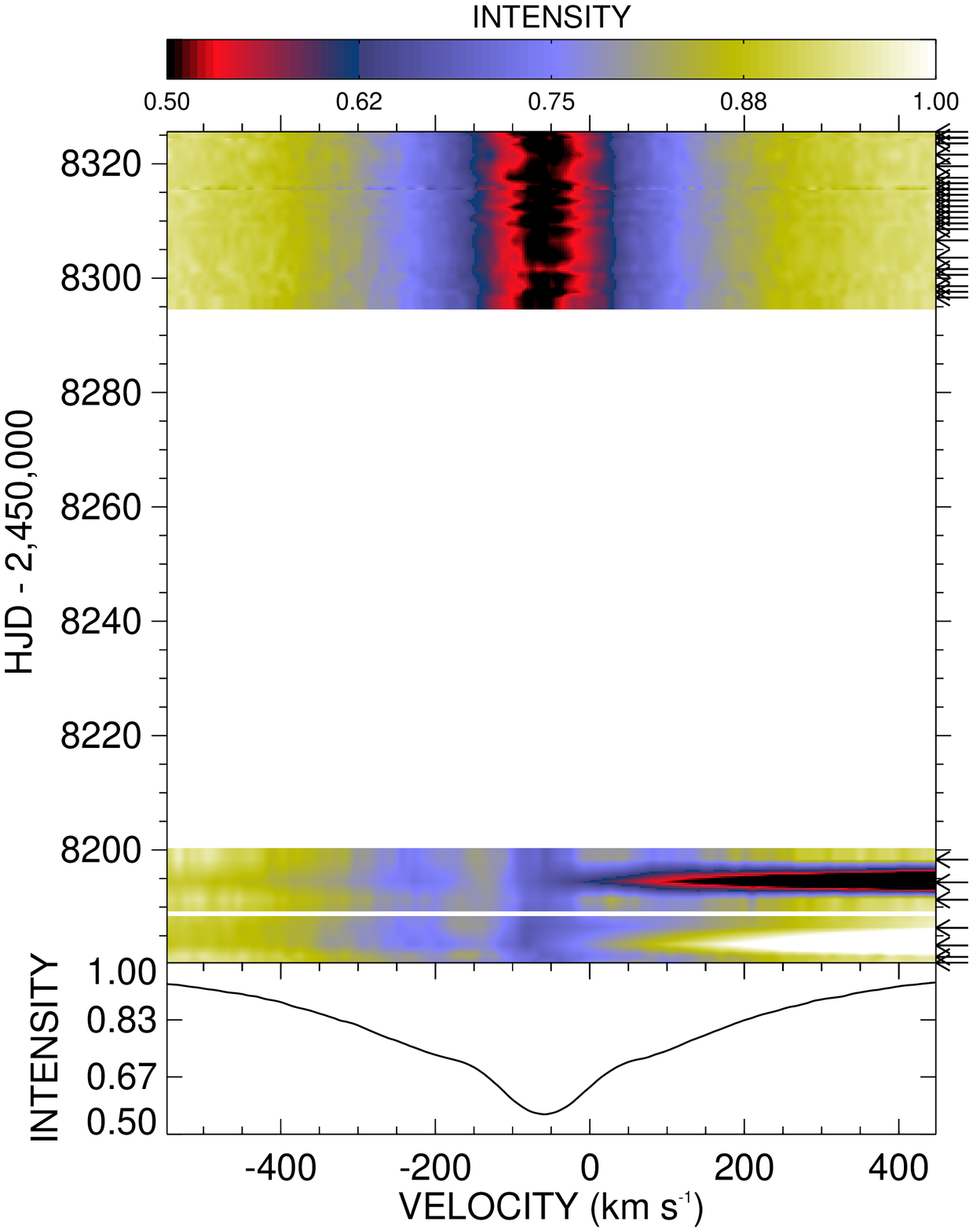}
\includegraphics[angle=0, width=5cm]{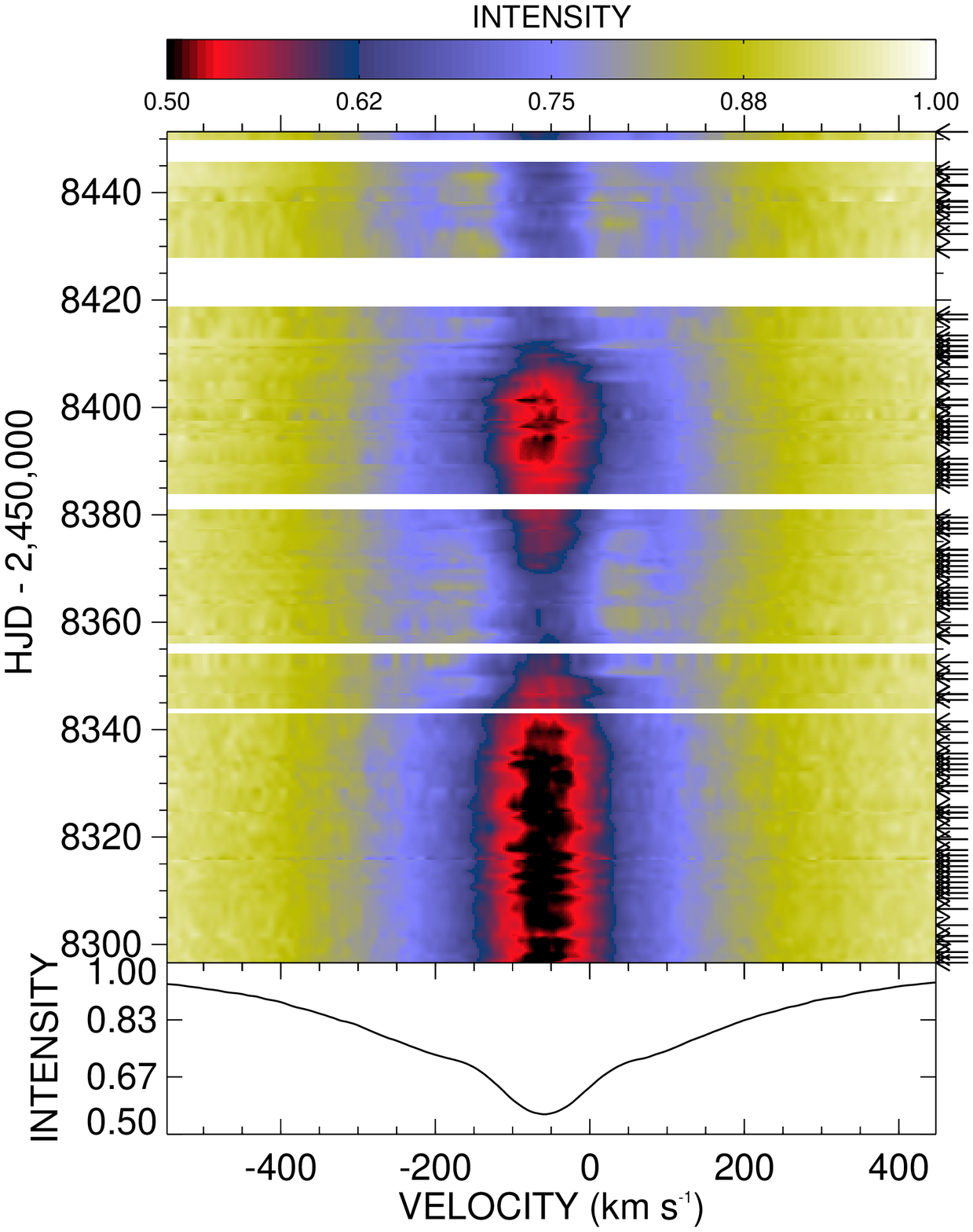}
\includegraphics[angle=0, width=5cm]{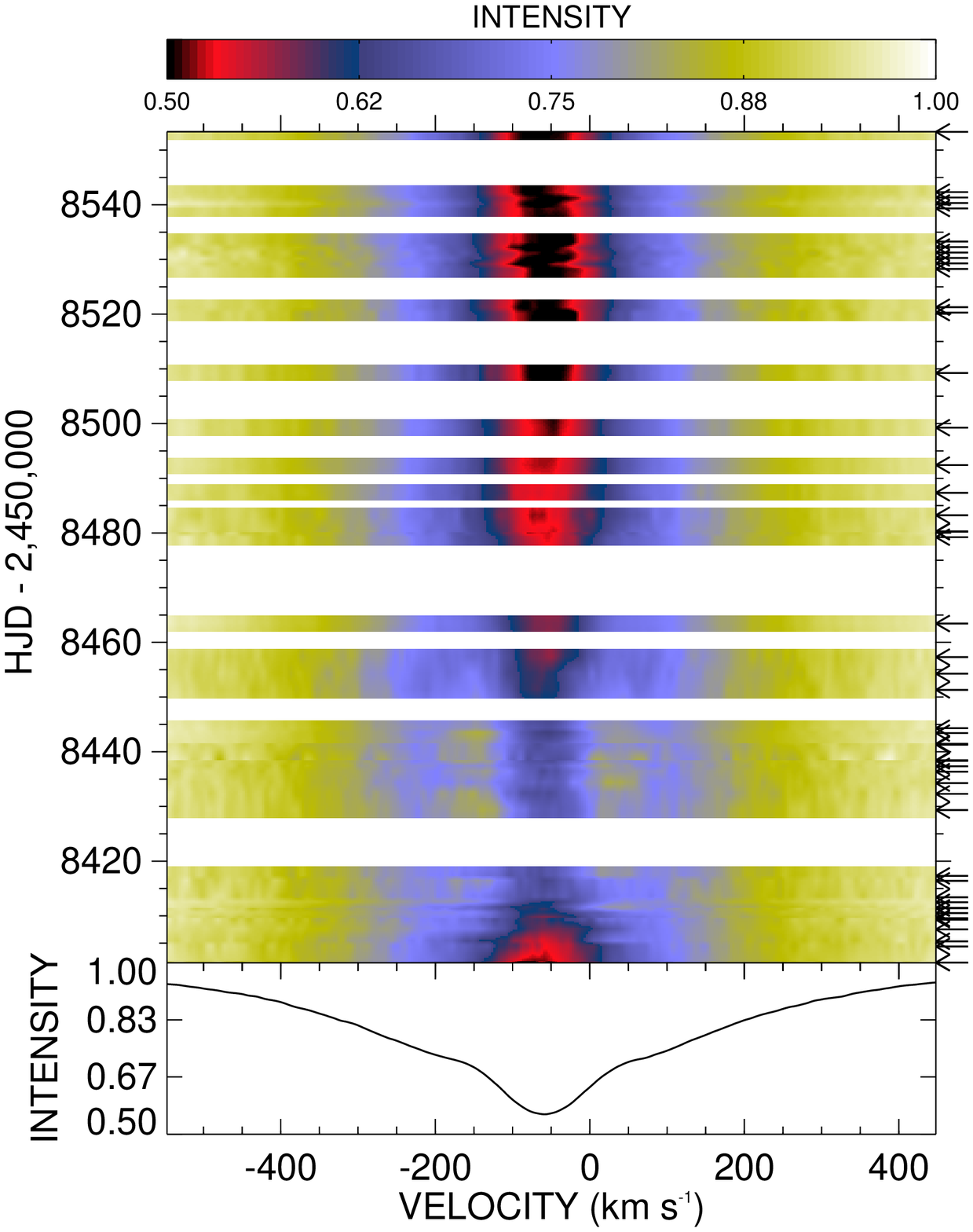}
\includegraphics[angle=0, width=5cm]{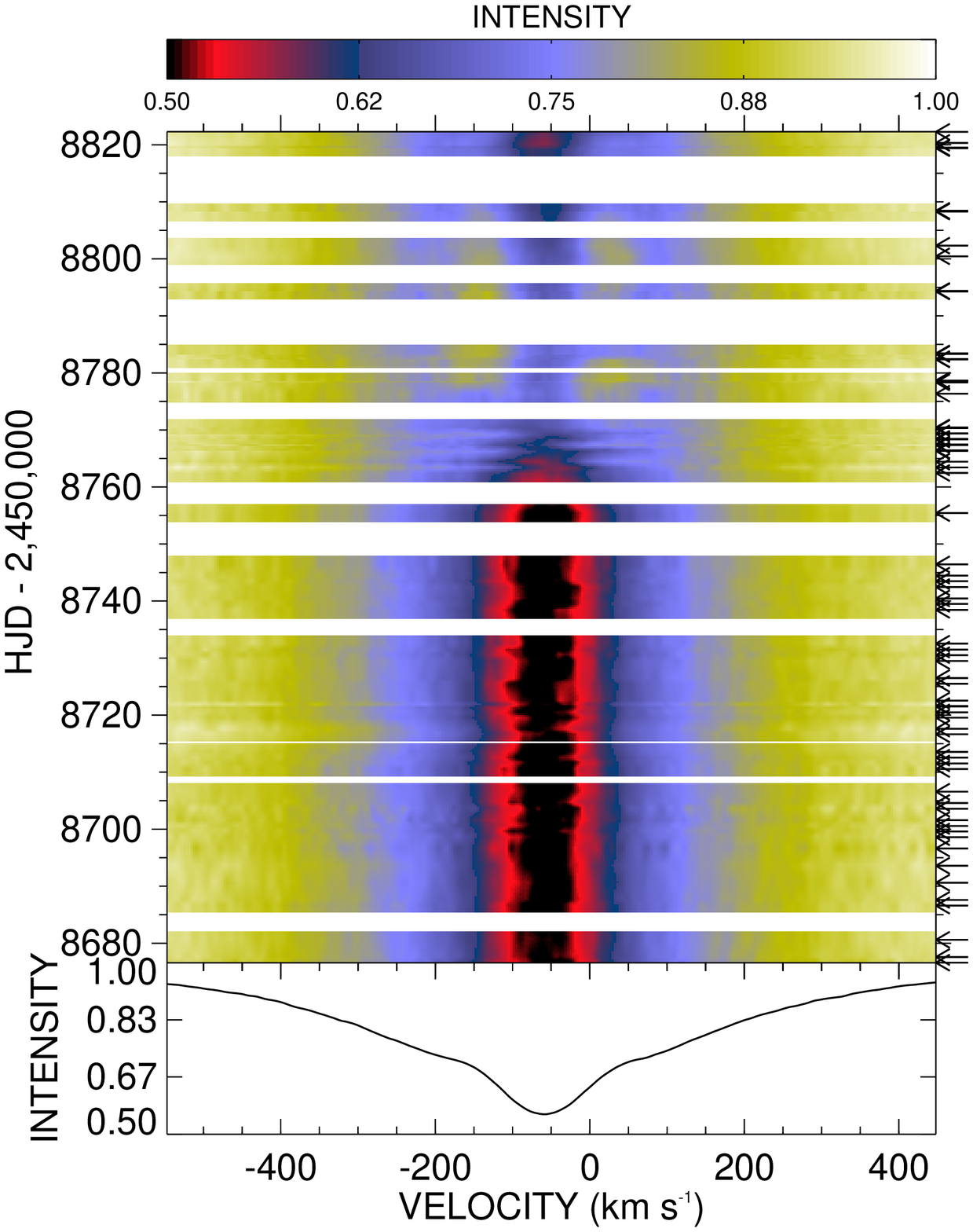}
\includegraphics[angle=0, width=5cm]{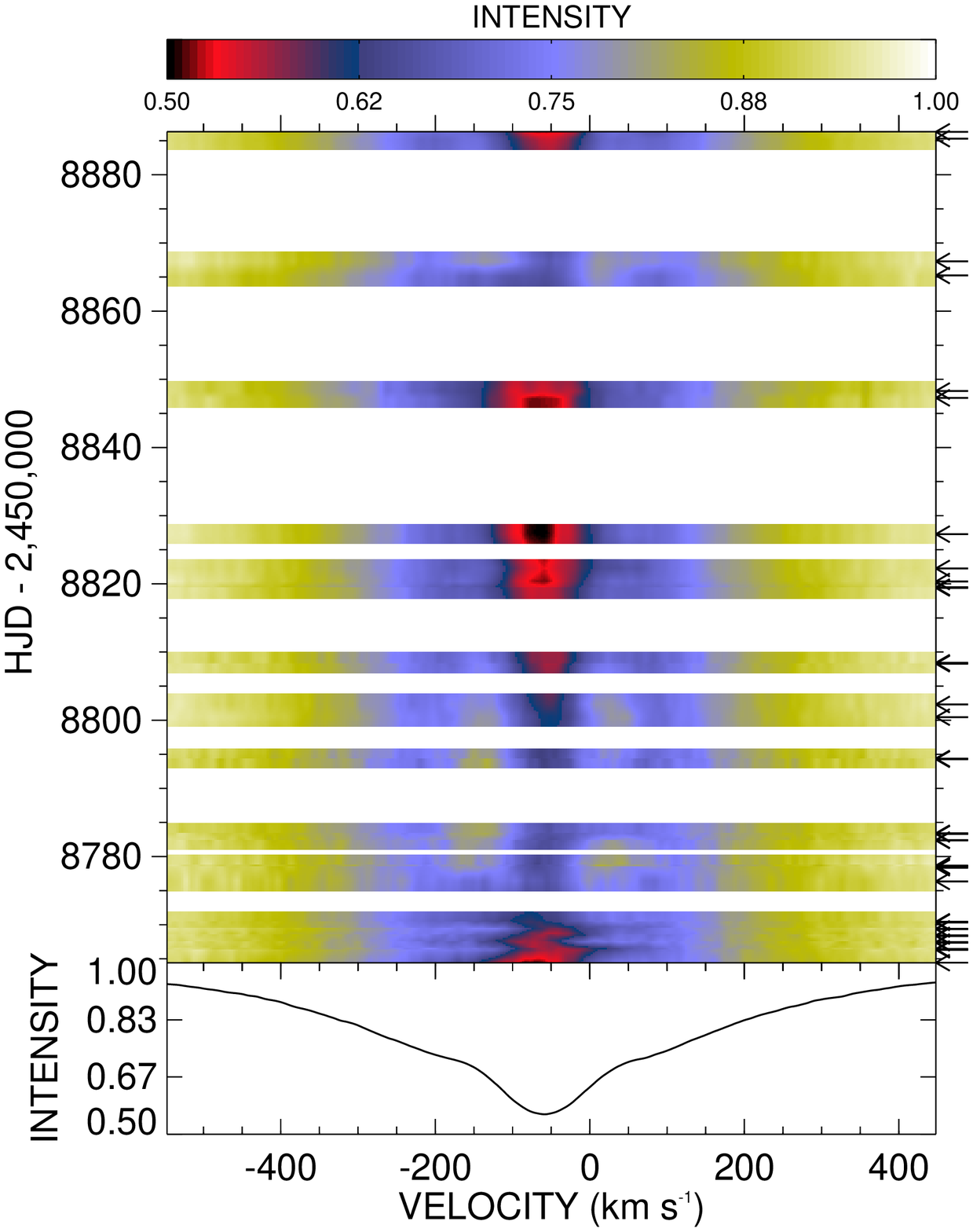}
\end{center} 
\caption{Dynamical representation of HD\,6226 H$\beta$ profiles spectra during our campaign. The average profile for the campaign is shown in the bottom panel of each plot and the vertical scale of each plot is made to be $\sim150$ d for each panel, allowing for some overlap between some of the dynamical spectra. Gaps are made to represent gaps in data $\gtrsim 2$ d.}
\label{dynam-halpha} 
\end{figure*} 
\clearpage

\begin{figure*} 
\begin{center} 
\includegraphics[angle=0, width=18cm]{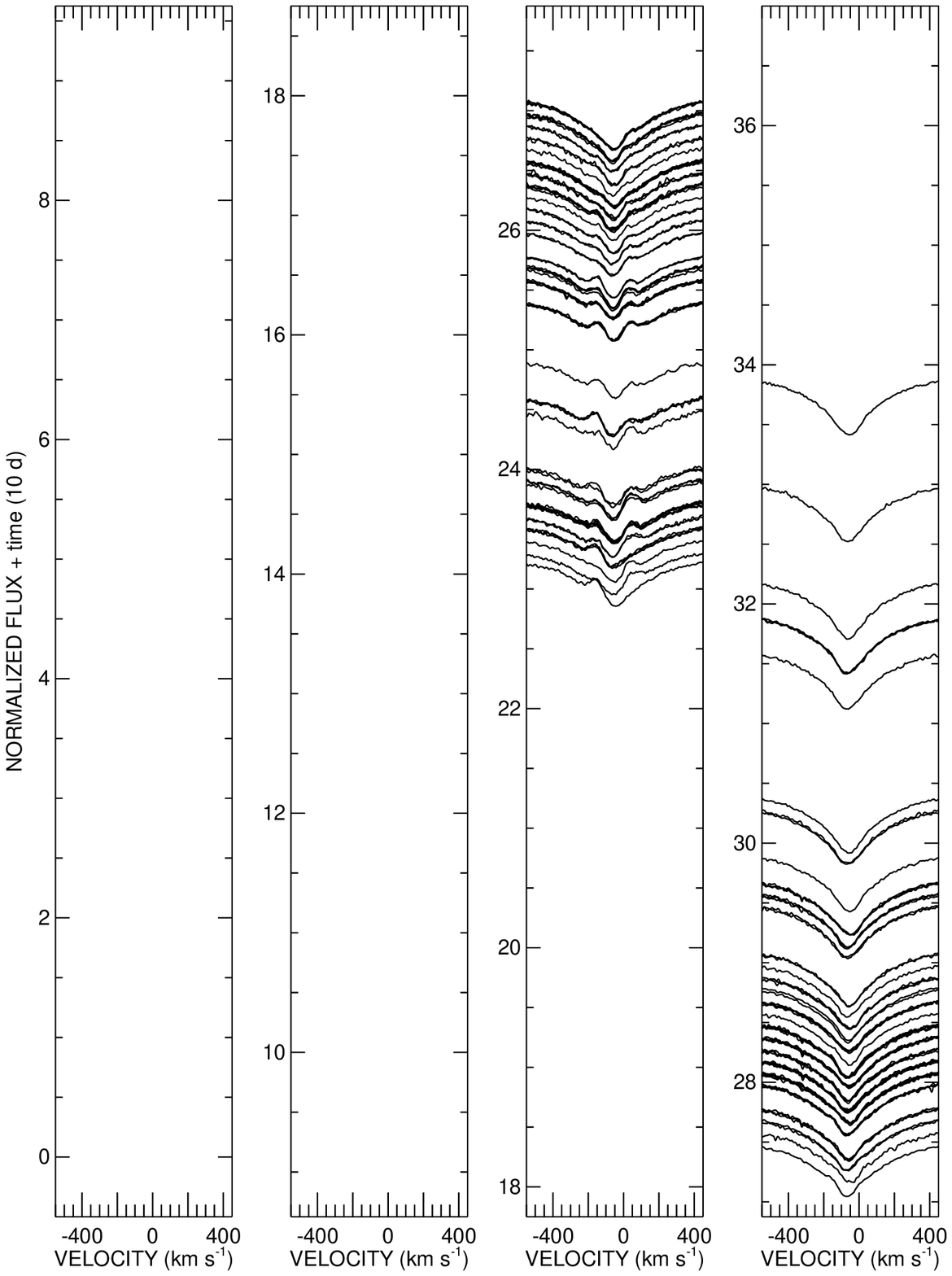}
\end{center} 
\caption{Line profiles of HD\,6226 H$\alpha$ profiles spectra during 2017. The offset for the spectra is representative of 10 d per each vertical value of 1. The time point of 0 represents 2017 January 1.}
\label{halpha-2017} 
\end{figure*} 
\clearpage

\begin{figure*} 
\begin{center} 
\includegraphics[angle=0, width=18cm]{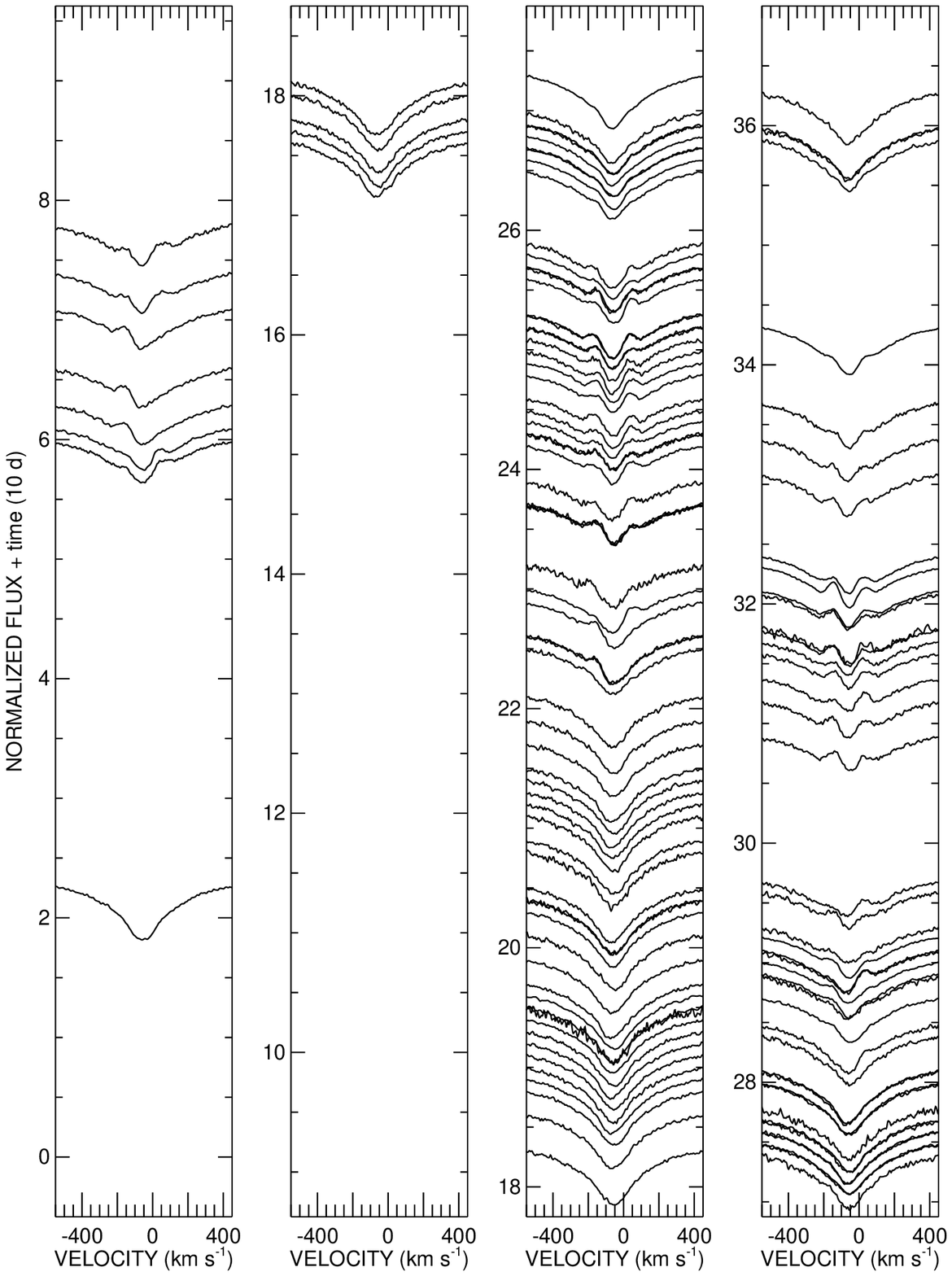}
\end{center} 
\caption{Line profiles of HD\,6226 H$\alpha$ profiles spectra during 2018. The offset for the spectra is representative of 10 d per each vertical value of 1. The time point of 0 represents 2018 January 1.}
\label{halpha-2018} 
\end{figure*} 
\clearpage

\begin{figure*} 
\begin{center} 
\includegraphics[angle=0, width=18cm]{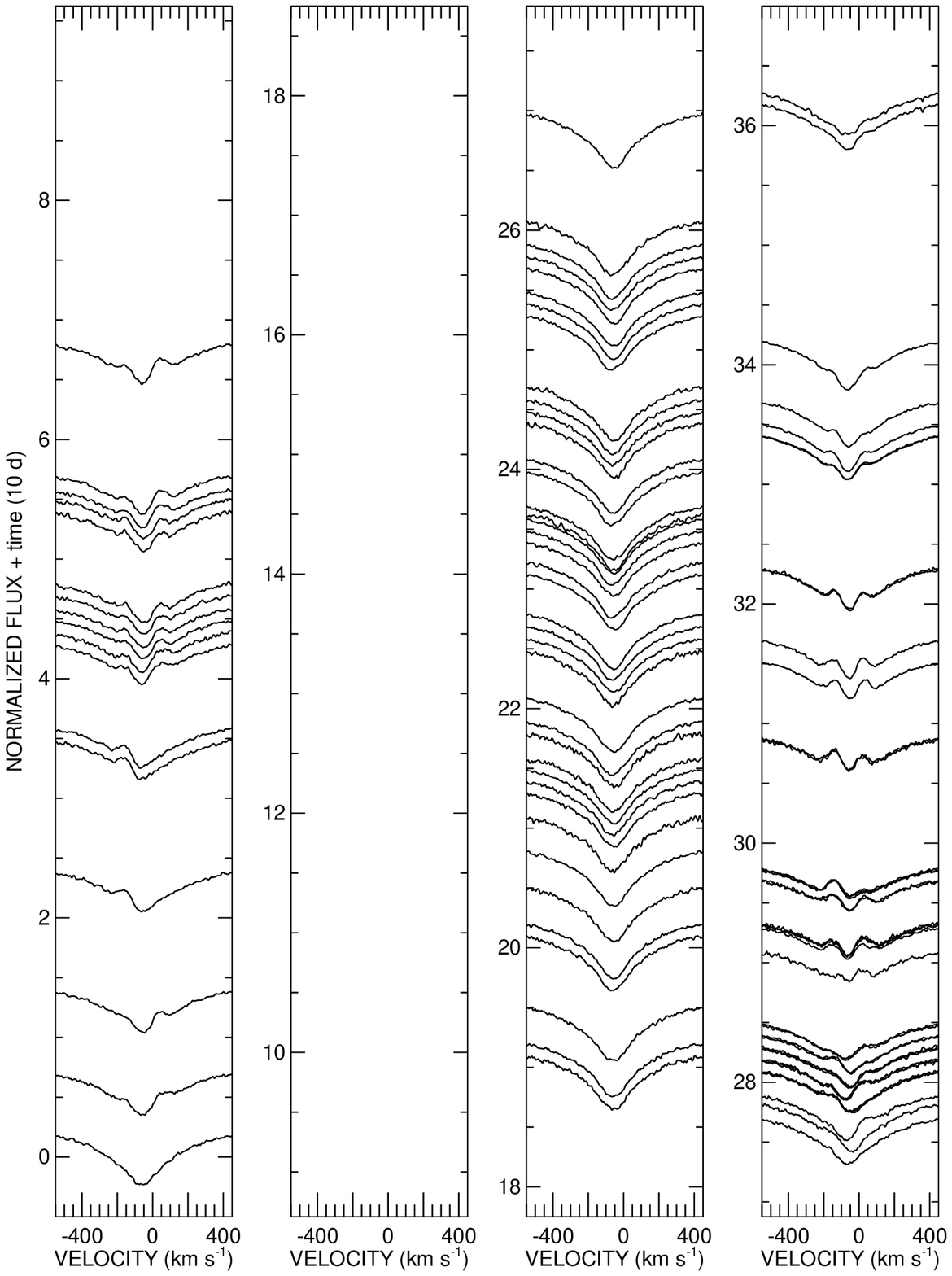}
\end{center} 
\caption{Line profiles of HD\,6226 H$\alpha$ profiles spectra during 2019. The offset for the spectra is representative of 10 d per each vertical value of 1. The time point of 0 represents 2019 January 1.}
\label{halpha-2019} 
\end{figure*} 
\clearpage

\begin{figure*} 
\begin{center} 
\includegraphics[angle=0, width=18cm]{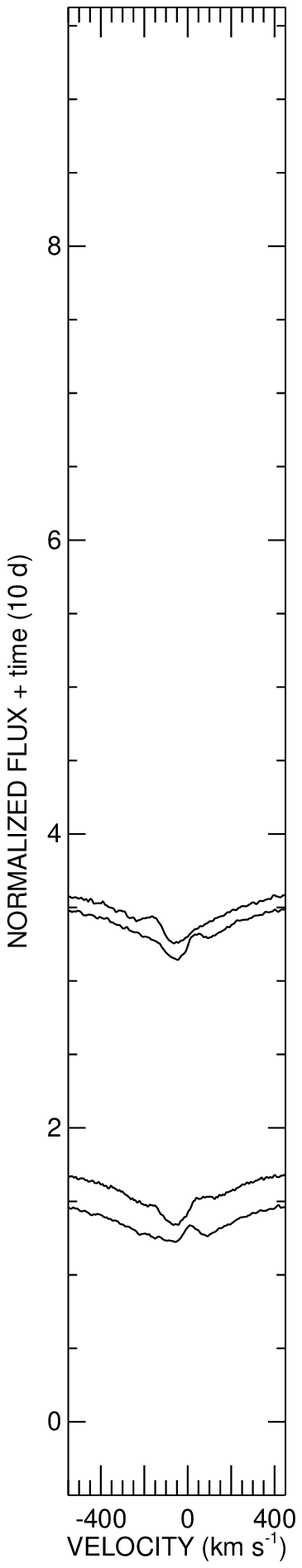}
\end{center} 
\caption{Line profiles of HD\,6226 H$\alpha$ profiles spectra during 2020 that were used in this analysis. The offset for the spectra is representative of 10 d per each vertical value of 1. The time point of 0 represents 2020 January 1.}
\label{halpha-2020} 
\end{figure*} 
\clearpage

\begin{figure*}

\begin{center} 
\includegraphics[angle=0, width=8cm]{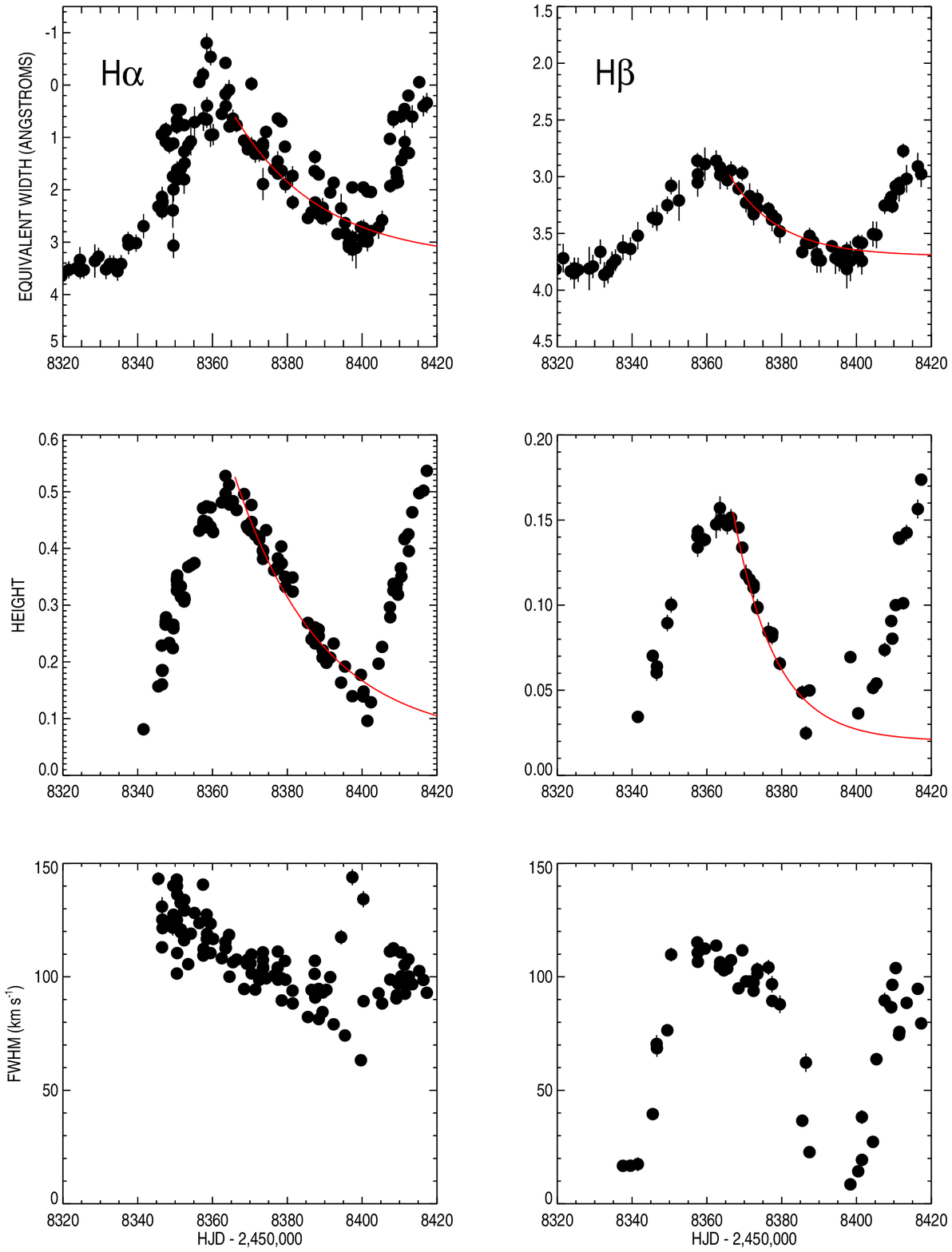}
\includegraphics[angle=0, width=8cm]{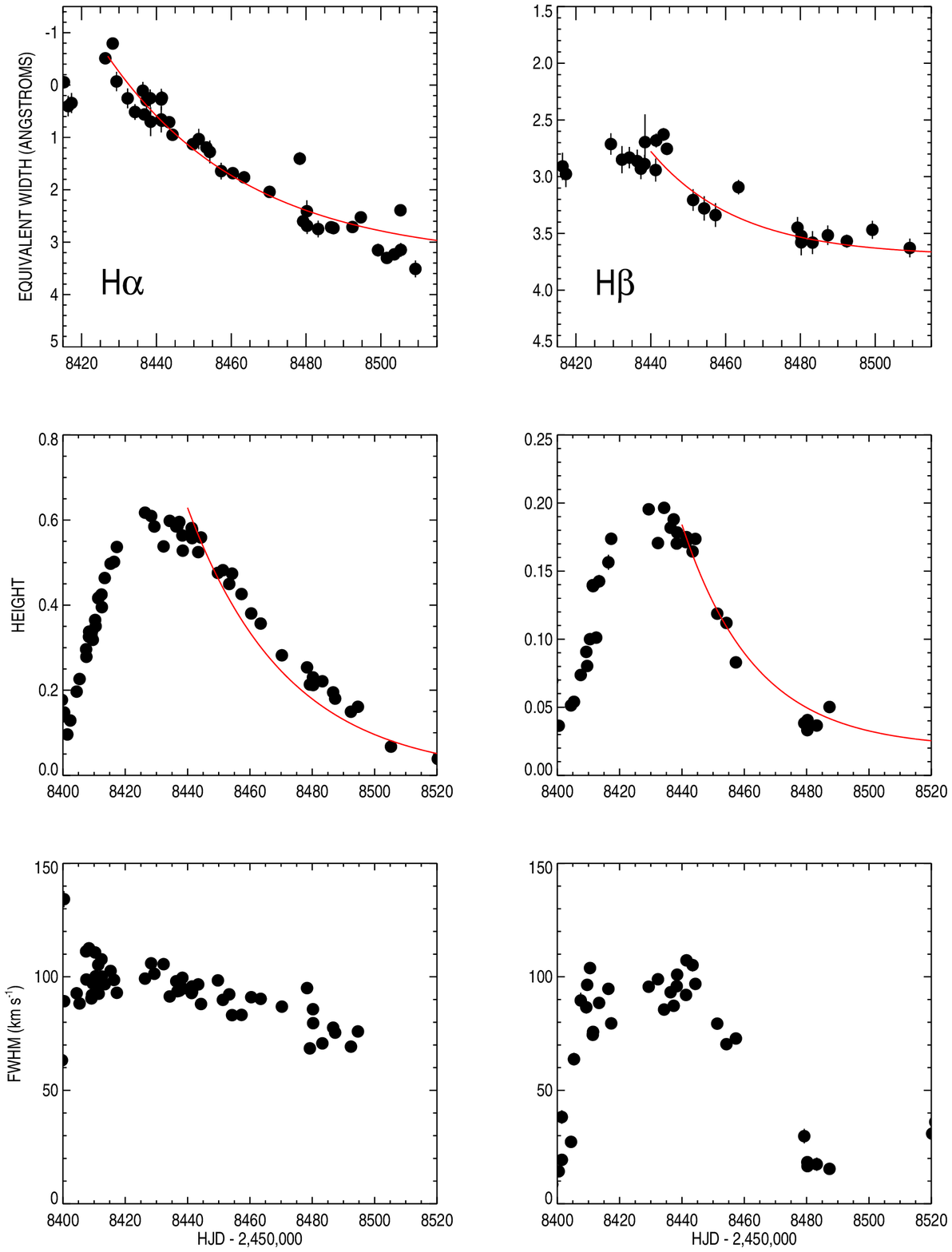}
\includegraphics[angle=0, width=8cm]{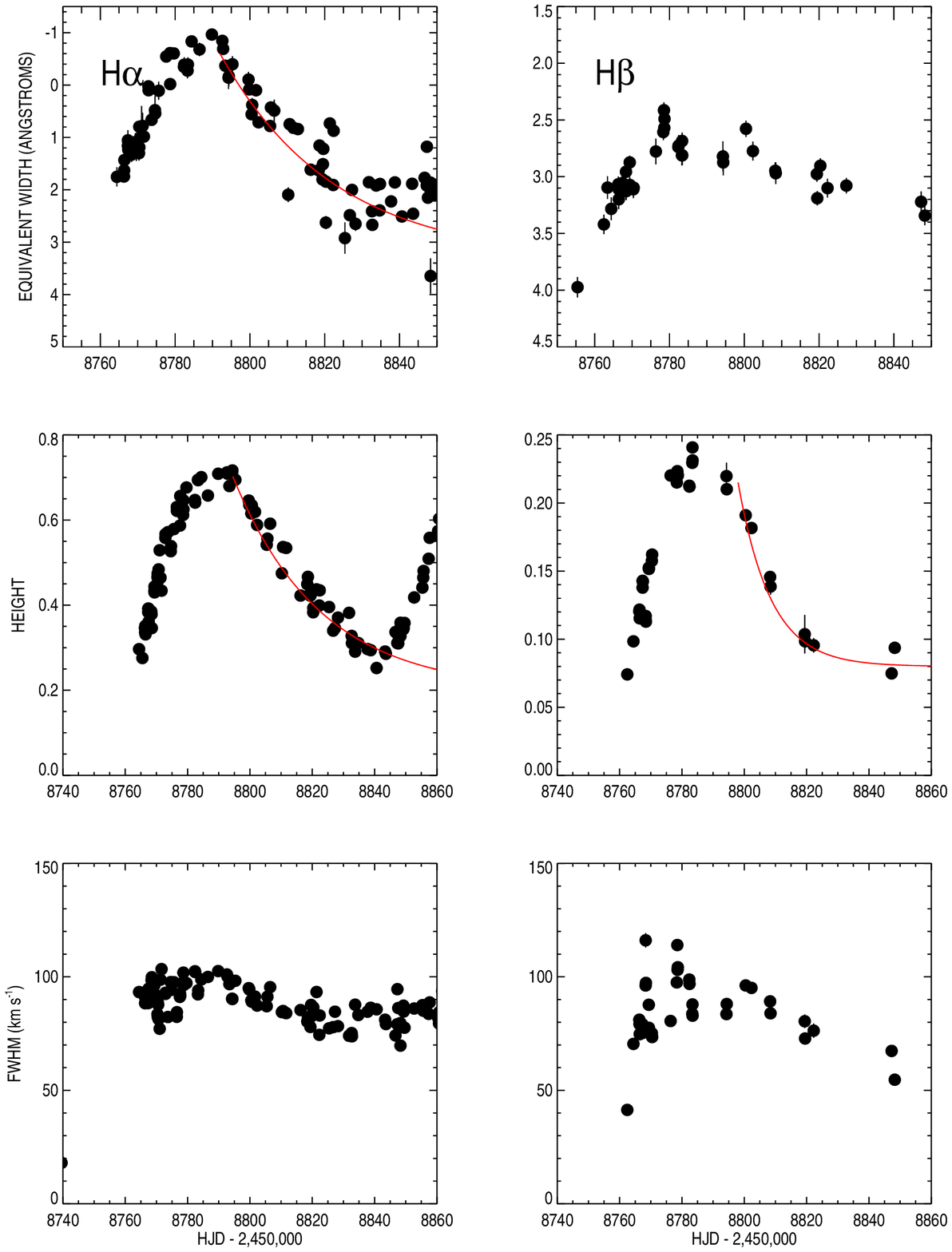}
\includegraphics[angle=0, width=8cm]{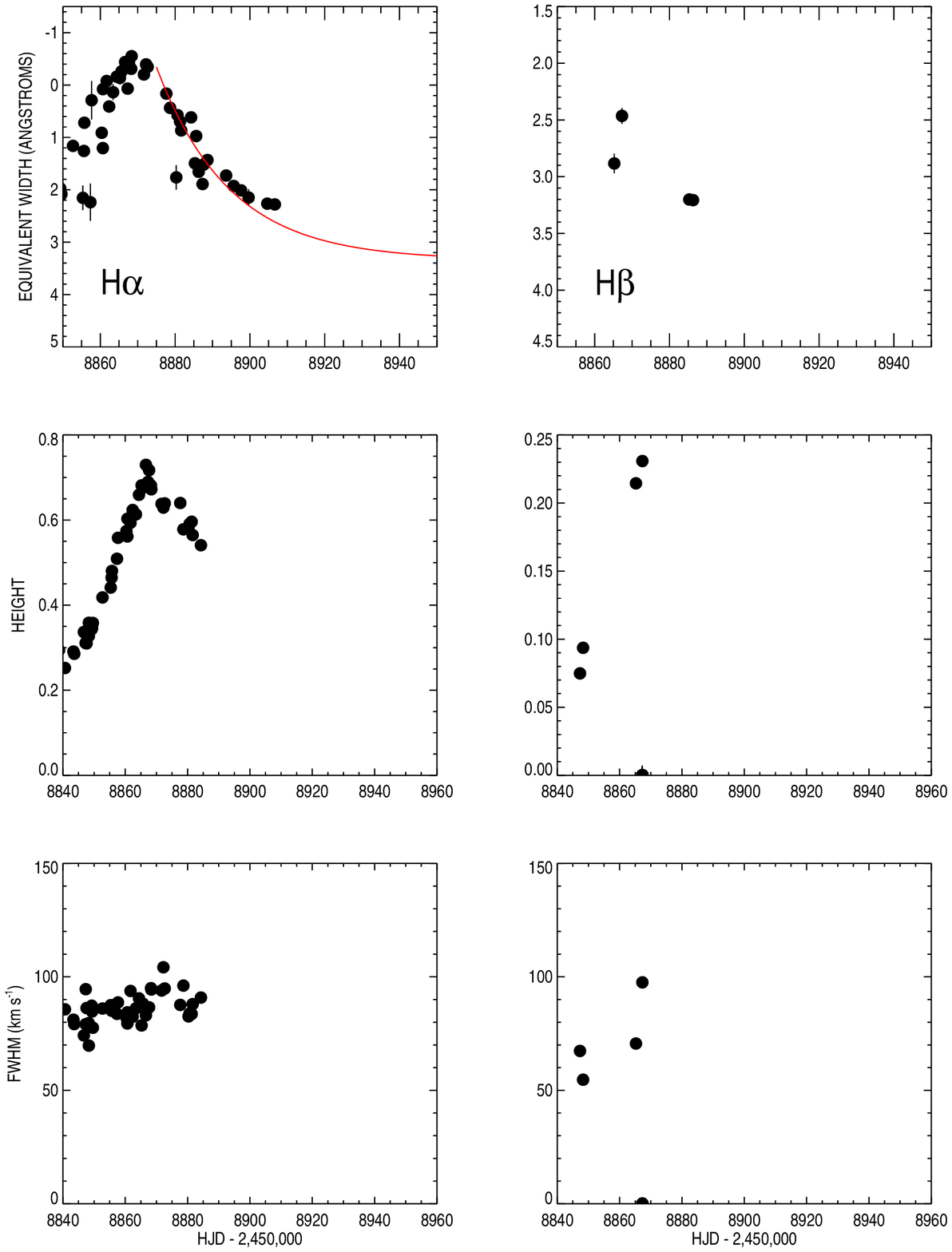}

\end{center} 
\caption{Modeling of the disk dissipation for four additional outbursts seen in our spectroscopic data, in the same format as Fig. \ref{outburst1}, with the fit parameters given in Table \ref{exponential}. } 
\label{dynam-puls-appendix} 
\end{figure*} 
\clearpage

\clearpage
\section{Data Tables}

\begin{table*}
\centering
\begin{minipage}{180mm}
\caption{Radial Velocity Measurements \label{radialvels}}
\begin{tabular}{lccccc}
\hline
HJD-2450000 &   He I 4471   &   He I 4721   & He I 4921 & He I 5876 &   Mg II 4481 \\
\hline
7987.5452	&	-56.13	$\pm$	0.94	&	-56.20	$\pm$	1.05	&	-51.78	$\pm$	1.07	&	-49.96	$\pm$	1.00	&	-57.55	$\pm$	1.53	\\
7988.4413	&	-57.60	$\pm$	0.65	&	-63.27	$\pm$	0.85	&	-59.11	$\pm$	0.82	&		\ldots		&	-65.28	$\pm$	0.98	\\
7989.4361	&	-59.95	$\pm$	0.80	&	-52.87	$\pm$	1.02	&	-54.21	$\pm$	1.26	&	-55.99	$\pm$	0.92	&	-49.94	$\pm$	0.97	\\
7990.4747	&	-59.26	$\pm$	0.87	&	-60.16	$\pm$	1.03	&	-56.01	$\pm$	0.95	&	-53.50	$\pm$	1.34	&	-60.14	$\pm$	1.33	\\
7990.5549	&	-58.91	$\pm$	0.83	&	-57.01	$\pm$	1.32	&	-55.26	$\pm$	1.12	&	-53.56	$\pm$	1.49	&	-65.39	$\pm$	1.54	\\
7990.6008	&	-58.82	$\pm$	1.09	&	-61.87	$\pm$	1.27	&	-58.24	$\pm$	1.20	&	-56.08	$\pm$	1.20	&	-66.63	$\pm$	1.41	\\
7991.4147	&	-60.44	$\pm$	0.71	&	-55.27	$\pm$	0.75	&	-55.79	$\pm$	0.87	&	-57.93	$\pm$	0.74	&	-61.21	$\pm$	0.79	\\
\end{tabular}
\end{minipage}
\end{table*}
\begin{table*}
\centering
\begin{minipage}{180mm}
\contcaption{Radial Velocity Measurements \label{radialvels}}
\begin{tabular}{lccccc}
\hline
HJD-2450000 &   He I 4471   &   He I 4721   & He I 4921 & He I 5876 &   Mg II 4481 \\
\hline
8025.3454	&	-60.95	$\pm$	0.73	&	-60.17	$\pm$	0.88	&	-56.76	$\pm$	0.88	&	-56.95	$\pm$	0.71	&	-62.62	$\pm$	0.96	\\
8025.3880	&	-61.85	$\pm$	0.82	&	-60.91	$\pm$	0.82	&	-55.28	$\pm$	0.88	&	-58.75	$\pm$	0.65	&	-60.96	$\pm$	0.98	\\
8025.4309	&	-60.04	$\pm$	0.74	&	-58.13	$\pm$	0.77	&	-53.33	$\pm$	0.92	&	-54.98	$\pm$	0.63	&	-60.80	$\pm$	1.05	\\
8025.4739	&	-59.74	$\pm$	0.85	&	-58.03	$\pm$	1.16	&	-53.13	$\pm$	1.14	&	-55.84	$\pm$	0.61	&	-56.91	$\pm$	1.34	\\
8026.3308	&	-55.28	$\pm$	0.89	&	-55.76	$\pm$	1.23	&	-48.69	$\pm$	1.14	&	-48.26	$\pm$	1.02	&	-53.56	$\pm$	1.88	\\
8026.3733	&	-55.21	$\pm$	1.04	&	-57.18	$\pm$	1.33	&	-47.25	$\pm$	1.19	&	-48.93	$\pm$	1.02	&	-50.56	$\pm$	1.47	\\
8026.4026	&	-46.67	$\pm$	0.68	&	-47.68	$\pm$	0.82	&	-49.02	$\pm$	1.11	&	-50.81	$\pm$	1.01	&	-45.19	$\pm$	1.19	\\
\end{tabular}
\end{minipage}
\end{table*}
\begin{table*}
\centering
\begin{minipage}{180mm}
\contcaption{Radial Velocity Measurements \label{radialvels}}
\begin{tabular}{lccccc}
\hline
HJD-2450000 &   He I 4471   &   He I 4721   & He I 4921 & He I 5876 &   Mg II 4481 \\
\hline
8052.2411	&	-54.99	$\pm$	0.95	&	-53.72	$\pm$	1.09	&	-48.33	$\pm$	1.18	&	-48.21	$\pm$	0.81	&	-49.34	$\pm$	1.41	\\
8052.3120	&	-53.27	$\pm$	0.91	&	-53.50	$\pm$	1.10	&	-50.35	$\pm$	1.07	&	-50.65	$\pm$	0.89	&	-50.67	$\pm$	1.38	\\
8052.3692	&	-57.33	$\pm$	0.96	&	-52.93	$\pm$	1.04	&	-49.92	$\pm$	0.94	&	-51.71	$\pm$	0.83	&	-53.33	$\pm$	1.44	\\
8054.4068	&	-58.92	$\pm$	0.81	&	-52.71	$\pm$	0.75	&	-50.31	$\pm$	0.76	&	-52.70	$\pm$	0.59	&	-53.45	$\pm$	0.87	\\
8058.2510	&	-65.43	$\pm$	0.82	&	-62.68	$\pm$	0.88	&	-58.21	$\pm$	1.07	&	-64.66	$\pm$	1.02	&	-69.12	$\pm$	1.42	\\
8058.3437	&	-60.29	$\pm$	0.78	&	-62.82	$\pm$	1.24	&	-61.02	$\pm$	1.06	&	-62.34	$\pm$	1.05	&	-60.70	$\pm$	1.58	\\
8059.3636	&	-61.54	$\pm$	0.74	&	-56.23	$\pm$	0.56	&	-53.77	$\pm$	0.69	&	-56.82	$\pm$	0.57	&	-58.45	$\pm$	0.94	\\
\end{tabular}
\end{minipage}
\end{table*}
\begin{table*}
\centering
\begin{minipage}{180mm}
\contcaption{Radial Velocity Measurements \label{radialvels}}
\begin{tabular}{lccccc}
\hline
HJD-2450000 &   He I 4471   &   He I 4721   & He I 4921 & He I 5876 &   Mg II 4481 \\
\hline
8352.4421	&	-50.81	$\pm$	0.60	&	-54.42	$\pm$	1.06	&		\ldots		&	-45.54	$\pm$	1.13	&	-51.53	$\pm$	1.11	\\
8352.5396	&	-46.92	$\pm$	1.69	&	-52.17	$\pm$	1.55	&	-48.07	$\pm$	1.53	&	-45.26	$\pm$	1.39	&	-63.12	$\pm$	2.15	\\
8353.4969	&	-60.32	$\pm$	0.69	&	-63.61	$\pm$	1.30	&		\ldots		&	-62.97	$\pm$	1.15	&	-63.72	$\pm$	1.59	\\
8357.4494	&	-51.11	$\pm$	0.49	&	-53.00	$\pm$	0.87	&	-54.25	$\pm$	0.79	&	-50.70	$\pm$	0.64	&	-52.38	$\pm$	0.85	\\
8357.5413	&	-54.69	$\pm$	0.83	&	-52.82	$\pm$	0.95	&	-52.33	$\pm$	1.00	&	-50.63	$\pm$	0.69	&	-53.58	$\pm$	1.48	\\
8357.5837	&	-56.57	$\pm$	1.10	&	-49.23	$\pm$	1.02	&	-48.12	$\pm$	0.98	&	-48.97	$\pm$	0.77	&	-45.06	$\pm$	1.39	\\
8358.4607	&	-57.64	$\pm$	0.84	&	-54.81	$\pm$	0.74	&		\ldots		&	-52.86	$\pm$	0.76	&	-55.82	$\pm$	1.37	\\
\end{tabular}
\end{minipage}
\end{table*}
\begin{table*}
\centering
\begin{minipage}{180mm}
\contcaption{Radial Velocity Measurements \label{radialvels}}
\begin{tabular}{lccccc}
\hline
HJD-2450000 &   He I 4471   &   He I 4721   & He I 4921 & He I 5876 &   Mg II 4481 \\
\hline
8437.3468	&	-63.77	$\pm$	0.97	&	-55.30	$\pm$	1.05	&	-58.69	$\pm$	1.20	&	-62.00	$\pm$	0.89	&	-62.63	$\pm$	1.33	\\
8438.3249	&	-60.60	$\pm$	0.75	&	-60.70	$\pm$	0.85	&	-59.04	$\pm$	1.01	&	-56.58	$\pm$	0.86	&	-60.99	$\pm$	1.13	\\
8438.4520	&	-52.80	$\pm$	1.91	&	-56.86	$\pm$	1.78	&	-56.99	$\pm$	1.26	&	-57.27	$\pm$	1.07	&	-63.54	$\pm$	2.05	\\
8441.2979	&	-59.39	$\pm$	0.92	&	-56.51	$\pm$	0.83	&	-54.77	$\pm$	0.85	&	-53.84	$\pm$	0.77	&	-57.97	$\pm$	1.33	\\
8441.4473	&	-50.76	$\pm$	0.38	&	-54.93	$\pm$	0.56	&	-52.27	$\pm$	0.88	&	-52.98	$\pm$	0.49	&	-52.57	$\pm$	0.71	\\
8443.4107	&	-59.65	$\pm$	0.60	&	-63.28	$\pm$	1.06	&	-58.90	$\pm$	0.95	&	-62.40	$\pm$	0.55	&	-61.31	$\pm$	1.36	\\
8444.3055	&	-51.97	$\pm$	0.44	&	-55.46	$\pm$	0.93	&	-49.62	$\pm$	1.01	&	-53.77	$\pm$	0.45	&	-49.79	$\pm$	0.81	\\
\end{tabular}
\end{minipage}
\end{table*}
\begin{table*}
\centering
\begin{minipage}{180mm}
\contcaption{Radial Velocity Measurements \label{radialvels}}
\begin{tabular}{lccccc}
\hline
HJD-2450000 &   He I 4471   &   He I 4721   & He I 4921 & He I 5876 &   Mg II 4481 \\
\hline
8739.4935	&	-60.49	$\pm$	0.68	&	-56.86	$\pm$	0.64	&	-53.48	$\pm$	0.84	&	-59.61	$\pm$	0.66	&	-58.26	$\pm$	0.89	\\
8740.4619	&	-59.24	$\pm$	0.94	&	-52.40	$\pm$	0.87	&	-52.70	$\pm$	1.06	&	-54.79	$\pm$	0.93	&	-53.06	$\pm$	1.18	\\
8742.4559	&	-58.87	$\pm$	0.67	&	-53.84	$\pm$	0.67	&	-51.67	$\pm$	0.91	&	-53.43	$\pm$	0.78	&	-52.47	$\pm$	0.82	\\
8743.4694	&	-63.36	$\pm$	1.04	&	-60.62	$\pm$	0.90	&	-58.40	$\pm$	1.02	&	-61.40	$\pm$	1.13	&	-60.70	$\pm$	1.64	\\
8744.4488	&	-64.31	$\pm$	0.77	&	-60.98	$\pm$	0.66	&	-58.33	$\pm$	0.74	&	-64.26	$\pm$	0.83	&	-64.45	$\pm$	0.93	\\
8746.4304	&	-64.14	$\pm$	1.23	&	-65.46	$\pm$	1.25	&	-62.37	$\pm$	1.09	&	-63.96	$\pm$	1.26	&	-65.45	$\pm$	1.89	\\
8755.4179	&	-59.12	$\pm$	0.79	&	-52.74	$\pm$	0.67	&	-51.58	$\pm$	0.93	&	-54.99	$\pm$	0.73	&	-54.21	$\pm$	1.15	\\
\end{tabular}
\end{minipage}
\end{table*}

\begin{table*}
\centering
\begin{minipage}{180mm}
\caption{H$\alpha$ Equivalent Widths. The HJD columns are HJD$-$2,450,000. \label{H-alpha-eqw-table}}
\begin{tabular}{lll lll lll l}
\hline
HJD & $W_\lambda$ (\AA) & HJD & $W_\lambda$ (\AA) & HJD & $W_\lambda$ (\AA) & HJD & $W_\lambda$ (\AA) & HJD & $W_\lambda$ (\AA) \\
\hline
7770.5334	&	1.22	$\pm$	0.04	&	8009.4070	&	0.75	$\pm$	0.12	&	8021.4226	&	1.94	$\pm$	0.13	&	8042.3477	&	2.56	$\pm$	0.14	&	8180.6390	&	0.41	$\pm$	0.16	\\
7777.6348	&	1.51	$\pm$	0.16	&	8009.4350	&	1.00	$\pm$	0.15	&	8021.4655	&	1.99	$\pm$	0.14	&	8042.4189	&	2.64	$\pm$	0.13	&	8181.2967	&	0.11	$\pm$	0.09	\\
7802.2705	&	2.66	$\pm$	0.11	&	8009.4496	&	0.79	$\pm$	0.17	&	8021.5084	&	2.11	$\pm$	0.14	&	8043.3350	&	2.79	$\pm$	0.12	&	8181.5151	&	-0.13	$\pm$	0.11	\\
7804.2711	&	2.82	$\pm$	0.09	&	8009.4922	&	0.64	$\pm$	0.14	&	8022.4587	&	2.05	$\pm$	0.14	&	8044.2988	&	2.78	$\pm$	0.12	&	8181.5399	&	-0.22	$\pm$	0.13	\\
7811.3601	&	2.81	$\pm$	0.16	&	8009.5823	&	0.73	$\pm$	0.08	&	8023.3396	&	2.19	$\pm$	0.21	&	8044.3433	&	3.24	$\pm$	0.12	&	8181.5640	&	-0.30	$\pm$	0.10	\\
7956.5319	&	2.94	$\pm$	0.35	&	8010.0691	&	1.01	$\pm$	0.30	&	8023.3975	&	2.44	$\pm$	0.13	&	8044.3556	&	2.74	$\pm$	0.14	&	8181.5879	&	-0.33	$\pm$	0.12	\\
7962.5612	&	3.14	$\pm$	0.38	&	8010.1136	&	0.57	$\pm$	0.24	&	8024.3935	&	2.04	$\pm$	0.14	&	8044.3556	&	2.67	$\pm$	0.14	&	8182.5423	&	-0.42	$\pm$	0.14	\\
\hline
\end{tabular}
\end{minipage}
\end{table*}

\begin{table*}
\centering
\begin{minipage}{180mm}
\contcaption{H$\alpha$ Equivalent Widths. The HJD columns are HJD$-$2,450,000. \label{H-alpha-eqw-table}}
\begin{tabular}{lll lll lll l}
\hline
HJD & $W_\lambda$ (\AA) & HJD & $W_\lambda$ (\AA) & HJD & $W_\lambda$ (\AA) & HJD & $W_\lambda$ (\AA) & HJD & $W_\lambda$ (\AA) \\
\hline
8347.5657	&	0.87	$\pm$	0.16	&	8385.5020	&	2.54	$\pm$	0.11	&	8441.4463	&	0.24	$\pm$	0.13	&	8720.4803	&	3.61	$\pm$	0.14	&	8812.8332	&	0.84	$\pm$	0.09	\\
8348.4560	&	1.16	$\pm$	0.15	&	8386.4720	&	2.43	$\pm$	0.09	&	8441.4473	&	0.68	$\pm$	0.10	&	8721.5096	&	3.68	$\pm$	0.14	&	8816.2459	&	1.62	$\pm$	0.08	\\
8349.3928	&	2.39	$\pm$	0.33	&	8387.3004	&	1.64	$\pm$	0.11	&	8443.4107	&	0.71	$\pm$	0.07	&	8721.6073	&	3.91	$\pm$	0.23	&	8818.3911	&	1.63	$\pm$	0.10	\\
8349.4227	&	1.75	$\pm$	0.12	&	8387.3740	&	1.37	$\pm$	0.14	&	8444.3055	&	0.95	$\pm$	0.09	&	8722.3977	&	4.41	$\pm$	0.13	&	8818.6059	&	1.16	$\pm$	0.11	\\
8349.5208	&	1.11	$\pm$	0.10	&	8387.4332	&	2.24	$\pm$	0.09	&	8449.7260	&	1.13	$\pm$	0.10	&	8722.5356	&	3.76	$\pm$	0.19	&	8819.4005	&	1.80	$\pm$	0.09	\\
8349.5560	&	3.07	$\pm$	0.24	&	8387.4548	&	2.42	$\pm$	0.11	&	8451.3159	&	1.03	$\pm$	0.19	&	8725.4398	&	3.53	$\pm$	0.14	&	8819.5133	&	1.51	$\pm$	0.09	\\
8349.5637	&	1.99	$\pm$	0.23	&	8388.3864	&	1.71	$\pm$	0.08	&	8453.3599	&	1.19	$\pm$	0.09	&	8725.4894	&	3.95	$\pm$	0.16	&	8819.6248	&	1.22	$\pm$	0.08	\\
\hline
\end{tabular}
\end{minipage}
\end{table*}

\begin{table*}
\centering
\begin{minipage}{180mm}
\contcaption{H$\alpha$ Equivalent Widths. The HJD columns are HJD$-$2,450,000. \label{H-alpha-eqw-table}}
\begin{tabular}{lll lll lll l}
\hline
HJD & $W_\lambda$ (\AA) & HJD & $W_\lambda$ (\AA) & HJD & $W_\lambda$ (\AA) & HJD & $W_\lambda$ (\AA) & HJD & $W_\lambda$ (\AA) \\
\hline
8881.6346	&	0.87	$\pm$	0.08	&	8885.6416	&	0.97	$\pm$	0.11	&	8887.6513	&	1.52	$\pm$	0.09	&	8895.6481	&	1.93	$\pm$	0.09	&	8904.6433	&	2.26	$\pm$	0.11	\\
8884.2778	&	0.62	$\pm$	0.09	&	8886.2977	&	1.66	$\pm$	0.08	&	8888.6345	&	1.43	$\pm$	0.08	&	8897.6450	&	2.01	$\pm$	0.12	&	8906.6573	&	2.28	$\pm$	0.10	\\
8885.2973	&	1.49	$\pm$	0.08	&	8887.3368	&	1.89	$\pm$	0.08	&	8893.6390	&	1.73	$\pm$	0.08	&	8899.6473	&	2.15	$\pm$	0.16	&		&				\\
\hline
\end{tabular}
\end{minipage}
\end{table*}

\begin{table*}
\centering
\begin{minipage}{180mm}
\caption{H$\beta$ Equivalent Widths. The HJD column is HJD - 2,450,000. \label{H-beta-eqw-table}}
\begin{tabular}{lll lll lll l}
\hline
HJD & $W_\lambda$ (\AA) & HJD & $W_\lambda$ (\AA) & HJD & $W_\lambda$ (\AA) & HJD & $W_\lambda$ (\AA) & HJD & $W_\lambda$ (\AA) \\
\hline
7987.5452	&	2.63	$\pm$	0.08	&	8025.3454	&	3.61	$\pm$	0.09	&	8052.2411	&	3.72	$\pm$	0.08	&	8362.4822	&	2.86	$\pm$	0.09	&	8454.2752	&	3.28	$\pm$	0.11	\\
7988.4413	&	2.80	$\pm$	0.08	&	8025.3880	&	3.58	$\pm$	0.10	&	8052.3120	&	3.80	$\pm$	0.08	&	8363.4887	&	2.92	$\pm$	0.10	&	8457.3021	&	3.34	$\pm$	0.10	\\
7989.4361	&	2.75	$\pm$	0.07	&	8025.4309	&	3.55	$\pm$	0.09	&	8052.3692	&	3.78	$\pm$	0.07	&	8363.5459	&	2.98	$\pm$	0.13	&	8463.4456	&	3.09	$\pm$	0.06	\\
7990.4747	&	2.85	$\pm$	0.08	&	8025.4739	&	3.60	$\pm$	0.11	&	8054.4068	&	3.73	$\pm$	0.08	&	8364.5478	&	2.99	$\pm$	0.06	&	8479.2329	&	3.45	$\pm$	0.09	\\
7990.5549	&	2.97	$\pm$	0.09	&	8026.3308	&	3.53	$\pm$	0.07	&	8058.2510	&	3.81	$\pm$	0.06	&	8365.4541	&	3.03	$\pm$	0.08	&	8480.2246	&	3.58	$\pm$	0.12	\\
7990.6008	&	2.84	$\pm$	0.08	&	8026.3733	&	3.53	$\pm$	0.09	&	8058.3437	&	3.69	$\pm$	0.08	&	8366.4522	&	2.94	$\pm$	0.08	&	8480.2827	&	3.52	$\pm$	0.09	\\
7991.4147	&	2.86	$\pm$	0.08	&	8026.4026	&	3.57	$\pm$	0.06	&	8059.3636	&	3.81	$\pm$	0.07	&	8368.4562	&	3.10	$\pm$	0.06	&	8483.2416	&	3.58	$\pm$	0.10	\\
\hline
\end{tabular}
\end{minipage}
\end{table*}

\begin{table*}
\centering
\begin{minipage}{180mm}
\contcaption{H$\beta$ Equivalent Widths. The HJD column is HJD - 2,450,000. \label{H-beta-eqw-table}}
\begin{tabular}{lll lll lll l}
\hline
HJD & $W_\lambda$ (\AA) & HJD & $W_\lambda$ (\AA) & HJD & $W_\lambda$ (\AA) & HJD & $W_\lambda$ (\AA) & HJD & $W_\lambda$ (\AA) \\
\hline
8762.4578	&	3.42	$\pm$	0.09	&	8768.3466	&	3.13	$\pm$	0.08	&	8778.5186	&	2.41	$\pm$	0.07	&	8794.3721	&	2.87	$\pm$	0.11	&	8827.2875	&	3.08	$\pm$	0.07	\\
8763.4293	&	3.10	$\pm$	0.10	&	8768.3660	&	2.96	$\pm$	0.08	&	8778.6023	&	2.57	$\pm$	0.09	&	8800.4499	&	2.58	$\pm$	0.07	&	8847.2837	&	3.22	$\pm$	0.09	\\
8764.4331	&	3.28	$\pm$	0.10	&	8768.4380	&	3.13	$\pm$	0.08	&	8778.6642	&	2.49	$\pm$	0.13	&	8802.3081	&	2.77	$\pm$	0.08	&	8848.2818	&	3.34	$\pm$	0.09	\\
8766.3172	&	3.07	$\pm$	0.07	&	8769.3947	&	3.07	$\pm$	0.07	&	8782.3610	&	2.74	$\pm$	0.08	&	8808.2901	&	2.95	$\pm$	0.08	&	8865.2392	&	2.88	$\pm$	0.09	\\
8766.3652	&	3.11	$\pm$	0.08	&	8769.3668	&	2.87	$\pm$	0.06	&	8782.4243	&	2.73	$\pm$	0.09	&	8808.4162	&	2.97	$\pm$	0.10	&	8867.2999	&	2.46	$\pm$	0.07	\\
8766.4357	&	3.20	$\pm$	0.09	&	8770.3383	&	3.11	$\pm$	0.08	&	8783.3183	&	2.81	$\pm$	0.09	&	8819.4005	&	2.98	$\pm$	0.07	&	8885.2973	&	3.20	$\pm$	0.05	\\
8767.3710	&	3.07	$\pm$	0.08	&	8770.4298	&	3.10	$\pm$	0.07	&	8783.3724	&	2.69	$\pm$	0.07	&	8819.5133	&	3.19	$\pm$	0.06	&	8886.2977	&	3.21	$\pm$	0.05	\\
8767.3740	&	3.06	$\pm$	0.11	&	8776.3572	&	2.78	$\pm$	0.11	&	8783.4027	&	2.81	$\pm$	0.09	&	8820.3363	&	2.90	$\pm$	0.07	&		&				\\
8767.4553	&	3.11	$\pm$	0.09	&	8778.3498	&	2.61	$\pm$	0.07	&	8794.2807	&	2.82	$\pm$	0.13	&	8822.2635	&	3.10	$\pm$	0.08	&		&				\\
\hline
\end{tabular}
\end{minipage}
\end{table*}

\begin{table*}
\centering
\begin{minipage}{180mm}
\caption{H$\alpha$ Gaussian Parameters for Excess Emission. The HJD column is HJD - 2,450,000. \label{H-alpha-gauss}}
\begin{tabular}{lll lll ll}
\hline
HJD & FWHM (km s$^{-1}$) & Center (km s$^{-1}$) & Height (Norm. Flux) &HJD & FWHM (km s$^{-1}$) & Center (km s$^{-1}$) & Height (Norm. Flux) \\
\hline
7770.5334	&	90.79	$\pm$	0.96	&	-51.61	$\pm$	0.96	&	0.4294	$\pm$	0.0040	&	8010.0691	&	87.70	$\pm$	1.35	&	-58.03	$\pm$	1.35	&	0.5309	$\pm$	0.0071	\\
7777.6348	&	88.78	$\pm$	1.22	&	-58.97	$\pm$	1.22	&	0.3583	$\pm$	0.0042	&	8010.1136	&	88.21	$\pm$	1.13	&	-59.86	$\pm$	1.13	&	0.5458	$\pm$	0.0060	\\
7802.2705	&	76.18	$\pm$	1.08	&	-19.72	$\pm$	1.08	&	0.1705	$\pm$	0.0021	&	8010.1596	&	90.32	$\pm$	1.23	&	-59.66	$\pm$	1.23	&	0.5382	$\pm$	0.0063	\\
7987.5452	&	112.29	$\pm$	1.32	&	-82.76	$\pm$	1.32	&	0.3056	$\pm$	0.0031	&	8010.2013	&	84.41	$\pm$	1.03	&	-57.95	$\pm$	1.03	&	0.5094	$\pm$	0.0053	\\
7988.4413	&	107.97	$\pm$	1.20	&	-28.86	$\pm$	1.20	&	0.3412	$\pm$	0.0033	&	8010.2544	&	88.47	$\pm$	1.07	&	-60.07	$\pm$	1.07	&	0.5346	$\pm$	0.0057	\\
7989.4361	&	98.24	$\pm$	1.28	&	-47.13	$\pm$	1.28	&	0.3789	$\pm$	0.0043	&	8010.2962	&	91.17	$\pm$	1.15	&	-59.89	$\pm$	1.15	&	0.5278	$\pm$	0.0058	\\
7990.4747	&	110.45	$\pm$	1.18	&	-71.94	$\pm$	1.18	&	0.3688	$\pm$	0.0034	&	8010.3788	&	86.44	$\pm$	0.86	&	-63.32	$\pm$	0.86	&	0.5329	$\pm$	0.0045	\\
\hline
\end{tabular}
\end{minipage}
\end{table*}

\begin{table*}
\centering
\begin{minipage}{180mm}
\contcaption{H$\alpha$ Gaussian Parameters for Excess Emission. The HJD column is HJD - 2,450,000. \label{H-alpha-gauss}}
\begin{tabular}{lll lll ll}
\hline
HJD & FWHM (km s$^{-1}$) & Center (km s$^{-1}$) & Height (Norm. Flux) &HJD & FWHM (km s$^{-1}$) & Center (km s$^{-1}$) & Height (Norm. Flux) \\
\hline
8023.3975	&	80.70	$\pm$	1.87	&	-56.59	$\pm$	1.87	&	0.2667	$\pm$	0.0055	&	8180.6390	&	98.62	$\pm$	0.67	&	-63.28	$\pm$	0.67	&	0.5474	$\pm$	0.0032	\\
8024.3935	&	84.62	$\pm$	1.20	&	-47.65	$\pm$	1.20	&	0.2990	$\pm$	0.0036	&	8181.2967	&	102.23	$\pm$	0.65	&	-61.22	$\pm$	0.65	&	0.5463	$\pm$	0.0030	\\
8024.4651	&	82.67	$\pm$	1.06	&	-48.12	$\pm$	1.06	&	0.3019	$\pm$	0.0034	&	8181.5151	&	104.28	$\pm$	0.65	&	-49.01	$\pm$	0.65	&	0.5378	$\pm$	0.0029	\\
8025.3454	&	71.55	$\pm$	0.87	&	-40.39	$\pm$	0.87	&	0.3224	$\pm$	0.0034	&	8181.5399	&	100.77	$\pm$	0.50	&	-47.49	$\pm$	0.50	&	0.5350	$\pm$	0.0023	\\
8025.3745	&	66.27	$\pm$	1.06	&	-42.03	$\pm$	1.06	&	0.3135	$\pm$	0.0044	&	8181.5640	&	102.85	$\pm$	0.52	&	-46.98	$\pm$	0.52	&	0.5357	$\pm$	0.0023	\\
8025.3880	&	71.35	$\pm$	0.85	&	-37.83	$\pm$	0.85	&	0.3204	$\pm$	0.0033	&	8181.5879	&	103.40	$\pm$	0.50	&	-45.86	$\pm$	0.50	&	0.5331	$\pm$	0.0022	\\
8025.4309	&	72.41	$\pm$	0.83	&	-39.87	$\pm$	0.83	&	0.3152	$\pm$	0.0031	&	8182.5423	&	104.44	$\pm$	0.60	&	-33.70	$\pm$	0.60	&	0.5236	$\pm$	0.0026	\\
\hline
\end{tabular}
\end{minipage}
\end{table*}

\begin{table*}
\centering
\begin{minipage}{180mm}
\contcaption{H$\alpha$ Gaussian Parameters for Excess Emission. The HJD column is HJD - 2,450,000. \label{H-alpha-gauss}}
\begin{tabular}{lll lll ll}
\hline
HJD & FWHM (km s$^{-1}$) & Center (km s$^{-1}$) & Height (Norm. Flux) &HJD & FWHM (km s$^{-1}$) & Center (km s$^{-1}$) & Height (Norm. Flux) \\
\hline
8363.4887	&	112.55	$\pm$	1.09	&	-56.35	$\pm$	1.09	&	0.4968	$\pm$	0.0042	&	8411.3838	&	96.49	$\pm$	1.29	&	-21.23	$\pm$	1.29	&	0.4171	$\pm$	0.0049	\\
8363.5459	&	112.85	$\pm$	1.09	&	-58.10	$\pm$	1.09	&	0.4867	$\pm$	0.0040	&	8411.4720	&	92.56	$\pm$	0.97	&	-21.61	$\pm$	0.97	&	0.4178	$\pm$	0.0038	\\
8364.4272	&	118.51	$\pm$	1.50	&	-40.44	$\pm$	1.50	&	0.5119	$\pm$	0.0056	&	8412.3715	&	107.74	$\pm$	1.35	&	-46.92	$\pm$	1.35	&	0.4249	$\pm$	0.0046	\\
8364.5478	&	100.07	$\pm$	0.53	&	-44.82	$\pm$	0.53	&	0.4773	$\pm$	0.0022	&	8412.4934	&	100.16	$\pm$	0.75	&	-52.13	$\pm$	0.75	&	0.3953	$\pm$	0.0026	\\
8365.4541	&	106.45	$\pm$	0.93	&	-46.99	$\pm$	0.93	&	0.4833	$\pm$	0.0036	&	8413.4110	&	96.75	$\pm$	1.09	&	-50.58	$\pm$	1.09	&	0.4637	$\pm$	0.0045	\\
8366.4522	&	107.63	$\pm$	0.94	&	-52.43	$\pm$	0.94	&	0.4676	$\pm$	0.0036	&	8415.2834	&	102.59	$\pm$	1.22	&	-54.02	$\pm$	1.22	&	0.4973	$\pm$	0.0051	\\
8368.4562	&	94.63	$\pm$	0.77	&	-44.10	$\pm$	0.77	&	0.4958	$\pm$	0.0035	&	8416.4004	&	98.64	$\pm$	0.96	&	-58.76	$\pm$	0.96	&	0.5017	$\pm$	0.0042	\\
\hline
\end{tabular}
\end{minipage}
\end{table*}
\begin{table*}
\centering
\begin{minipage}{180mm}
\contcaption{H$\alpha$ Gaussian Parameters for Excess Emission. The HJD column is HJD - 2,450,000. \label{H-alpha-gauss}}
\begin{tabular}{lll lll ll}
\hline
HJD & FWHM (km s$^{-1}$) & Center (km s$^{-1}$) & Height (Norm. Flux) &HJD & FWHM (km s$^{-1}$) & Center (km s$^{-1}$) & Height (Norm. Flux) \\
\hline
8772.8639	&	92.95	$\pm$	0.87	&	-52.88	$\pm$	0.87	&	0.5580	$\pm$	0.0045	&	8848.2818	&	79.60	$\pm$	0.62	&	-38.90	$\pm$	0.62	&	0.3463	$\pm$	0.0023	\\
8772.9158	&	92.50	$\pm$	0.82	&	-51.86	$\pm$	0.82	&	0.5670	$\pm$	0.0043	&	8848.3183	&	77.83	$\pm$	0.73	&	-36.81	$\pm$	0.73	&	0.3586	$\pm$	0.0029	\\
8773.6991	&	82.31	$\pm$	0.61	&	-46.58	$\pm$	0.61	&	0.5716	$\pm$	0.0036	&	8849.2524	&	87.28	$\pm$	0.84	&	-75.06	$\pm$	0.84	&	0.3437	$\pm$	0.0029	\\
8774.6085	&	97.70	$\pm$	1.01	&	-56.33	$\pm$	1.01	&	0.5268	$\pm$	0.0047	&	8849.3407	&	84.72	$\pm$	0.67	&	-70.49	$\pm$	0.67	&	0.3464	$\pm$	0.0024	\\
8774.6842	&	94.22	$\pm$	0.61	&	-55.84	$\pm$	0.61	&	0.5386	$\pm$	0.0030	&	8849.6030	&	77.59	$\pm$	0.56	&	-67.39	$\pm$	0.56	&	0.3581	$\pm$	0.0022	\\
8775.6183	&	97.63	$\pm$	0.76	&	-46.97	$\pm$	0.76	&	0.5788	$\pm$	0.0039	&	8852.6931	&	86.04	$\pm$	0.82	&	-50.43	$\pm$	0.82	&	0.4181	$\pm$	0.0035	\\
8777.6303	&	91.71	$\pm$	0.80	&	-56.52	$\pm$	0.80	&	0.6565	$\pm$	0.0049	&	8855.3400	&	87.51	$\pm$	0.54	&	-63.69	$\pm$	0.54	&	0.4413	$\pm$	0.0023	\\
\hline
\end{tabular}
\end{minipage}
\end{table*}

\begin{table*}
\centering
\begin{minipage}{180mm}
\caption{H$\beta$ Gaussian Parameters for Excess Emission. The HJD column is HJD - 2,450,000. \label{H-beta-gauss}}
\begin{tabular}{lll lll ll}
\hline
HJD & FWHM (km s$^{-1}$) & Center (km s$^{-1}$) & Height (Norm. Flux) &HJD & FWHM (km s$^{-1}$) & Center (km s$^{-1}$) & Height (Norm. Flux) \\
\hline
7987.5452	&	107.95	$\pm$	4.20	&	-103.93	$\pm$	4.20	&	0.1232	$\pm$	0.0043	&	8026.3733	&	30.18	$\pm$	2.21	&	-44.47	$\pm$	2.21	&	0.0466	$\pm$	0.0031	\\
7988.4413	&	119.90	$\pm$	2.73	&	-12.40	$\pm$	2.73	&	0.1075	$\pm$	0.0021	&	8026.4026	&	23.57	$\pm$	1.43	&	-53.75	$\pm$	1.43	&	0.0524	$\pm$	0.0027	\\
7989.4361	&	139.79	$\pm$	3.78	&	-42.80	$\pm$	3.78	&	0.1092	$\pm$	0.0026	&	8026.4160	&	31.84	$\pm$	2.31	&	-46.20	$\pm$	2.31	&	0.0432	$\pm$	0.0028	\\
7990.4747	&	108.54	$\pm$	2.39	&	-81.45	$\pm$	2.39	&	0.1279	$\pm$	0.0024	&	8032.3226	&	22.51	$\pm$	1.92	&	-31.91	$\pm$	1.92	&	0.0449	$\pm$	0.0027	\\
7990.5549	&	106.35	$\pm$	2.32	&	-77.11	$\pm$	2.32	&	0.1315	$\pm$	0.0025	&	8033.3153	&	4.89	$\pm$	1.00	&	-310.56	$\pm$	1.01	&	-0.0370	$\pm$	0.0063	\\
7990.6008	&	119.97	$\pm$	3.10	&	-72.90	$\pm$	3.10	&	0.1277	$\pm$	0.0028	&	8036.2965	&	17.23	$\pm$	1.71	&	-55.31	$\pm$	1.71	&	0.0370	$\pm$	0.0035	\\
7991.4147	&	112.91	$\pm$	1.91	&	-27.30	$\pm$	1.91	&	0.1401	$\pm$	0.0020	&	8037.2970	&	3.00	$\pm$	1.41	&	-307.36	$\pm$	1.41	&	-0.0423	$\pm$	0.0067	\\
\hline
\end{tabular}
\end{minipage}
\end{table*}

\begin{table*}
\centering
\begin{minipage}{180mm}
\contcaption{H$\beta$ Gaussian Parameters for Excess Emission. The HJD column is HJD - 2,450,000. \label{H-beta-gauss}}
\begin{tabular}{lll lll ll}
\hline
HJD & FWHM (km s$^{-1}$) & Center (km s$^{-1}$) & Height (Norm. Flux) &HJD & FWHM (km s$^{-1}$) & Center (km s$^{-1}$) & Height (Norm. Flux) \\
\hline
8437.3468	&	87.24	$\pm$	1.12	&	-37.61	$\pm$	1.12	&	0.1880	$\pm$	0.0021	&	8768.4380	&	97.33	$\pm$	1.61	&	-55.57	$\pm$	1.61	&	0.1129	$\pm$	0.0016	\\
8438.3249	&	95.88	$\pm$	1.54	&	-52.19	$\pm$	1.54	&	0.1702	$\pm$	0.0024	&	8769.3668	&	87.70	$\pm$	1.97	&	-70.35	$\pm$	1.97	&	0.1525	$\pm$	0.0029	\\
8438.4520	&	100.96	$\pm$	2.48	&	-42.19	$\pm$	2.48	&	0.1784	$\pm$	0.0038	&	8769.3947	&	77.55	$\pm$	0.95	&	-66.21	$\pm$	0.95	&	0.1516	$\pm$	0.0016	\\
8441.2979	&	92.03	$\pm$	1.38	&	-54.03	$\pm$	1.38	&	0.1713	$\pm$	0.0022	&	8770.3383	&	75.15	$\pm$	0.91	&	-26.44	$\pm$	0.91	&	0.1575	$\pm$	0.0016	\\
8441.4473	&	107.27	$\pm$	2.16	&	-47.56	$\pm$	2.16	&	0.1749	$\pm$	0.0030	&	8770.4298	&	73.47	$\pm$	0.88	&	-22.98	$\pm$	0.88	&	0.1621	$\pm$	0.0017	\\
8443.4107	&	105.18	$\pm$	2.78	&	-58.25	$\pm$	2.78	&	0.1644	$\pm$	0.0038	&	8776.3572	&	80.54	$\pm$	1.23	&	-37.84	$\pm$	1.23	&	0.2202	$\pm$	0.0029	\\
8444.3055	&	96.90	$\pm$	2.17	&	-45.77	$\pm$	2.17	&	0.1736	$\pm$	0.0033	&	8778.3498	&	97.55	$\pm$	1.46	&	-41.55	$\pm$	1.46	&	0.2151	$\pm$	0.0028	\\
\hline
\end{tabular}
\end{minipage}
\end{table*}

\bsp	
\label{lastpage}
\end{document}